\documentclass[reqno, 11pt]{amsart}

\usepackage[english]{babel}
\usepackage[T1]{fontenc}
\usepackage[utf8]{inputenc}

\usepackage{amsfonts,amssymb,amsbsy,amsmath,amsthm,enumerate}

\usepackage{epigraph}

\usepackage{hyperref}
\usepackage{eucal}
\usepackage{color}
\usepackage{mathrsfs}
\usepackage[small,nohug,
]{diagrams}
\diagramstyle[labelstyle=\scriptstyle]
\newarrow{Corresponds}<--->

\usepackage{lxfonts}

\theoremstyle{plain}
\newtheorem{theorem}{Theorem}[section]
\theoremstyle{plain}
\newtheorem{corollary}{Corollary}[section]
\theoremstyle{plain}
\newtheorem{proposition}{Proposition}[section]
\theoremstyle{plain}
\newtheorem{lemma}{Lemma}[section]
\theoremstyle{definition}
\newtheorem{definition}{Definition}[section]
\theoremstyle{definition}

\theoremstyle{definition}

\theoremstyle{definition}

\theoremstyle{definition}

\theoremstyle{definition}
\newtheorem{remark}{Remark}[section]

\numberwithin{equation}{section}
\numberwithin{figure}{section}
\numberwithin{table}{section}

\newcommand{\R}{\mathbb{R}}
\newcommand{\N}{\mathbb{N}}
\newcommand{\C}{\mathbb{C}}                           

\newcommand{\Z}{\mathbb{Z}}
\newcommand{\T}{\mathbb{T}}

\newcommand{\s}[1]{\CMcal{#1}}
\newcommand{\bb}[1]{\mathscr{#1}}
\newcommand{\rr}[1]{\mathfrak{#1}}
\newcommand{\n}[1]{\mathbb{#1}}

\newcommand{\ketbra}[2]{|#1\rangle\langle#2|}

\newcommand{\expo}[1]{\,\mathrm{e}^{#1}\,}                 
\newcommand{\dd}{\,\mathrm{d}}
\newcommand{\ii}{\,\mathrm{i}\,}

\newcommand{\virg}[1]{\lq\lq#1\rq\rq}                
\newcommand{\ie}{\textsl{i.\,e.\,}}
\newcommand{\eg}{\textsl{e.\,g.\,}}
\newcommand{\cf}{\textsl{cf}.\,}

\title[The Geometry of (non-Abelian) Landau levels]{The Geometry of (non-Abelian) Landau levels}

\author[G. De~Nittis]{Giuseppe De Nittis}

\address[G. De~Nittis]{Facultad de Matem\'aticas \& Instituto de F\'{\i}sica,
  Pontificia Universidad Cat\'olica de Chile,
  Santiago, Chile.}
\email{gidenittis@mat.uc.cl}

\author[K. Gomi]{Kyonori Gomi}

\address[K. Gomi]{Department of Mathematics, Tokyo Institute of Technology,
2-12-1 Ookayama, Meguro-ku, Tokyo, 152-8551, Japan.}
\email{kgomi@math.titech.ac.jp}

\author[M. Moscolari]{Massimo Moscolari}

\address[M. Moscolari]{Department of Mathematical Sciences, Aalborg University, Skjernvej 4A, 9220 Aalborg, Denmark}

\email{massimomoscolari@math.aau.dk}


\thanks{
\virg{Real} and \virg{Quaternionic} vector bundles,  Dixmier trace.}
\begin{document}

\begin{abstract}
The purpose of this paper is threefold: First of all the topological aspects of the  Landau Hamiltonian 
are reviewed  in the light  (and with the jargon) of theory  of topological insulators.
In particular it is shown that the Landau Hamiltonian has a generalized even time-reversal symmetry (TRS). Secondly, a new tool for the 
computation of the topological numbers associated with each Landau level is introduced. The latter is obtained by combining the Dixmier trace and the (resolvent of the) harmonic oscillator. Finally,  these results are extended to models with non-Abelian magnetic fields. Two models are investigated in details: the Jaynes-Cummings model and the \virg{Quaternionic} model.
\end{abstract}

\maketitle

\vspace{-5mm}
\tableofcontents

\epigraph{{\itshape{``Que vivan los estudiantes\\
Jard\'{\i}n de nuestra alegr\'{\i}a\\
Son aves que no se asustan\\
De animal ni polic\'{\i}a.''}}}{\textit{Violeta Parra}}

\section{Introduction}\label{sect:intro}

The quantum motion of a two-dimensional electron subjected to a perpendicular and uniform magnetic field $B$ is  determined by the differential operator
\begin{equation}\label{eq:intro_001}
H_B\;:=\;\frac{\epsilon_B}{2}\left[\left(-{\ii }\ell_B\frac{\partial}{\partial x_1}-\frac{x_2}{2\ell_B}\right)^2+\left(-{\ii }\ell_B \frac{\partial}{\partial x_2}+\frac{x_1}{2\ell_B}\right)^2\right]
\end{equation}
where $\epsilon_B\propto B$ fixes the magnetic energy scale and $\ell_B\propto B^{-\frac{1}{2}}$ defines the typical magnetic length.
The equation \eqref{eq:intro_001}
defines a self-adjoint operator on the (position) Hilbert space $L^2(\R^2)$. The operator $H_B$  is known as  \emph{Landau Hamiltonian.}
The spectral theory of  $H_B$ is  well established since the dawn of Quantum Mechanics
\cite{fock-28,landau-30} especially for its connection with the elementary theory
of the harmonic oscillator (see also \cite{avron-herbst-simon-78} for a more modern point of view). The spectrum of $H_B$ is pure point and is given by
$$
\sigma(H_B)\;=\;\left.\left\{E_{j}:=\epsilon_B\left(j+\frac{1}{2}\right)\ \right| \ j\in\N_0\right\}
$$
where $\N_0:=\N\cup\{0\}$.
The energy $E_j$ is usually called  the $j$-th Landau level. Each Landau level is infinitely degenerate due to the high symmetry of $H_B$.
Indeed, the Landau Hamiltonian commutes with the perpendicular component of the angular momentum 
$L_3$ and with the generators of the magnetic translations $G_1$ and $G_2$ (see Section \ref{subsec:LH}). The infinite dimensional eigenspace associated to each Landau level $E_j$ is completely characterized by the related spectral projection $\Pi_j$, called the $j$-th Landau projection. The topology of the Landau Hamiltonian is encoded exactly in the Landau projections.

\medskip

As it is well known, the magnetic field breaks the time reversal symmetry (TRS). In Quantum Mechanics the \virg{standard} time reversal symmetry is implemented by the complex conjugation $C$ and a simple computation shows that $CH_BC=H_{-B}$.
In the context of the classification scheme for topological insulators \cite{altland-zirnbauer-97,schnyder-ryu-furusaki-ludwig-08,kitaev-09,ryu-schnyder-furusaki-ludwig-10} the Landau Hamiltonian  is usually considered as the prototype model for \emph{class A} systems. The latter are the systems which break all the fundamental (pseudo-)symmetries.
As a consequence, the topological phases of $H_B$ are predicted to be labelled by integers according to the celebrated periodic table for topological insulators. From a physical point of view these topological invariants are associated with the quantized values of the transverse Hall conductivity \cite{thouless-kohmoto-nightingale-nijs-82,bellissard-elst-schulz-baldes-94}. On the other hand they are  mathematically associated with the Chern numbers of suitable vector bundles \cite{dubrovin-novikov-80-1,dubrovin-novikov-80-2,novikov-81,lyskova-85} that are constructed from the  Landau projections $\Pi_j$ by exploiting the
invariance of 
$H_B$ under the magnetic translations \cite{zak1,zak2}.
Indeed the invariance under the $\Z^2$-action induced by the magnetic translations  allows to define a magnetic version of the Bloch-Floquet-Zak transform \cite{kuchment-93}
which  decomposes $H_B$, along with all its spectral functions,  into a fibered operator over the Brillouin (two dimensional) torus $\n{B}_B\simeq\n{T}^2:=\R^2/\Z^2$ (see Section \ref{subsect:magn_BFZ_tras} for more details). At level of the projections $\Pi_j$ this procedure defines  continuous maps $\n{T}^2\ni k\mapsto \Pi_j(k)$  which provide   complex vector bundles $\bb{E}_{j}\to\n{T}^2$. 
The construction of these (so called) \emph{spectral bundles}  is classical and 
depicts an incarnation of the Serre-Swan duality \cite{serre-55,swan-62}. There is a rich literature devoted to a rigorous definition of the spectral bundles (see \eg \cite{panati-07} or \cite[Lemma 4.5]{denittis-lein-11} or \cite[Section 2]{denittis-gomi-14})
and the main aspects  of the construction will be summarized in Section \ref{subsect:VB-R-landau}. The crucial point we want to emphasize  here
 is that the topology of the Landau levels $E_j$ is encoded in the spectral bundles 
$\bb{E}_{j}$ associated with the Landau projections $\Pi_j$. In absence of extra symmetries these vector bundles must be classified in the category of complex vector bundles over the two-dimensional torus $\n{T}^2$ and a  classical result by Peterson \cite{peterson-59} provides the isomorphism
\begin{equation}\label{eq:intro_01}
{\rm Vec}^m_\C\big(\n{T}^2\big) \;\stackrel{c_1}{\simeq}\;H^2\big(\n{T}^2,\Z\big)\;\simeq\;\Z 
\end{equation}
where ${\rm Vec}^m_\C(\n{T}^2)$ is the set of equivalence classes of rank $m$ complex vector bundles over $\n{T}^2$, $H^2(\n{T}^2,\Z)$
is the second cohomology group of $\n{T}^2$ (with integer coefficients) and the map $c_1$ is the (first) Chern class.

\medskip

The chain of isomorphisms \eqref{eq:intro_01}, that
provides the vector bundle interpretation of the topological classification of the Landau projections $\Pi_j$, is justified on the basis of the absence of further fundamental symmetries. Nevertheless, a deeper look at the structure of the  operator $H_B$ shows that it is not totally correct to claim that the Landau Hamiltonian  does not have (pseudo-)symmetries. 
To this end we need the following notion:
\begin{definition}[Generalized TRS]\label{def:introG-TRS}
Let $\s{H}$ be a complex Hilbert space endowed with an anti-unitary involution $C$ which defines a complex structure. A (non necessary bounded) linear operator $H$ has a \emph{generalized} time reversal symmetry (TRS) if there is an auxiliary unitary operator $F$ such that  
$$
\Theta\;H\;\Theta^{-1}\;=\;H\;,\qquad\quad \Theta\;:=\;F  C\;.
$$
The symmetry $\Theta$ is called \emph{\virg{Real}} (or even) if $\Theta^2=+{\bf1}$ and \emph{\virg{Quaternionic}} (or odd) if $\Theta^2=-{\bf 1}$. 
\end{definition}
\medskip

\noindent
{At the algebraic level a generalized symmetry $\Theta$ of the type described in Definition \ref{def:introG-TRS}
implements the inversion of the time evolution generated by the operator $H$, and for this reason it is fair to refer to  $\Theta$ as a TRS. Indeed, this is mainly due to the anti-linearity of $\Theta$ in accordance with the Wigner's theorem \cite{wigner-59}. Interestingly,
the Wigner's theorem point of view has been used in various recent works \cite{freed-moore-13,thiang-15,kubota-17} to generalize and \virg{algebrize} the classification program for topological insulators to  abstract Hilbert spaces  endowed with suitable group actions. Within such  algebraic framework,  Definition \ref{def:introG-TRS} provides the natural formulation for a \virg{genuine} TRS.  
It is also worth mentioning that the generalized structure of the TRS presented in 
Definition \ref{def:introG-TRS} has already been considered since the first works on the classification of topological insulators (see \eg \cite[p. 3]{schnyder-ryu-furusaki-ludwig-08} or \cite[eq. (3)]{ryu-schnyder-furusaki-ludwig-10}). On the other hand, in application to model  of  physical inspiration \cite{altland-zirnbauer-97,schnyder-ryu-furusaki-ludwig-08,ryu-schnyder-furusaki-ludwig-10} the choice of a \virg{physical} TRS is often subordinate to (usually undeclared) extra conditions like the invariance of the position operators under the time inversion. However, these additional constraints are not usually introduced in the abstract algebraic setting, especially when the notion of spatial coordinates is not definable from the context. This fact opens a space of discussion  about  the best possible definition of TRS in the context of topological insulators. We do not pretend to solve this semantic and/or philosophic diatribe in this paper. And, independently of the terminology used, it is a fact that the Landau Hamiltonian admits  a generalized symmetry of the type described in Definition \ref{def:introG-TRS}.}
\begin{proposition}\label{prop:intro_01}
The Landau Hamiltonian $H_B$  admits a  generalized TRS of \virg{Real} type implemented  by $\Theta=FC$ where $F$ is the flip operator defined by
$$
(F\psi)(x_1,x_2)\;=\;\psi(x_2,x_1)\;,\qquad\quad\psi\in L^2(\R^2)
$$ 
and $C$ is the usual complex conjugation $C\psi=\overline{\psi}$.
\end{proposition}
\medskip

\noindent 
The proof of Proposition \ref{prop:intro_01} will be discussed in Section \ref{subsec:sym-LH}. Even though the proof of the latter result does not present major complications,  it deserves some considerations. 
As already mentioned, the Landau Hamiltonian $H_B$ is usually considered as the prototypical example for topological insulators in class A, that is
the class of systems that break all the fundamental symmetries, and two dimensional systems in this class have $\Z$ distinguished topological phases
(interpreted as Chern numbers). 
{
However, such classification is based 
on the (tacit!) assumption that the complex conjugation provides 
the natural, and the unique, realization for the TRS.
On the other hand,  Proposition \ref{prop:intro_01} tells us that the Landau Hamiltonian $H_B$ possesses the generalized \virg{Real} TRS  $\Theta$, hence it may be considered as   a topological system of \emph{class AI} (according to
the accepted nomenclature) in the classification scheme that chooses $\Theta$ as fundamental symmetry. 
This fact generates an apparent contradiction, since 2-dimensional systems in class AI are predicted to exist  only in the topologically
trivial phase \cite{schnyder-ryu-furusaki-ludwig-08,kitaev-09,ryu-schnyder-furusaki-ludwig-10}. Even though it may seem that this contradiction exists only at a level of use (or abuse) of terminology, it inspires a related question: Is the symmetry $\Theta$ responsible for a different, maybe finer, topological classification of the Landau Hamiltonian?
The answer to the latter question is the  first  main result of this paper (Theorem \ref{theo:main1}) and it is obtained by
an accurate analysis of the correct category of vector 
bundles underlying the Landau Hamiltonian. As a payoff, we get also the reconciliation of the (apparent) contradiction stated above.
Our analysis aims also to point out the importance of the (a priori) choice of the fixed fundamental symmetries used for the classification of topological systems. 
Without extra constraints (\eg the action of the symmetries on a bunch of special operators), 
different choices of fundamental symmetries 
are possible  and appropriate choices
can lead  in principle to a   finer classification tables.}
Before  passing to the description of the first main result,  two brief final  comments are necessary: First of all the  TRS $\Theta$ in 
Proposition \ref{prop:intro_01} is pretty \virg{fragile} in the sense that it can be broken  by the effect of a generic background potential (\cf Remark \ref{rk:top-class2}); Secondly, the study of topological systems with generalized symmetries of the type  of $\Theta$  can be included in the more general analysis
of topological systems with \emph{point group symmetries} (see \cite{gomi-13,cornfeld-chapman-19,G-T} and reference therein).

\medskip

The operator $\Theta$ described in Proposition \ref{prop:intro_01} combines together with the 
Bloch-Floquet-Zak transform and endows the  spectral bundles 
$\bb{E}_{j}$ associated with the projections $\Pi_j$ with an additional \virg{Real} structure. 
In a nutshell, this means that the vector bundle $\bb{E}_{j}\to \T^2$ acquires an involutive and fiberwise anti-linera map on the total space 
$\Theta:\bb{E}_{j}\to\bb{E}_{j}$ which covers an involution $\rr{f}:\T^2\to\T^2$ on the base space. The notion of \virg{Real} vector bundle has been introduced for the first time in \cite{atiyah-66}. The construction of  \virg{Real} vector bundles from gapped systems with an even TRS is explained in detail in \cite{denittis-gomi-14} (see also Section \ref{subsect:VB-R-landau} for a more concise description)  and the topological (cohomology based) classification of these structures has been investigated in \cite{denittis-gomi-14,denittis-gomi-15}. In order to properly describe the type of \virg{Real} vector bundle associated with each pair $(\Pi_j,\Theta)$ one needs to specify the type of involution that $\Theta$ induces on the Brillouin torus $\n{B}_B\simeq\n{T}^2$. It turns out that (see Section \ref{subsect:VB-R-landau})  the 
relevant  involution  is the following:
\begin{definition}[Flip involution]\label{def_FI}
Let $\n{T}^{2}$ be a two dimensional torus with coordinates $(k_1,k_2)$. The \emph{flip involution} is the involutive homeomorphism  
$\rr{f}:\n{T}^{2}\to \n{T}^{2}$ defined by
$\rr{f}:(k_1,k_2)\mapsto (-k_2,-k_1)$. The pair
$(\n{T}^2,\rr{f})$ defines an \emph{involutive space}
with 
 a non-empty fixed-point set $(\n{T}^2)^{\rr{f}}:=\{k\in\n{T}^2\ |\ \rr{f}(k)=k\}
 \simeq \n{S}^1
 $   made by anti-diagonal points  $(k,-k)$. 
 \end{definition}

\medskip

In summary, each pair $(\Pi_j,\Theta)$ defines a 
\virg{Real} vector bundle $(\bb{E}_{j}, \Theta)$ over the involutive space $(\n{T}^2,\rr{f})$. 
The latter claim 
is proved in detail in Proposition \ref{prop:LL-VB}.
Let ${\rm Vec}^m_{\rr{R}}(\n{T}^2,\rr{f})$
be the set of isomorphism classes of 
rank $m$ \virg{Real} vector bundles over $(\n{T}^2,\rr{f})$. The topological classification of $(\Pi_j,\Theta)$ amounts to the classification of ${\rm Vec}^m_{\rr{R}}(\n{T}^2,\rr{f})$ and the latter boils down to the following chain of isomorphisms
\begin{equation}\label{eq:intro_kan_iso}
{\rm Vec}^1_{\rr{R}}\big(\n{T}^2,\rr{f}\big)\;\stackrel{c_1^{\rr{R}}}{\simeq}\;
H^2_{\Z^2}\big(\n{T}^2,\Z(1)\big)\;\stackrel{\imath}{\simeq}\;H^2\big(\n{T}^2,\Z\big)\;.
\end{equation}
In the first isomorphism,
  $H^2_{\Z^2}(\n{T}^2,\Z(1))$ is the second equivariant cohomology group of the involutive space $(\n{T}^2,\rr{f})$ with local system of coefficients $\Z(1)$ (\cf \cite[Section 5.1]{denittis-gomi-14} and references therein for a brief introduction to the equivariant cohomology)
  and the map $c_1^{\rr{R}}$  is called (first) \virg{Real} Chern class. This first
isomorphism is known as Kahn's isomorphism and has been proved in \cite{kahn-59}; see also 
\cite[Section 5.2]{denittis-gomi-14} and \cite[Appendix A]{gomi-13}. 
The second isomorphism in \eqref{eq:intro_kan_iso} is the content of Proposition \ref{prop:cohom-R}.
in this case
 $\imath$ is the map which  forgets the $\Z_2$ action induced by the  involution $\rr{f}$. Since  $c_1=\imath\circ c_1^{\rr{R}}$ coincides with the (usual) Chern class it follows from \eqref{eq:intro_kan_iso} that the \virg{Real} \emph{line bundles}\footnote{Line bundle is used as a synonym for rank-one vector bundle.} 
over $(\n{T}^2,\rr{f})$ are completely specified by the Chern class of the underlying complex line  bundle.
This fact is preparatory to present the first main result of this paper.
\begin{theorem}\label{theo:main1}
At each Landau level $E_j$ of the Hamiltonian $H_B$ it is associated a \virg{Real} line bundle $(\bb{E}_{j},\Theta)$ over the involutive (Brillouin) torus $(\n{T}^2_B,\rr{f})$. The topology of this line bundle  is completely classified by its Chern class $c_1(\bb{E}_{j})\in\Z$. It turns out that
$$
c_1(\bb{E}_{j})\;=\;1\;,\qquad\quad \forall\; j\in\N_0
$$
implying accordingly that all the Landau levels are topologically equivalent.
\end{theorem}

\medskip
\noindent
A complete proof of Theorem \ref{theo:main1} will be presented in Section \ref{subsect:top_class_LL}. Let us point out here that Theorem \ref{theo:main1}  adds two new information to the isomorphisms \eqref{eq:intro_kan_iso}: First of all the rank  of the (complex) vector bundles $\bb{E}_{j}$ is  ${\rm rk}(\bb{E}_{j})=1$; Secondly the Chern number associated to each vector bundle $\bb{E}_{j}$ can be computed and  is $c_1(\bb{E}_{j})=1$. These information can be extracted from the Bloch-Floquet-Zak representation of the Landau projections $\Pi_j$, but it is often useful (especially for generalizations beyond the periodic case) to have available formulas that allow the computation of the same quantities in the position space $L^2(\R^2)$ directly in terms of the projections $\Pi_j$.
The existing formulas of this type involve the use of the \emph{trace per unit volume} (see Section \ref{subsect:ttrace_unit_N}). However, it is also possible to compute the same quantities in a more (noncommutative) \virg{geometric} way by using the \emph{Dixmier trace} (see Appendix  \ref{sec:Limit_formula}): This is the second main result of this work.

\medskip

Let us consider the harmonic oscillator
\begin{equation}\label{eq:landau_harmonic_int}
Q_B\;=\;\sum_{j=1}^2\left(-\ell_B^2\frac{\partial^2}{\partial x_j^2}+\frac{1}{4}\frac{x_j^2}{\ell_B^2}\right)\;.
\end{equation}
The relation between $Q_B$ and the Landau operator $H_B$, made manifest by the presence of the magnetic length $\ell_B^2$, will be clarified in Section \ref{subsect:ttrace_unit_N}.
The operator \eqref{eq:landau_harmonic_int} has a pure point spectrum given by
$$
\sigma(Q_B)\;=\;\big\{\lambda_j:=j+2\ |\ j\in\N_0\big\}
$$
and each eigenvalue has finite degeneracy $\text{Mult}[\lambda_j]=j+1$. 
The resolvent
\begin{equation}\label{eq:Q_op2_int}
Q_{B,\xi}^{-1}\;:=\;(Q_B+2\xi{\bf 1})^{-1}
\end{equation}
is certainly well defined for  $\xi\geqslant 0$ and is a compact operator. It turns out that (Corollary \ref{corol:rank} and Corollary \ref{corol:chern})
\begin{equation}\label{eq:top_dix_intro}
\begin{aligned}
{\rm rk}(\bb{E}_{j})\;&=\;{\rm Tr}_{\rm Dix}\big(Q_{B,\xi}^{-1}\Pi_j\big)\\
c_1(\bb{E}_{j})\;&=\; \frac{\ii}{\ell_B^2}{\rm Tr}_{\rm Dix}\big(Q_{B,\xi}^{-1}\Pi_j\big[\partial_1(\Pi_j),\partial_2(\Pi_j)\big]\big)
\end{aligned}
\end{equation}
where ${\rm Tr}_{\rm Dix}$ denotes the Dixmier trace and $\partial_i(A)$ is a shorthand for the commutators (when defined)
$$
\partial_i(A)\;:=\;-\ii\big[X_i,A\big]\;,\qquad\quad i=1,2
$$
where $X_1$ and $X_2$ are the position operators. The Dixmier trace is the prototype of a singular (non-normal) trace. It was introduced for the first time by Dixmier in \cite{dixmier-66}. The theory of Dixmier trace is reviewed in Appendix \ref{sec:Limit_formula} along with a list of selected references.
Formulas \eqref{eq:top_dix_intro} provide the link between the topology of the spectral bundle associated with the Landau projection $\Pi_j$ and some \virg{numerical indexes} of $\Pi_j$  calculated 
directly in the position space through the functional $T\mapsto{\rm Tr}_{\rm Dix}(Q_{B,\xi}^{-1}T)$. These equations are indeed obtained as a special application of the following more general result:
\begin{theorem}\label{teo:Dix_Tr_for_main2_intro}
For all $T\in\s{M}^1_B$ the following equality
\begin{equation}\label{eq:intro_dix_TUV}
\frac{1}{2\Omega_B} {\rm Tr}_{\rm Dix}\big(Q_{B,\xi}^{-1} T\big)\;=\;\lim_{n\to\infty}\frac{1}{|\Lambda_n|}{\rm Tr}_{L^2(\R^2)}\big(\chi_{\Lambda_n} T \chi_{\Lambda_n}\big)
\end{equation}
 holds true independently of $\xi\geqslant0$. 
\end{theorem}

\medskip
\noindent
Theorem \ref{teo:Dix_Tr_for_main2_intro} is proved in full detail in Appendix \ref{subsect:ttrace_unit_N_dix}. In order to make the claim understandable, let us add few information. The quantity $\Omega_B:=\pi\ell_B^2$ provides the area of the \emph{magnetic disk} of radius $\ell_B$. 
The von Neumann algebra generated by the spectral projections of $H_B$ is denoted with $\s{M}_B$, and $\s{M}^1_B\subset \s{M}_B$ is a suitable (weakly dense) ideal. The 
  $\Lambda_n\subseteq\R^2$  
 must provide an 
 increasing sequence of compact subsets such that $\Lambda_n\nearrow\R^2$ and which satisfies the \emph{F{\o}lner condition} (see \eg \cite{greenleaf-69} for more details). The (Lebesgue) volume of $\Lambda_n$ is denoted with $|\Lambda_n|$ and $\chi_{\Lambda_n}$ is the
characteristic function of the set $\Lambda_n$
which acts as a (multiplication) self-adjoint projection
 on $L^2(\R^2)$. The right-hand side of \eqref{eq:intro_dix_TUV} defines the {trace per unit of volume} of the operator $T$ and Theorem \ref{teo:Dix_Tr_for_main2_intro} provides the link between the Dixmier trace and the trace per unit of volume through the mediation of the  regularizing operator $Q_{B,\xi}^{-1}$.

\medskip

The use of the Dixmier trace combined with the resolvent of the harmonic oscillator $Q_{B}$  offers some advantages with respect to  the use of the trace per unit of volume. First of all, 
 the computation of the trace per unit volume always implies the  choice (a priori) of a suitable approximating sequence $\{\Lambda_n\}$ of bounded regions 
  of the plane to obtain the desired thermodynamic limit. This election could in principle affect the result depending of the nature of the operator whose trace is to be calculated. One of the main question in this business is to understand the class of operators that admit a well-defined trace per unit of volume independently of  the election of the approximating sequence $\{\Lambda_n\}$. On the other hand, the use of the 
left-hand side of \eqref{eq:intro_dix_TUV} circumvents this problem by proposing an intrinsic way to calculate the thermodynamic quantities.
From a physical point of view  the operator  
$Q_{B}$ is responsible for a \virg{natural} quantization of the space as a consequence of the quantization of the oscillation frequencies
(or equivalently the quantization of the \emph{cyclotronic orbits}). The thermodynamic limit is then  recovered  through the computation of the  Dixmier trace. This argument provides the \virg{physical justification} for equality \eqref{eq:intro_dix_TUV}. Even from a computation point of view, the use of the left-hand side of \eqref{eq:intro_dix_TUV} presents some advantages. Indeed, the Dixmier trace can be easily computed on the basis which diagonalizes $Q_{B}$ (in view of Lemma \ref{lemma:base_trace}) and the latter is known explicitly: For instance it can be described in terms of the Laguerre functions \eqref{eq:lag_pol}.
Finally, the content of Theorem \ref{teo:Dix_Tr_for_main2_intro}   has also a noncommutative geometric flavor.
Indeed, the  left-hand side of \eqref{eq:intro_dix_TUV} has the structure of the  Connes' noncommutative  integral (\cf \cite[eq. (7.83)]{gracia-varilly-figueroa-01})
which provides the noncommutative version of the  
Wodzicki's residue (see \eg \cite[Section 7.6]{gracia-varilly-figueroa-01} for more details). Moreover, equation  \eqref{eq:intro_dix_TUV} also provides the extension to the continuum of an analogous
  formula proved by Bellissard for two-dimensional discrete systems \cite[Section 2.6]{bellissard-03}. These analogies suggest that
the resolvent $Q_{B,\xi}^{-1}$ should be related to a Dirac operator. Indeed, it can be proven that $Q_{B,\xi}^{-1}$ is proportional to $|D_B|^{-d}$  (with $d=2$) where $D_B$ is a suitable \emph{magnetic Dirac operator}\footnote{\label{note1} For the interested readers we anticipate that  $D_B$ acts on $L^2(\R^2)\otimes\C^4$ and is given by
$$
D_B\;:=\;\frac{1}{\sqrt{2}}\big(K_1\otimes\gamma_1+K_2\otimes\gamma_2+
G_1\otimes\gamma_3+
G_1\otimes\gamma_4\big)
$$
where the operators $K_1,K_2,G_1,G_2$ are defined in Section \ref{subsec:LH} and the $4\times4$ matrices $\gamma_1,\gamma_2,\gamma_3,\gamma_4$ are Clifford generators which
satisfy the canonical anti-commutation relations (CAR) on $\C^4$.}. The magnetic Dirac operator $D_B$ enters in the construction of a \emph{bounded} (magnetic) spectral triple for the Landau operator which provides the \virg{natural} geometric object for the noncommutative geometry of the Quantum Hall effect in the continuum. An accurate analysis of these aspects, motivated by the exigency to conclude the program started in \cite{bellissard-elst-schulz-baldes-94} twenty-five years ago, will be presented in a separated work \cite{denittis-moscolari-20}.
{Finally, let us cite  the recent work \cite{azamov-mcdonald-sukochev-zanin-19} where a similar use of the Dixmier trace has been proposed.}

\medskip

The last achievement of this paper concerns the 
extension of the topological analysis performed for the Landau Hamiltonian $H_B$ to models with \emph{non-Abelian} magnetic fields.
A sufficiently detailed explanation of what is meant with {non-Abelian} magnetic field is postponed to
 Section \ref{sec:mSo-no-ab}. Let us only anticipate here that we will focus mainly on models for particles with spin $s=\frac{1}{2}$. 
In this case, the resulting non-Abelian magnetic Hamiltonians act on the space $L^2(\R^2)\otimes\C^2$ and have the typical structure 
\begin{equation}
\s{H}_{B,b}\;:=\; H_B\otimes {\bf 1}_{2}\;+\;c_b\epsilon_B\s{W}\;+\;C_{B,b}{\bf 1}
\end{equation}
where $H_B$ is the Landau Hamiltonian,
$c_b:=\frac{b}{B\ell_B}$ is a coupling  constant for the {non-Abelian} magnetic field,
 $C_{B,b}\propto c_b^2\epsilon_B$ is a suitable energy constant and $\s{W}$ is the  term that specifies the type of coupling between the particle  and the non-Abelian magnetic field. 
We will focus on two models in particular. In Section \ref{subsect:Jaynes-Cummings} we will study the \emph{Jaynes-Cummings model}
specified by the (Rashba-type) coupling potential
$$
\s{W}_{JC}\;:=\; \big(K_1\otimes \sigma_2-K_2\otimes \sigma_1\big)\;.
$$ 
Here $\sigma_1,\sigma_2,\sigma_3$ denote the three Pauli matrices and $K_1,K_2$ are the {kinetic momenta} \eqref{eq:Kms} associated with the Landau Hamiltonian  $H_B$. 
It turns out that the Jaynes-Cummings model has a pure point spectrum which  
can be calculated explicitly, as shown by \eqref{eq:JC_spec}. Moreover, this model possesses a generalized TRS of even type (denoted by $\Xi$).
Mimicking  the analysis  of the Landau Hamiltonian we will prove in Section \ref{subsect:Jaynes-Cummings} the following result:
\begin{theorem}\label{theo:main3}
At each energy level $E_j^{\pm}$ of the Jaynes-Cummings model it is associated a \virg{Real} line bundle $(\bb{E}_{j}^\pm,\Xi)$ over the involutive (Brillouin) torus $(\n{T}^2_B,\rr{f})$. The topology of this line bundle  is completely classified by its Chern class $c_1(\bb{E}_{j}^\pm)\in\Z$. It turns out that
$$
c_1(\bb{E}_{j}^\pm)\;=\;1\;,\qquad\quad \forall\; j\in\N_0
$$
implying accordingly that all the energy levels are topologically equivalent.
\end{theorem}

\medskip
\noindent
It is worth pointing out that also in this case the topological numbers  (the rank and the  Chern number) associated with the energy levels can be computed with the help of the Dixmier trace.

\medskip

The second model of {non-Abelian} magnetic Hamiltonian studied in  Section \ref{eq:non-Ab-Q}  is specified by the coupling potential
\begin{equation}\label{eq:pot_WQ}
\s{W}_{Q}\;:=\; r_0(K_1-K_2)\otimes{\bf 1}_2\;+\; (K_1+K_2)\otimes (r_1\sigma_1+r_2\sigma_3)
\end{equation}
where $r_0,r_1,r_2$ are real constants subject to the normalization $r_0^2+r_1^2+r_2^2=1$. 
{The detailed study of the spectrum of this operator is beyond the scope of this work and this is left for future investigations. However, only on the basis of the analysis of the symmetries of this model it is possible to anticipate some interesting topological properties. 
First of all, let us note that the Hamiltonian $\s{H}_{Q}$ associated with the potential $\s{W}_{Q}$ has a 
positive spectrum as a consequence of the general structure \eqref{eq:LH-NA}. Secondly,
the potential $\s{W}_{Q}$ endows $\s{H}_{Q}$ with 
  generalized TRS of odd type. If one assumes the existence of a gap around the energy $E>0$ in the spectrum of $\s{H}_{Q}$, then the related spectral projection $\s{P}_E:=\chi_{(-\infty,E]}(\s{H}_{Q})$ will be endowed  with a \virg{Quaternionic} structure according to \cite{denittis-gomi-14-gen}.
This fact justifies the choice of the name
\emph{\virg{Quaternionic} model} for $\s{H}_{Q}$. Moreover, it permits to relate the topological properties of $\s{P}_E$
 to the classification of 
\virg{Quaternionic} vector bundles over the 
 involutive space $(\n{T}^2,\rr{f})$, namely to the description of the set  
 ${\rm Vec}^{2m}_{\rr{Q}}(\n{T}^2,\rr{f})$
 of isomorphism classes of 
rank $2m$ \virg{Quaternionic} vector bundles over $(\n{T}^2,\rr{f})$. The description of 
 ${\rm Vec}^{2m}_{\rr{Q}}(\n{T}^2,\rr{f})$
 boils down to the following 
chain of maps
\begin{equation}\label{eq:intro_quat_iso}
{\rm Vec}^{2m}_{\rr{Q}}\big(\n{T}^2,\rr{f}\big)\;\stackrel{\kappa}{\simeq}\;
H^2_{\Z_2}\big(\T^2|(\T^2)^{\rr{f}},\Z(1)\big)\;\stackrel{\jmath}{\to}\; H^2_{\Z_2}\big(\T^2, \Z(1)\big)\;\stackrel{c_1}{\simeq}\;\Z\end{equation}
where the first isomorphism is induced by the \emph{FKMM-invariant} $\kappa$ \cite{denittis-gomi-14-gen,denittis-gomi-18,denittis-gomi-18-b}, the second map $\jmath$  amounts to the injection $\jmath:\Z\to\Z$ given by $j:n\mapsto 2n$ (Lemma \ref{lemm_eq_cohom_Q} and Lemma \ref{lemma:quat2Z}) and the last isomorphism is the same as described in \eqref{eq:intro_kan_iso}.
In conclusion, one has that ${\rm Vec}^{2m}_{\rr{Q}}(\n{T}^2,\rr{f})\simeq 2\Z$ and the isomorphism classes are completely determined by the (even) values of the Chern classes of the underlying complex vector bundles (Corollary \ref{corol:appen_fin}). The latter  general result is central to prove:
\begin{theorem}\label{theo:main4}
Assume that the spectrum of the \virg{Quaternionic} model $\s{H}_{Q}$ has a gap
around the energy $E>0$ and let $\s{P}_E$ be the associated Fermi projection. To $\s{P}_E$ is 
associated an even rank  \virg{Quaternionic} vector bundle $(\bb{E}_{E},\Xi')$ over the involutive (Brillouin) torus $(\n{T}^2,\rr{f})$. Moreover the topology of this vector bundle  is completely classified by its Chern class $c_1(\bb{E}_{j})\in2\Z$ which can only take even values.
\end{theorem}
}

\medskip
\noindent

 Section \ref{eq:non-Ab-Q}
 is devoted to the proof of Theorem \ref {theo:main4}. Also in this   case the relevant topological invariants  can be computed by formulas involving the Dixmier trace of the projection  $\s{P}_E$.

\medskip

In conclusion this paper contains a detailed study of the topology and the geometry of the Landau levels which  takes into account the role of extra structures  deriving from possible generalized TRS. It is shown  that these structures can be of \virg{Real} type or even of \virg{Quaternionic} type in the regime of non-Abelian magnetic fields. This implies that the topology of these systems should be studied inside the correct category of \virg{Real} or \virg{Quaternionic} vector bundles. As additional, but not less relevant result, we proved that the topological numbers which specify the topology of the energy spectrum of the various models under analysis can be computed directly in the position space using the resolvent of the harmonic oscillator and the Dixmier trace. 
This paper contains several new results but the investigation initiated here  is far from being considered completed. The case of magnetic systems in  presence of background potentials (periodic, aperiodc, random, $\ldots$)
is already subject of ongoing investigations.

\medskip
\noindent
{\bf Structure of the paper.}
Section \ref{sect:flip-exch} is devoted to the connection between the {flip} involution and the particle exchange symmetry. Instead, the geometry and the topology associated with the energy levels of the Landau Hamiltonian are studied in Section \ref{sec:mSo}. 
Then, the results obtained in this section are generalized to the case of non-Abelian magnetic fields
in Section \ref{sec:mSo-no-ab}. In particular the Jaynes-Cummings model and the \virg{Quaternionic} model are studied in detail here. Appendix \ref{sec:Eq-co-flip} contains the computations of the equivariant cohomology groups of the two-dimensional torus endowed with the flip involution. Finally, Appendix \ref{sec:Limit_formula} provides a \virg{crash course} on the Dixmier trace as well as the proof of the crucial Theorem \ref{teo:Dix_Tr_for_main2_intro}.


\medskip
\noindent
{\bf Acknowledgements.} 
GD's research is supported
 by
the  grant \emph{Fondecyt Regular} -  1190204.
KG's research is supported by 
the JSPS KAKENHI Grant Number 15K04871. 
MM's research is supported by the
Grant 8021-00084B of the Danish
Council for Independent Research - Natural Sciences.
GD is indebted to Jean Bellissard for suggesting him the idea of a magnetic spectral triple based on the Dirac operator $D_B$ described in Note \ref{note1}.
GD wish to thank Fernanda Florido Calvo 
for checking the consistency of the main results during the writing of her Bachelor's thesis. MM is grateful to Horia Cornean for inspiring discussions. MM gratefully acknowledge the support of the National Group of Mathematical Physics (GNFM-INdAM).
{The largest part of this document was edited in Santiago del Chile and the first version of the draft was completed during the days when "Chile despert\'o". It is a fact that the first to wake up were the Chilean students. GD loved and admired them for for their courage and their desire to make the Chile a better place.}


\section{Flip operator and particle exchange symmetry}\label{sect:flip-exch}

The aim of this preliminary section is to introduce the notion of \emph{flip operator} which will play a crucial role in the subsequent part of the work.  We find instructive to relate this  operator to a fundamental physical symmetry: the \emph{particle exchange symmetry}.

\medskip

Let $\s{H}:=L^2(\R)\otimes\C^\ell$ be the Hilbert space of a one-dimensional particle of spin $(\ell-1)/2$. A system made of \emph{two} 
 of these particles is described in the Hilbert space
$$
\s{H}^{(2)}\;:=\;\bigotimes_{j={1,2}}\s{H}_j\;\simeq\;L^2(\R^2)\otimes\C^{\ell^2}
$$
where the label $j=1,2$ is used to distinguish the two particles.
Let
$$
\Psi(x_1,x_2)\;:=\;\left(\begin{array}{c}\psi_1(x_1,x_2) \\\vdots \\\psi_{\ell^2}(x_1,x_2)\end{array}\right)
$$
be a generic vector of $\s{H}^{(2)}$.

\medskip

The \emph{flip operator} $F:\s{H}^{(2)}\to \s{H}^{(2)}$ is defined by
$$
(F\Psi)(x_1,x_2)\;:=\;\Psi(x_2,x_1)\;,\qquad\quad \Psi\in\s{H}^{(2)}\;.
$$
It is a unitary involution, \ie ${F}={F}^{-1}={F}^*$.
Clearly the role of $F$ is to exchange particle 1 with particle 2 and vice versa. Sometimes, in the physical literature $F$ is  known as the \emph{particle exchange operator}.

\medskip

The \emph{complex conjugation} operator is naturally defined on  $\s{H}^{(2)}$
by
$$
({C}\Psi)(x_1,x_2)\;:=\;\overline{\Psi}(x_1,x_2)\;=\;\left(\begin{array}{c}\overline{\psi_1}(x_1,x_2) \\\vdots \\\overline{\psi_{\ell^2}}(x_1,x_2)\end{array}\right)\;,\qquad\quad \Psi\in\s{H}^{(2)}\;.
$$
It is anti-linear and verifies ${C}={C}^{-1}={C}^*$. Moreover one can easily check that
$$
{F}{C}\;=\;{C}{F}.
$$
In the physics literature the operator $C$ implements the time reversal symmetry for bosonic particles.

\medskip

Let 
$$
{p}_j\;:=\;-\ii\frac{\partial}{\partial x_j}\;\otimes\;{\bf 1}_{\ell^2}\;=\;\left(\begin{array}{ccc}
-\ii\frac{\partial}{\partial x_j} &  &  \\
 & \ddots &  \\
 &  & -\ii\frac{\partial}{\partial x_j}\end{array}\right)\;,\qquad\quad j=1,2
$$
be the momentum operator of the $j$-th particle. In absence of any interaction between the two particles the  kinetic (total) energy of the system is described (in a suitable system of physical units) by the Hamiltonian
$$
H_0\;:=\;{p_1}^2\;+\;{p_2}^2\;.
$$
Simple calculations show that
$$
Fp_jF\;=\;p_{j+1}\;,\qquad Cp_jC\;=\;-p_j\;,\qquad\quad j=1,2
$$
where $j+1$  is meant modulo 2. As a consequence, one has that
$$
FH_0F\;=\;H_0\;=\;CH_0C\;,
$$
namely $H_0$ is left invariant under the independent action of $F$ and $C$.

\medskip

Let us suppose now that the kinetic momentum of the particle 1 is changed by a gauge potential $A_1$ produced by the particle 2 according to 
$p_1\mapsto p_1+A_1$ where $A_1\in \s{C}(\R)\otimes{\rm Mat}_{\ell^2}(\C)$ acts  as 
$$
(A_1\Psi)(x_1,x_2)\;:=\;\left(\begin{array}{ccc}
a_{1,1}(x_2) & \ldots & a_{1,\ell^2}(x_2) \\
\vdots & \ddots & \vdots \\
 a_{\ell^2,1}(x_2)& \ldots & a_{\ell^2,\ell^2}(x_2)\end{array}\right)\;
  \left(\begin{array}{c}\psi_1(x_1,x_2) \\\vdots \\
 \psi_{\ell^2}(x_1,x_2)\end{array}\right)\;,
$$
namely as the matrix-valued operator of multiplication by functions $a_{n,m}$ in the \emph{only} variable $x_2$. Let us assume that also the kinetic momentum of the particle 2 is changed in a similar way by a gauge potential $A_2$ produced by the particle 2.
The new total energy of the system is then given by
$$
{H}_A\;:=\;\big({p}_1+A_1({x}_2)\big)^2\;+\;\big({p}_2+A_2({x}_1)\big)^2\;.
$$
The action of $F$  on the gauge potentials is given by
$$
FA_1(x_2)F\;:=\;A_1(x_1)\;,\qquad\quad FA_2(x_1)F\;:=\;A_2(x_2)\;,
$$
while $C$ acts as
$$
CA_1(x_2)C\;:=\;\overline{A_1}(x_2)\;,\qquad\quad CA_2(x_1)C\;:=\;\overline{A_2}(x_1)\;.
$$

\medskip

In general ${H}_A$ is not invariant under the separate action of $F$ and $C$.
However, we are interested in the case in which $H_A$ is left invariant by the composed operator 
$$
\Theta\;:=\;FC\;=\;CF\;.
$$
By definition $\Theta$ is an anti-unitary involution, namely
$
\Theta=\Theta^{-1}=\Theta^*
$. Moreover,  a direct calculation shows that
$$
\Theta{H}_A\Theta\;=\;\big({p}_1-\overline{A}_2({x}_2)\big)^2\;+\;\big({p}_2-\overline{A}_1({x}_1)\big)^2
$$ 
hence the symmetry condition 
\begin{equation}\label{eq:flip_01}
\Theta{H}_A\Theta\;=\:H_A
\end{equation}
is guaranteed by the constraint 
$$
A_1\;=\;-\overline{A}_2\;.
$$

\medskip

In the next sections we will see how 
a two-dimensional particle in a uniform magnetic field provides a particular example of a system where the symmetry \eqref{eq:flip_01}  is realized.

\section{The geometry of the Landau levels}
\label{sec:mSo}

The quantum dynamics of a particle of {mass} $m$   and {charge} $q$  (for electrons $q=-e$ with $e>0$) is generated  by the 
 \emph{magnetic} Schr\"{o}dinger operator %
\begin{equation}\label{eq:mSo1}
H_A\;:=\;\dfrac{1}{2m}\left(-\ii \hslash \nabla-\frac{q}{c}\ A\right)^2
\end{equation}
defined on the Hilbert space $L^2(\R^d)$. The Hamiltonian \eqref{eq:mSo1} is expressed in the  CGS system of units:
here $c$ is the speed of  light and $h:=2\pi
\hslash$ is the {Planck constant}. The vector potential $A$ is responsible for 
the coupling of the particle with the magnetic field $B:=\nabla\times A$. Under quite general assumptions on the vector potential $A$ the operator \eqref{eq:mSo1}
turns out to be self-adjoint with cores $\s{C}_c(\R^d)\subset \s{S}(\R^2)$ given by the compactly supported continuous functions and the Schwartz functions, respectively \cite[Theorem 3]{leinfelder-simader-81}. 

\subsection{The Landau Hamiltonian}
\label{subsec:LH}
The Landau Hamiltonian $H_B$ is the magnetic Schr\"{o}dinger operator on $L^2(\R^2)$ with vector potential describing a uniform perpendicular magnetic field $B$. It is well known that such a vector potential 
is not unique but needs a choice of  gauge. In this work we will use the vector potential in the \emph{symmetric gauge}, namely
\begin{equation}\label{eq:vp}
A_{\rm L}(x)\;:=\;\frac{B}{2}{\rm e}_\bot\wedge x\;=\;\frac{B}{2}(-x_2,x_1),\qquad x\;:=\;(x_1,x_2)\in\R^2
\end{equation}
where ${\rm e}_\bot:=(0,0,1)$ is the unit vector orthogonal to the plan $\R^2$ where the particle is confined and $B\in\R$ describes the strength and the orientation with respect to ${\rm e}_\bot$ of the magnetic field. The \emph{Landau Hamiltonian} $H_B:=H_{A_L}$  is then defined as the two-dimensional {magnetic} Schr\"{o}dinger operator \eqref{eq:mSo1} with the vector potential \eqref{eq:vp}. In the following we will assume  $B>0$ which means that the magnetic field is positively oriented with respect to the direction of ${\rm e}_\bot$.

\medskip

Froom the constants which appear in the definition of $H_A$ and $A_L$ one can defines the  \emph{magnetic energy}
$$
\epsilon_B\;:=\;\frac{|q|B\hslash}{mc}\;,
$$
and the  \emph{magnetic length}
$$
\ell_B\;:=\;\sqrt{\frac{c\hslash}{|q|B}}\;.
$$
Henceforward, we will assume $q<0$ which corresponds to the case of electrons.

\medskip

With the notation introduced above the Landau Hamiltonian can be written as
\begin{equation}\label{eq:LH}
H_B\;:=\; \frac{\epsilon_B}{2}\left(K_{1}^2+K_{2}^2\right)
\end{equation}
where the (magnetic) \emph{kinetic momenta} $K_{1}$ and $K_{2}$ are defined by
\begin{equation}\label{eq:Kms}
K_{1}\;:=\;\left(-{\ii }\ell_B\frac{\partial}{\partial x_1}-\frac{x_2}{2\ell_B}\right),\qquad K_{2}\;:=\;\left(-{\ii }\ell_B \frac{\partial}{\partial x_2}+\frac{x_1}{2\ell_B}\right)\;.
\end{equation}
The expressions  \eqref{eq:Kms}  define
essentially self-adjoint operators with core $\s{C}_c(\R^d)$ 
(see \eg \cite[Proposition 9.40]{hall-13})
Therefore, we will use the symbols $K_{1}$ and $K_{2}$ to denote   the unique self-adjoint extensions of the operator initially  defined by \eqref{eq:LH}.

\medskip

The spectral theory of the 
Landau Hamiltonian $H_B$ is  a well established topic \cite{avron-herbst-simon-78} and it is strictly related with the elementary theory
of the harmonic oscillator \cite{fock-28,landau-30}. In order to compute the spectrum of $H$ let us introduce the \emph{dual} momenta
\begin{equation}\label{eq:Dms}
G_{1}\;:=\;\left(-{\ii }\ell_B\frac{\partial}{\partial x_2}-\frac{x_1}{2\ell_B}\right),\qquad G_{2}\;:=\;\left(-{\ii }\ell_B \frac{\partial}{\partial x_1}+\frac{x_2}{2\ell_B}\right)\;.
\end{equation}
Also the \eqref{eq:Dms} defines a pair of self-adjoint operators with core $\s{C}_c(\R^2)$. Moreover, the  commutation relations 
\begin{equation}\label{eq:cc1}
\begin{aligned}
&[K_{1},K_{2}]\;=\;  -\ii  {\bf 1}\;=\;[G_{1};G_{2}]\\
&[K_{i},G_{j}]\;=\;0\ \ \ \ \qquad i,j=1,2
\end{aligned}
\end{equation}
can be easily proved  by a direct computation on the   core $\s{C}_c(\R^2)$.
In view of the \eqref{eq:cc1} one can define two pairs of \emph{creation-annihilation operators}
\begin{equation}\label{eq:cr-an-op}
\rr{a}^{\pm}\;:=\;\frac{1}{\sqrt{2}}\left(K_{1}\pm\ii K_{2}\right)\;,\qquad \rr{b}^{\pm}:=\frac{-1}{\sqrt{2}}\left(G_{1}\pm\ii G_{2}\right)\;.
\end{equation}
The $\rr{a}^{\pm}$ and $\rr{b}^{\pm}$, are closable operators initially defined on the dense domains $\s{C}_c(\R^2)\subset\s{S}(\R^2)$. 
Moreover, $\rr{a}^{-}$ and $\rr{b}^{-}$ are the  adjoint of $\rr{a}^{+}$ and $\rr{b}^{+}$, respectively. These operators are
subjected to the following commutation rules: 
\begin{equation}\label{eq:cc2}
[\rr{a}^{\pm},\rr{b}^{\pm}]\;=\;0\;, \qquad \qquad [\rr{a}^{-},\rr{a}^{+}]\;=\;{\bf 1}\;=\;[\rr{b}^{-},\rr{b}^{+}]\;.
\end{equation}

\medskip

 Let $\psi_{0}\in \s{S}(\R^2)$ be the normalized function 
\begin{equation}\label{eq:herm1}
\psi_{0}(x)\;:=\;\frac{1}{\ell_B\sqrt{2\pi}}\ \expo{-\frac{|x|^2}{4\ell_B^2}}.
\end{equation}
 A direct computation shows that $\rr{a}^{-}\psi_{0}=0=\rr{b}^{-}\psi_{0}$. Acting on $\psi_{0}$ with the creation operators one defines 
\begin{equation}\label{eq:herm2}
\psi_{n}\;:=\;\frac{1}{\sqrt{n_1!n_2!}}\ (\rr{a}^{+})^{n_1}(\rr{b}^{+})^{n_2}\psi_{0},\qquad n\;:=\;(n_1,n_2)\in\N^2_0\;.
\end{equation}
Evidently $\psi_{n}\in \s{S}(\R^2)$ for any $n\in\N^2_0$. Moreover, by a recursive application of the commutation rules \eqref{eq:cc2}, one can prove  that $(\psi_{n},\psi_{n'})=\delta_{n,n'}$. The set $\{\psi_{n}\ |\ n\in\N^2_0\}\subset \s{S}(\R^2)$ provides a complete orthonormal system for $L^2(\R^2)$ called (magnetic) \emph{Laguerre basis}.  
The set of the finite linear combinations of elements of this  basis defines a dense subspace $\s{L}_B\subset \s{S}(\R^2)$ which is left invariant by the action of  $\rr{a}^{\pm}$ and $\rr{b}^{\pm}$. 
Moreover, $\rr{a}^{\pm}$ and  $\rr{b}^{\pm}$ are closable on $\s{L}_B$. 
The normalized eigenfunctions \eqref{eq:herm2} can be expressed as \cite{johnson-lippmann-49,raikov-warzel-02}

\begin{equation}\label{eq:lag_pol}
\psi_{n}(x)\;:=\;\psi_0(x)\ \sqrt{\frac{n_1!}{n_2!}}\left[\frac{x_1+\ii x_2}{\ell_B\sqrt{2}}\right]^{n_2-n_1}L_{n_1}^{(n_2-n_1)}\left(\frac{|x|^2}{2\ell_B^2}\right)
\end{equation}
where
$$
L_m^{(\alpha)}\left(\zeta\right)\;:=\;\sum_{j=0}^{m}\frac{(\alpha+m)(\alpha+m-1)\ldots(\alpha+j+1)}{j!(m-j)!}\left(-\zeta\right)^j\;,\quad\alpha,\zeta\in \R
$$
are the \emph{generalized Laguerre polynomial} of degree $m$ (with the usual convention $0!=1$) \cite[Sect. 8.97]{gradshteyn-ryzhik-07}.  

\medskip

By 
using the definitions \eqref{eq:cr-an-op} and the commutation relations \eqref{eq:cc1} one obtains after a straightforward calculation that
\begin{equation}\label{eq:Lh_no}
H_B\;=\;\epsilon_B\left(\rr{a}^{+}\rr{a}^{-}+\frac{1}{2}{\bf 1}\right)\;=\;\epsilon_B\left(\rr{a}^{-}\rr{a}^{+}-\frac{1}{2}{\bf 1}\right)\;. 
\end{equation}
The first consequence  is that any Laguerre vector $\psi_{n}$ is an eigenvector of $H_B$. This implies that the Laguerre  basis  provides an orthonormal system which diagonalizes $H_B$ according to
$$
H_B\psi_{n}\; =\; \epsilon_B\left(n_1+\frac{1}{2}\right)\psi_{n},\qquad\qquad n\;=\;(n_1,n_2)\in\N_0^2.
$$
Hence, the spectrum of $H_B$ is a sequence of eigenvalues given by
\begin{equation}\label{eq:spec_H_B}
\sigma(H_B)=\left.\left\{E_{j}:=\epsilon_B\left(j+\frac{1}{2}\right)\ \right| \ j\in\N_0\right\}
\end{equation}
and $H_B$ turns out to be essentially self-adjoint also on the   core 
$\s{L}_B$  \footnote{The closure of the cores $\s{L}_B\subset\s{S}(\R^2)$ or $\s{C}^\infty_c(\R^2)$  with respect to the operator graph norm of $H_B$ defines the domain $\s{D}(H_B)\subset L^2(\R^2)$ which is called \emph{(second) magnetic Sobolev space} \cite{lieb-loss-97}.}. 
We refer to the eigenvalue  $E_{j}$ as the $j$-th \emph{Landau level}. 

\medskip

Each Landau level  is infinitely degenerate. 
A simple computation shows that the orthogonal component of the angular moment can be written as
$$
\begin{aligned}
L_3\;:&=\;-\ii\hslash\left(x_1\frac{\partial}{\partial x_2}-x_2\frac{\partial}{\partial x_1}\right)\\
&=\;\hslash\left(\frac{K_{1}^2+K_{2}^2}{2}-\frac{G_{1}^2+G_{2}^2}{2}\right)\;=\;\hslash\left(\rr{a}^{+}\rr{a}^{-}-\rr{b}^{+}\rr{b}^{-}\right)\;.
\end{aligned}
$$
This implies that $[H_B,L_3]=0$.
Then, the Laguerre functions $\psi_{(n_1,n_2)}$ are simultaneous  eigenfunctions of $H_B$ and $L_3$ with eigenvalues $E_{n_1}$ and  $l:=\hslash(n_1-n_2)$, respectively. This shows that the possible eigenvalues of the angular momentum $L_3$ for a particle in the energy level $n_1$  are $\ell=\hslash m$ with $-\infty < m\leqslant n_1$.

\medskip

Let $\s{H}_j\subset L^2(\R^2)$ be the eigenspace relative to the $j$-th  eigenvalue of $H_B$. Clearly,  $\s{H}_j$ is spanned by $\psi_{(j,m)}$ with $m\in\N_0$ and the spectral projection $\Pi_{j}:L^2(\R^2)\to\s{H}_j$ is described (in Dirac notation) by
\begin{equation}\label{eq:Lan_proj}
\Pi_{j}\;:=\;\sum_{m=0}^{\infty}\ketbra{\psi_{(j,m)}}{\psi_{(j,m)}}\;.
\end{equation}
One infers from \eqref{eq:herm2} the  recursive relations
\begin{equation}\label{eq:shift_P}
\Pi_{j}\;=\;\frac{1}{j}\;\rr{a}^{+} \Pi_{j-1}  \rr{a}^{-}\;,\qquad\quad \Pi_{j}\;=\;\frac{1}{j+1}\rr{a}^{-}\Pi_{j+1}\rr{a}^{+}
\end{equation}
and after an iteration  one gets
$$
\Pi_{j}\;=\;\frac{1}{j!}\;(\rr{a}^{+})^j  \Pi_{0} (\rr{a}^{-})^j\;.
$$
The \emph{Landau projections} $\Pi_{j}$ are integral kernel operators with kernel given by \cite{raikov-warzel-02}
\begin{equation}\label{L_proj_ker}
\Pi_{j}(x,y)\;:=\;\frac{1}{2\pi\ell_B^2}\expo{-\frac{|x-y|^2}{4\ell_B^2}} \expo{-\ii\frac{x\wedge y}{2\ell_B^2}} L_{j}^{(0)}\left(\frac{|x-y|^2}{2\ell_B^2}\right)
\end{equation}
where $x \wedge y:=x_1y_2-x_2y_1$.

\subsection{Discrete symmetries of the Landau Hamiltonian}
\label{subsec:sym-LH}
We have already mentioned  in Section \ref{sect:flip-exch} that the \virg{standard}
{time reversal symmetry} of a spinless particle is implemented by the  {complex conjugation} $C\psi=\overline{\psi}$.  

\medskip

The effect of the operator $C$ on the  dynamics of a two-dimensional charged particle subjected to a uniform magnetic field can be deduced by observing that
\begin{equation}\label{eq:disc_sym_1_abo1}
C\;K_1\;C\;=\;-G_2\;,\qquad\quad C\;K_2\;C\;=\;-G_1\;
\end{equation}
which in turn implies
\begin{equation}\label{eq:disc_sym_1}
C\;\rr{a}^{\pm}\;C\;=\;\pm\ii\rr{b}^{\mp}\;.
\end{equation}
From the above relations it follows that the Landau Hamiltonian  and the angular momentum are not left invariant by the action of $C$. In particular one has that 
$$
\begin{aligned}
C\;H_B\;C\;&=\; H_{-B}\;,\\
C\;L_3\;C\;&=\; -L_3\;,
\end{aligned}
$$
namely the effect of the transformation implemented by $C$ is to invert the sign of the magnetic filed $B$ and of the angular momentum $L_3$.

\medskip

The Landau Hamiltonian, although is not left invariant by the complex conjugation $C$, admits a generalized TRS in the sense of Definition \ref{def:introG-TRS}. Consider the \emph{flip operator} (\cf Section \ref{sect:flip-exch}) 
${F}:L^2(\R^2)\to L^2(\R^2)$ defined by
\begin{equation}\label{eqq:flip_ones}
({F}\psi)(x_1,x_2)\;:=\;\psi(x_2,x_1)\;,\qquad\quad \psi\in L^2(\R^2)\;.
\end{equation}
Clearly ${F}={F}^{-1}={F}^*$, namely $F$ is a linear unitary involution. Moreover $F$ is real in the sense that $CF=FC$. The operator $F$ implements the exchange of the components of
position and momentum. This implies that 
\begin{equation}\label{eq:disc_sym_1_abo2}
F\;K_1\;F\;=\;G_1\;,\qquad\quad F\;K_2\;F\;=\;G_2\;
\end{equation}
and in turn
$$
F\;\rr{a}^{\pm}\;F\;=\;-\;\rr{b}^{\pm}\;.
$$
Let $\Theta:=FC$. This is an anti-unitary involution since $\Theta^2=FCFC=F^2={\bf 1}$. Moreover, by combining together \eqref{eq:disc_sym_1_abo1} and \eqref{eq:disc_sym_1_abo2} one gets
\begin{equation}\label{eq:disc_sym_1_abo3} 
\Theta\;K_1\;\Theta\;=\;-K_2\;,\qquad\quad \Theta\;K_2\;\Theta\;=\;-K_1\;.
\end{equation}
Equations \eqref{eq:disc_sym_1_abo2} along with \eqref{eq:LH} imply 
\begin{equation}\label{eq:TRS_HB}
\Theta\;H_B\;\Theta\;=\;H_B\;
\end{equation}
and this completes 
the proof of Proposition \ref{prop:intro_01}.

\medskip

\begin{remark}[Perturbation by  potentials]\label{rk:top-class2}{\upshape
In usual physical applications the Landau Hamiltonian $H_B$ is perturbed by an \emph{electrostatic potential} $V$ (a multiplication operator on $L^2(\R^2)$) which describes the interaction of the electrons with  the nuclei arranged in the space. The perturbed Hamiltonian is usually denoted by $H_{B,V}:=H_B+V$. 
Since $H_{B,V}$ is required to be self-adjoint, $V$ has to be real, namely $CV=VC$.
 Therefore, $H_{B,V}$  has the odd TRS
implemented by $\Theta$
 if and only if $V$ commutes with the flip operator, \ie $FV=VF$. The latter  condition is quite strong. For instance a potential of the type
$$
V(x_1,x_2)\;=\;a\;\cos(x_1)\;+\;b\;\cos(x_2)\;,\qquad\quad a,b\in\R
$$ 
 commutes with $F$ if and only if $a=b$. As soon as $a\neq b$ the associated 
perturbed operator $H_{B,V}$ breaks the generalized TRS induced by $\Theta$.
}\hfill $\blacktriangleleft$
\end{remark}

\begin{remark}[Spectral projections and noncommutative vector bundles]\label{rk:top-class3}{\upshape
The spectral projection $\Pi_j$ associated to the $j$-th Landau level inherits the same symmetries of the Landau operator. In particular one has that $\Theta \Pi_j \Theta=\Pi_j$ for all $j\in\N_0$. For reasons that will be clarified in  Section \ref{subsect:VB-R-landau}, it is appropriate to refer to the pair $(\Pi_j,\Theta)$ as the (non-commutative) \virg{Real} line bundle associated to the $j$-th Landau level $E_j$. This description still survives if a suitable, sufficiently weak perturbation $V$  is added. More precisely, let us assume that 
the Landau Hamiltonian is   
perturbed by a bounded {electrostatic potential} $V$ of norm $\|V\|<+\infty$. By observing that the separation between two consecutive Landau levels (gap size) is
$$
\Delta E\;:=\;E_{j+1}\;-\;E_j\;=\;\epsilon_B
$$
one concludes with a simple perturbative argument \cite[V \S-3, Theorem 4.10]{kato-95} that if 
$$
2 \|V\|\;<\; \epsilon_B
$$
then the points $g_j:=j\epsilon_B$, $j\in\N_0$ are still in the resolvent set of $H_{B,V}$. This allows to define a perturbed spectral projection by means of the Riesz-Dunford integral
$$
\Pi_j^V\;:=\;\frac{\ii}{2\pi }\oint_{C_j}\dd z\;\frac{1}{H_{B,V}- z{\bf 1}}\;
$$
where $C_j:=\{g_j+\frac{\epsilon_B}{2}(\expo{\ii\theta}-1)\ |\ \theta\in[0,2\pi]\}$ is the circle in the complex plane which crosses the real axis in $g_{j-1}$ and $g_j$. When the perturbation $V$ meets the conditions described in Remark \ref{rk:top-class2} to ensure that $\Theta H_{B,V}\Theta=H_{B,V}$ it also follows that 
$\Theta \Pi_j^V\Theta=\Pi_j^V$ and the pair $(\Pi_j^V,\Theta)$ still defines a  (non-commutative) \virg{Real} line bundle associated to the $j$-th perturbed Landau level. However, 
it is worth to emphasize that  the case of admissible electrostatic perturbations  is beyond the scope of this work.
}\hfill $\blacktriangleleft$
\end{remark}

\subsection{The {magnetic} Bloch-Floquet-Zak transform}
\label{subsect:magn_BFZ_tras}
Let us define the 
family of unitary operators
$$
T_m\;:=\;(-1)^{m_1m_2}\expo{-\ii\sqrt{2\pi}(m_1G_2+m_2G_1)}\;,\qquad\quad m:=(m_1,m_2)\in\Z^2\;.
$$
The commutations relations \eqref{eq:cc1} imply that 
$$
[T_m,H_B]\;=\;0\;,\qquad\quad \forall\; m\in\Z^2 
$$
namely the operators $T_m$ are symmetries of the Landau Hamiltonian $H_B$. Moreover, an application of the Baker-Campbell-Hausdorff formula shows that
$$
T_{m+m'}\;=\;
T_mT_{m'}\;,\qquad\quad\forall\; m,m'\in\Z^2\;, 
$$ 
namely the mapping $m\mapsto T_m$   provides a unitary representation of $\Z^2$ on $L^2(\R^2)$ which leaves invariant $H_B$. 
An explicit computation provides
$$
(T_m\psi)(x)\;=\;(-1)^{\sigma(m)}\expo{-\ii\sqrt{\frac{\pi}{2}}\frac{m\wedge x}{\ell_B}}\;\psi(x-\sqrt{{2\pi}}\ell_Bm)\;,\qquad\psi\in L^2(\R^2)
$$
where $m\wedge x:=m_1x_2-m_2x_1$ and $\sigma(m)=m_1m_2$.
The operators $T_1$ and $T_2$ are therefore called \emph{magnetic translations} \cite{zak1,zak2}.

\medskip

The $\Z^2$-action implemented by the magnetic translations  can be used to define   the \emph{magnetic} Bloch-Floquet-Zak transform 
which is a (natural) generalization of the usual Bloch-Floquet transform \cite{kuchment-93}. For that we need to introduce the magnetic \emph{unit cell}
$$
\n{Y}_B\;:=\;\left[0, \sqrt{2{\pi}}\ell_B\right]^2
$$
and the magnetic \emph{Brillouin torus}
$$
\n{B}_B\;:=\;\R^2/\left(\frac{\sqrt{2\pi}}{\ell_B}\Z\right)^2\;\simeq\; \left[0, \frac{\sqrt{2\pi}}{\ell_B}\right]^2\;.
$$
Topologically, $\n{B}_B$ is a rescaled version of the standard torus $\T^2=\n{S}^1\times\n{S}^1$. 
For every $\psi\in\s{C}_c(\R^2)$ let us define  the transform
\begin{equation}\label{eq:BF_01}
(\bb{U}_B\psi)_k(y)\;:=\;\sum_{m\in\Z^2}\expo{-\ii k\cdot(y-\sqrt{2\pi}
\ell_Bm)}(T_m\psi)(y)
\end{equation}
where $k:=(k_1,k_2)$, $y:=(y_1,y_2)$ and  $k\cdot y:=k_1y_1+k_2y_2$.
From \eqref{eq:BF_01}, one immediately infers the (pseudo-)periodicity properties
\begin{equation}\label{eq:BF_02}
\left\{
\begin{aligned}
&T_n(\bb{U}_B\psi)_k(y)\;=\;(\bb{U}_B\psi)_k(y)\\
&(\bb{U}_B\psi)_{k+n\frac{\sqrt{2\pi}}{\ell_B}}(y)\;=\;\expo{-\ii\frac{\sqrt{2\pi}}{\ell_B} n\cdot y}(\bb{U}_B\psi)_k(y)
\end{aligned}
\right.\qquad\forall\; n\in\Z^2\;.\\
\end{equation}
The first of \eqref{eq:BF_02} can be equivalently written as
$$
(\bb{U}_B\psi)_k(y-n\sqrt{2\pi}{\ell_B})\;=\;\expo{\ii\sqrt{\frac{\pi}{2}}\frac{n_1y_2-n_2y_1}{\ell_B}}\;(\bb{U}_B\psi)_k(y)\;,\qquad\forall\; n\in\Z^2\;
$$
and shows that the function $(\bb{U}_B\psi)_k$ is completely determined by its restriction on the unit cell $\n{Y}_B$. Equivalently, one can think to 
 $(\bb{U}_B\psi)_k$ as an element of the \emph{fiber} Hilbert space
$$
\rr{h}_B\;:=\;\{\phi\in L^2_{\rm loc}(\R^2)\;|\; T_n\phi=\phi\;,\quad \forall\; n\in\Z^2\}.
$$
endowed with the scalar product 
$$
\langle\phi_1,\phi_2\rangle_{\rr{h}_B}\;:=\;\int_{\n{Y}_B}\dd y\; \overline{\phi_1}(y)\phi_2(y)\;.
$$
Clearly, one has the isomorphism of Hilbert spaces
\begin{equation}\label{eq:BF_04}
\rr{h}_B\;\simeq\; L^2(\n{Y}_B)\;.
\end{equation}
From the second of \eqref{eq:BF_02} one can  argue that $\bb{U}_B$ establishes a unitary transformation $\bb{U}_B:L^2(\R^2)\to \s{H}^{B}_{\rm eq}$ where the space of equivariant functions
$$
\s{H}^{B}_{\rm eq}\;:=\;\left\{\varphi\in L^2_{\rm loc}(\R^2, \rr{h}_B)\; \big|\; \varphi_{k+n\frac{\sqrt{2\pi}}{\ell_B}}(y)=\expo{-\ii\frac{\sqrt{2\pi}}{\ell_B} n\cdot y}\varphi_k(y)\;,\quad \forall\; n\in\Z^2 \right\}
$$
is made into a Hilbert space by the scalar product
$$
\langle\varphi_1,\varphi_2\rangle_{\s{H}^{B}_{\rm eq}}\;:=\;\int_{\n{B}_B}\dd\mu(k)\; \langle{\varphi_1}_k,{\varphi_2}_k\rangle_{\rr{h}_B}
$$
where $\dd\mu(k):= \frac{\ell_B^2}{2\pi}\dd k$ is the normalized (Haar) measure of $\n{B}_B$. The Hilbert space $\s{H}^{B}_{\rm eq}$ is indeed a direct integral $\int^\oplus_{\n{B}_B} \dd\mu(k)\; \rr{h}_B$ (see \cite[Part II, Chapter 6]{dixmier-81} for the theory of direct integrals) and in view of the isomorphism \eqref{eq:BF_04} one infers that the {magnetic} Bloch-Floquet-Zak transform  defines a unitary map
$$
\bb{U}_B\;:\;L^2(\R^2)\;\longrightarrow\;
\int^\oplus_{\n{B}_B} \dd\mu(k)\; L^2(\n{Y}_B)\;.
$$

\medskip

The main feature of the magnetic Bloch-Floquet-Zak transform is that it diagonalizes the 
magnetic translations, \ie
$$
(\bb{U}_BT^m\psi)_k(y)\;=\;\expo{-\ii \sqrt{2\pi}{\ell_B}\;k\cdot m}\; \psi_k(y)\;,\qquad\quad m\in\Z^2\;.
$$
Said differently, the
 operators $T^m$ acquire a fibered representation on the direct integral given by
$$
T^m\;\longmapsto\;\bb{U}_BT^m\bb{U}_B^{-1}:=\int^\oplus_{\n{B}_B} \dd\mu(k)\;  {T}^m(k)
$$
where  ${T}^m(k)$ acts on the fiber Hilbert space $L^2(\n{Y}_B)$ as the multiplication by the phase $\expo{-\ii \sqrt{2\pi}{\ell_B}\;k\cdot m}$. 

\medskip

It is worth noting that the generators 
 $G_1$ and $G_2$ of the  magnetic translations 
 do not commute with the magnetic translations; indeed  $[T_j,G_j]=(-1)^{j+1}\sqrt{2\pi}T_j$, $j=1,2$. This implies that the generators 
 $G_1$ and $G_2$ are not decomposable operators
  in the sense that they do not respect the direct integral decomposition induced by the magnetic Bloch-Floquet-Zak transform. The same  happens to the operators $\rr{b}^+$ and $\rr{b}^-$.

\subsection{\virg{Real} line bundles associated to Landau levels}
\label{subsect:VB-R-landau}

The operators $K_1$ and $K_2$, and in turn the raising-lowering operators $\rr{a}^\pm$ and the Landau Hamiltonian $H_B$, commute with the magnetic translations and so they admit a direct integral decomposition. Let 
$$
K_j\;\longrightarrow\;\bb{U}_BK_j\bb{U}_B^{-1}\;:=\;\int^\oplus_{\n{B}_B} \dd\mu(k)\; {K}_j(k)\;,\qquad\quad j=1,2
$$
be the direct integral decomposition of $K_j$. 
A direct computation shows that the operators ${K}_j(k)$ act on a suitable dense domain of $L^2(\n{Y}_B)$ as 
$$
\begin{aligned}
&K_1(k)\;:=\;-\ii\ell_B\frac{\partial}{\partial y_1}\;-\;\frac{y_2}{2\ell_B}\;+\;\ell_Bk_1\;,\\
&K_2(k)\;:=\;-\ii\ell_B\frac{\partial}{\partial y_2}\;+\;\frac{y_1}{2\ell_B}\;+\;\ell_Bk_2\;.
\end{aligned}
$$
In view of the linear dependence of ${K}_j(k)$ by $k$,  it follows that the domain of definition of ${K}_j(k)$ is independent of $k$. A similar description holds for the Landau Hamiltonian $H_B$. One has that 
$$
H_B\;\longrightarrow\;\bb{U}_BH_B\bb{U}_B^{-1}\;:=\;\int^\oplus_{\n{BT}_B} \dd\mu(k)\; H_B(k)\;,\qquad\quad j=1,2
$$
where
\begin{equation}\label{eq:k-HB}
H_B(k)\;:=\; \frac{\epsilon_B}{2}\left(K_1(k)^2+K_2(k)^2\right)\;.
\end{equation}
The domain of $H_B(k)$ turns out to be independent of $k$ and the map
$$
\n{B}_B\;\ni\;k\;\longmapsto\; \frac{1}{H_B(k)- z{\bf 1}}\;\in\;\s{K}(L^2(\n{Y}_B))\;,\qquad\quad z\in\C\setminus\R 
$$
is norm-continuous\footnote{Indeed one can prove that the mapping
$k\mapsto H_B(k)$ defines an entire analytic family in the sense of Kato with compact resolvent. For more details we refer to \cite[Section 3.3]{monaco-panati-pisante-teufel-18}.}
 with values in the algebra of the compact operators  $\s{K}(L^2(\n{Y}_B))$.
This immediately implies that the spectrum of $H_B(k)$ is purely  discrete and changes continuously with respect to $k$.
A general result in the theory of fibered operators states that
$$
\sigma(H_B)\;=\;\bigcup_{k\in\n{B}_B}\sigma(H_B(k))\;.
$$
This equality, along with the description of $\sigma(H_B)$ given by \eqref{eq:spec_H_B} and the continuity of the spectra of $H_B(k)$ implies that
$$
\sigma(H_B(k))\;=\;\sigma(H_B)\;=\;\{E_j\; |\; j\in\N_0\}\;,\qquad\quad \forall\ k\in\n{T}^2_B\;.
$$
In other words
 the energy bands $k\mapsto E_j(k)$ associated to the fibered Hamiltonian $H_B(k)$ are \emph{flat}, namely they are   constant and coincide with the Landau levels, \ie $E_j(k)=E_j$. The only missing information concerns the multiplicity of these eigenvalues. To answer this question we need to examine the Landau projections.

\medskip

The continuity of the resolvent of $H_B(k)$ leads to a continuous family of spectral projections
$$
\n{B}_B\;\ni\;k\;\longmapsto\; \Pi_j(k)\;:=\;\frac{\ii}{2\pi }\oint_{C_j}\dd z\;\frac{1}{H_B(k)- z{\bf1}}\;\in\;\s{K}(L^2(\n{Y}_B))\;, 
$$
where the circles $C_j$ have been defined in Remark \ref{rk:top-class3}. Moreover, the compactness of $(H_B(k)-z {\bf 1})^{-1}$ implies that $\Pi_j(k)$ is finite rank, and therefore trace class,
 for all $k\in\n{B}_B$.
By applying  functional calculation
 one gets
\begin{equation}\label{eq:diag_proj}
\Pi_j\;\longrightarrow\;\bb{U}_B\Pi_j\bb{U}_B^{-1}\;:=\;\int^\oplus_{\n{B}_B} \dd\mu(k)\; \Pi_j(k)\;,\qquad\quad j\in\N_0
\end{equation}
where $\Pi_j$ is the Landau projection  \eqref{eq:Lan_proj}. 
\begin{lemma}\label{lemma:deg-eigen_XX}
Let $k\mapsto \Pi_j(k)$ be the continuous family of projections associated to the Landau projection $\Pi_j$ by the Bloch-Floquet-Zak transform. It holds true that 
$$
{\rm Tr}_{L^2(\n{Y}_B)}\big(\Pi_j(k)\big)\;=\;1\;,\qquad\quad \forall\; k\in \n{B}_B\;.
$$
\end{lemma}
\proof
Since the $\Pi_j(k)$ are trace class it follows that ${\rm Tr}_{L^2(\n{Y}_B)}(\Pi_j(k))=r(k)\in\N$
for all $k\in \n{T}^2_B$. However, the continuity of $k\mapsto \Pi_j(k)$ implies the continuity of $k\mapsto r(k)$ which in turn implies $r(k)=r_0\in\N$ for all $k\in \n{T}^2_B$.
Since the measure $\dd\mu$ is normalized one gets
$$
r_0\;=\;\int_{\n{B}_B} \dd\mu(k)\; {\rm Tr}_{L^2(\n{Y}_B)}\big(\Pi_j(k)\big)\;=\;1
$$
where the second equality will be proved in  Lemma \ref{lemma:trac1} and Corollary \ref{corol:rank}. 
\qed

\medskip
\noindent
Since the multiplicity of  $E_j(k)$ is measured by the trace of $\Pi_j(k)$ one immediately gets:
\begin{corollary}\label{prop:deg-eigen_X}
For each $k\in\n{B}_B$ and $j\in \N_0$ the Landau level $E_j(k)=E_j$ is a non degenerate eigenvalue of the fiber Hamiltonian $H_B(k)$.
\end{corollary}

\medskip

The the map $k\mapsto \Pi_j(k)$  defines a complex vector bundle $\bb{E}_{j}\to\n{T}^2\simeq\n{B}_B$ with total space given by 
\begin{equation}\label{eq_E_j_bund}
\bb{E}_{j}\;:=\;\bigsqcup_{k\in\n{T}_B^2}\text{Ran}\big[\Pi_j(k)\big]\;.
\end{equation}
The construction of the \emph{spectral bundle} (also called Bloch-bundle \cite{panati-07}) $\bb{E}_{j}\to\n{T}^2$ is standard (see \cite[Lemma 4.5]{denittis-lein-11} or \cite[Section 2]{denittis-gomi-14} for more details) and provides a manifestation of the Serre-Swan duality. In view of 
Lemma \ref{lemma:deg-eigen_XX} one infers that
$$
{\rm dim}\big(\text{Ran}\big[\Pi_j(k)\big]\big)\;=\;{\rm Tr}_{L^2(\n{Y}_B)}\big(\Pi_j(k)\big)\;=\;1\;,
$$
namely the $\bb{E}_{j}$ are \emph{line} bundles.
We will refer to 
 $\bb{E}_{j}$ as the $j$-th \emph{Landau line bundle}.

\medskip

 The complex 	conjugation $C$ and the flip operator $F$ do not have a nice behavior under the Bloch-Floquet-Zak transform $\bb{U}_B$. However, their composition $\Theta:=FC$ acts in a nice way. Indeed from \eqref{eq:disc_sym_1_abo1} and \eqref{eq:disc_sym_1_abo2} one deduces $\Theta T_1 \Theta=T_2$. By combining this relation with the definition \eqref{eq:BF_01} one gets 
$$
\Theta[(\bb{U}_B\psi)_{k}]\;=\;(\bb{U}_B\Theta\psi)_{\rr{f}(k)}\;=\;(\bb{U}_BF\overline{\psi})_{\rr{f}(k)}\;,\qquad\quad \psi\in L^2(\R^2)\;
$$
where $\rr{f}:\n{B}_B\to\n{B}_B$ is
 the \emph{flip-involution} on the torus $\n{B}_B\simeq\n{T}^2$ described in Definition \ref{def_FI}.
 Then, with a little abuse of notation, we can think of $\Theta$ as an anti-unitary map which acts on the direct integral intertwining the fiber over $k$ with the fiber over $\rr{f}(k)$.

\medskip

From the TRS invariance \eqref{eq:TRS_HB}
one can reconstruct the action of $\Theta$ on the fiber Hamiltonians $H_B(k)$ defined by \eqref{eq:k-HB}. A straightforward calculation shows that
\begin{equation}
\Theta\;H_B(k)\;\Theta\;=\;H_B(\rr{f}(k))\;,\qquad\quad \forall\ k\in\n{B}_B\;.
\end{equation}
 By functional calculus one concludes that the same relation is inherited by the Landau projections, \ie 
\begin{equation}
\Theta\;\Pi_j(k)\;\Theta\;=\;\Pi_j(\rr{f}(k))\;,\qquad\quad \forall\ k\in\n{B}_B\;.
\end{equation}
This symmetry translates at level of the Landau line bundle $\bb{E}_{j}$ and defines a homeomorphism of the total space still (with a little abuse of notation) denoted by $\Theta:\bb{E}_{j}\to \bb{E}_{j}$ with the property that $\Theta$ is an anti-linear involution between the conjugate fibers over $k$ and $\rr{f}(k)$, \ie
$$
\left\{
\begin{aligned}
&\Theta\;:\:\bb{E}_{j}|_k\;\longrightarrow\; \bb{E}_{j}|_{\rr{f}(k)}\\
&\Theta^2|_k\;=\;{\rm Id}_{\bb{E}_{j}}
\end{aligned}
\right.\;,\qquad\quad \forall\ k\in\n{B}_B\;.
$$
A complex vector bundle endowed with such a symmetry is called \virg{Real} \cite{atiyah-66,denittis-gomi-14}. Summing up the considerations stated above, and using the topological identification  $\n{B}_B\simeq\T^2$, we can conclude that:
\begin{proposition}\label{prop:LL-VB}
To each Landau level $E_{j}$ of the Landau Hamiltonian  $H_B$  is  associated a \virg{Real} line bundle $(\bb{E}_{j},\Theta)$ over the involutive torus $(\n{T}^2,\rr{f})$.
\end{proposition}
%

\subsection{The topology of the  Landau levels}
\label{subsect:top_class_LL}
In view of Proposition \ref{prop:LL-VB} 
the  topological properties of the $j$-th Landau level $E_j$   can be read from the \virg{Real} line bundle $(\bb{E}_{j},\Theta)$. 

\medskip

Let ${\rm Vec}^m_{\rr{R}}(\n{T}^2,\rr{f})$ be the set of equivalence classes of rank $m$ \virg{Real} vector bundles over the involutive torus $(\n{T}^2,\rr{f})$. Proposition \ref{prop:LL-VB} implies that each Landau level $E_j$ defines an element of ${\rm Vec}^1_{\rr{R}}(\n{T}^2,\rr{f})$. Therefore, the study of the topological properties of $E_j$  amounts to the topological classification of ${\rm Vec}^m_{\rr{R}}(\n{T}^2,\rr{f})$. The latter is provided by the following crucial result:
\begin{proposition}\label{prop:cohom-R}
Let $\n{T}^2$ be a two dimensional torus endowed with the flip involution $\rr{f}$.
There are isomorphisms
$$
{\rm Vec}^1_{\rr{R}}\big(\n{T}^2,\rr{f}\big)\;\stackrel{c_1^{\rr{R}}}{\simeq}\;H^2_{\Z^2}\big(\n{T}^2,\Z(1)\big)\;\stackrel{\imath}{\simeq}\; H^2\big(\n{T}^2,\Z\big)\;{\simeq}\;\Z
$$
where $\imath$ is the map which  forgets the $\Z_2$ action of the involution and $c_1=\imath\circ c_1^{\rr{R}}$ coincides with the (usual) Chern class.
\end{proposition}
\medskip

\noindent
The first isomorphism ${\rm Vec}^1_{\rr{R}}(\n{T}^2,\rr{f})\simeq H^2_{\Z^2}(\n{T}^2,\Z(1))$ is known as Kahn's isomorphism. It has been proved in \cite{kahn-59}; see also 
\cite[Section 5.2]{denittis-gomi-14} and \cite[Appendix A]{gomi-13}.
The map $c_1^{\rr{R}}$ establishing the isomorphism is called (first) \virg{Real} Chern class and the target space $H^2_{\Z^2}(\n{T}^2,\Z(1))$ is the second equivariant cohomology group of the involutive space $(\n{T}^2,\rr{f})$ with local system of coefficient $\Z(1)$ (for an introduction to the equivariant cohomology we refer to \cite[Section 5.1]{denittis-gomi-14} and references therein).
The last isomorphism $H^2\big(\n{T}^2,\Z\big) \simeq\Z$ is very well known in literature. 
The equality $c_1=\imath\circ c_1^{\rr{R}}$ follows from the definition of $c_1^{\rr{R}}$ \cite{kahn-59}. The only missing new ingredient  to complete the proof of Proposition \ref{prop:cohom-R} is the justification of the second isomorphism established by the forgetting map $\imath$. This is supplied by Lemma \ref{lemma:iso_forg}.

\medskip

Proposition \ref{prop:cohom-R} says that each complex line bundle over $\n{T}^2$ admits 
a unique (up to isomorphisms) 
 \virg{Real} structure compatible with the flip involution $\rr{f}$. As a consequence,
since the \virg{Real} structure does not introduce any additional topological information,
the 
(usual) Chern class provides a complete topological characterization for the line bundle.

\medskip

We are now in position to sketch the  proof of Theorem 
\ref{theo:main1}.

\proof[{Proof of Theorem \ref{theo:main1}}]
From Proposition \ref{prop:LL-VB} we know that   each Landau level $E_{j}$ defines a \virg{Real} line bundle $(\bb{E}_{j},\Theta)$
over   $(\n{T}^2_B,\rr{f})$, and therefore an element of ${\rm Vec}^1_{\rr{R}}(\n{T}^2,\rr{f})$.
Proposition \ref{prop:cohom-R} clarifies that  the topology of the \virg{Real} line bundle $(\bb{E}_{j},\Theta)$ is completely described by the Chern class $c_1(\bb{E}_{j})\in\Z$
of the underlying complex line bundle. The computation 
 $c_1(\bb{E}_{j})=1$ for all $j\in\N_0$ is not new in literature (see \eg
\cite[Lemma 5]{bellissard-elst-schulz-baldes-94} or \cite{kunz-87}).
 However, we will present a  different computation  in Section \ref{subsect:VB-Chern-N}.
\qed

\subsection{The rank of the Landau projections}
\label{subsect:ttrace_unit_N}
The purpose of this section is to complete the proof of Lemma
\ref{lemma:deg-eigen_XX} which provides the computation of the rank of the 
(complex) vector bundle $\bb{E}_{j}$ associated to the Landau projection $\Pi_j$ by the construction described in Section \ref{subsect:VB-R-landau}. Indeed, we will do a little more, proving the first of formulas \eqref{eq:top_dix_intro}, \ie
\begin{equation}\label{eq:top_dix_intro_01}
{\rm rk}(\bb{E}_{j})\;=\;{\rm Tr}_{\rm Dix}\big(Q_{B,\xi}^{-1}\Pi_j\big)\;=\;1\;.
\end{equation}

\medskip

The symbol ${\rm Tr}_{\rm Dix}$ in \eqref{eq:top_dix_intro_01}
denotes the  \emph{Dixmier trace} (see Appendix \ref{sec:Limit_formula},and references therein for a crash course
on the subject). The operator 
$Q_{B,\xi}^{-1}$ is the resolvent of
\begin{equation}\label{eq:Q_op1}
Q_B\;:=\;\frac{1}{2}\left(
K_1^2+K_2^2+G_1^2+G_2^2\right) 
\end{equation}
where the operators $K_j$ and $G_j$ are defined by \eqref{eq:Kms}.
From  \eqref{eq:cr-an-op} one obtains that
$$
Q_B\;=\;\rr{a}^+\rr{a}^-+\rr{b}^+\rr{b}^-+2{\bf 1}\;.
$$
The last equality 
suggests that $Q_B$ is a self-adjoint operator diagonalized by the Laguerre basis \eqref{eq:herm2}. Indeed one has that 
$$
Q_B \psi_{(n,m)}\;=\;(n+m+2)\ \psi_{(n,m)},\qquad\qquad\ \forall\ (n,m)\in\N^2_0\;.
$$
Then, $Q_B$ has a pure point positive spectrum given by
$$
\sigma(Q_B)\;=\;\left\{\lambda_j:=j+2\ |\ j\in\N_0\right\}
$$
and every eigenvector $\lambda_j$ has a finite {multiplicity} $\text{Mult}[\lambda_j]=j+1$. The eigenspace associated to 
$\lambda_j$ is spanned by $\{\psi_{(n,m)}\ |\ n+m=j\}$. 
Finally 
\begin{equation}\label{eq:Q_op2}
Q_{B,\xi}^{-s}\;:=\;(Q_B+2\xi{\bf 1})^{-s}\;\in\;\s{K}(L^2(\R^2))
\end{equation}
is a compact operator for every $s>0$ and  $\xi\geqslant 0$.

\begin{remark}[Relation with the harmonic oscillator]{\upshape
Starting from  \eqref{eq:Kms} and \eqref{eq:Dms}, one can rewrite $Q_B$ 
in the form \eqref{eq:landau_harmonic_int}, namely as a 
two-dimensional harmonic oscillators in the dimensionless variable ${x_j}/{\sqrt{2}\ell_B}$  and  frequency $\omega={1}/{\sqrt{2}}$.
}\hfill $\blacktriangleleft$
\end{remark}

\medskip

Lemma \ref{lemma:meas_Q} states that $Q_{B,\xi}^{-s}$ is trace class for  for all $s>2$ and that  
$Q_{B,\xi}^{-2}$ is a measurable operator in the Dixmier ideal.
However, this measurability properties  change when  $Q_B^{-s}$ is multiplied by the Landau projection $\Pi_j$. Lemma \ref{lemma:meas_Q_2} shows that $Q_{B,\xi}^{-s}\Pi_j$ is trace class for  for all $s>1$ and that  
$Q_{B,\xi}^{-1}\Pi_j$ is a measurable operator in the Dixmier ideal. 
In particular Lemma \ref{lemma:meas_Q_2} (ii) provides the proof of the second equality in \eqref{eq:top_dix_intro_01}.
Therefore, the remaining part of this section will be devoted to the proof of the first equality in \eqref{eq:top_dix_intro_01}.

\medskip
The rank of the vector bundle $\bb{E}_{j}$ can be computed as
$$
{\rm rk}(\bb{E}_{j})\;=\;  {\rm Tr}_{L^2(\n{Y}_B)}\big(\Pi_j(k)\big)
\;=\;\int_{\n{B}_B} \dd\mu(k)\; {\rm Tr}_{L^2(\n{Y}_B)}\big(\Pi_j(k)\big)
$$
in view of the independence of ${\rm Tr}_{L^2(\n{Y}_B)}(\Pi_j(k))$ and the normalization of the measure $\dd\mu(k)$. The next result needs the trace per unit volume $\s{T}_B$ defined in Appendix \ref{subsect:ttrace_unit_N_dix}.
\begin{lemma}\label{lemma:trac1}
Let $\Pi_j$ be the $j$-th Landau projection. It holds true that
$$
\int_{\n{B}_B}\dd\mu(k)\; {\rm Tr}_{L^2(\n{Y}_B)}\big(\Pi_j(k)\big)\;=\;|\n{Y}_B|\;\s{T}_B(\Pi_j)\;,
$$
where the proportionality factor is the volume of the unit cell $\n{Y}_B$.
\end{lemma}
\proof
Let
$\chi_{\n{Y}_B}$ be the characteristic function of the unit cell $\n{Y}_B$. The set of functions $g_m(x):=\frac{1}{2\sqrt{\pi}\ell_B}\chi_{\n{Y}_B}(x)\expo{\ii\frac{\sqrt{\pi}}{\ell_B}m\cdot x}$ provides an orthonormal basis of $L^2(\n{Y}_B)$. Moreover, one has that
$$
{\rm Tr}_{L^2(\R^2)}\big(\chi_{\n{Y}_B} \Pi_j \chi_{\n{Y}_B}\big)\;=\;\sum_{m\in\Z^2}\langle g_m,\Pi_j g_m\rangle_{L^2(\R^2)}\;.
$$
in view of Lemma \ref{lemma:trace_unit_vol}.
Since the Bloch-Floquet-Zak is unitary it follows that 
$$
\langle g_m,\Pi_j g_m\rangle_{L^2(\R^2)}\;=\;\int_{\n{B}_B}\dd\mu(k)\; \langle{(\bb{U}_B g_m)}_k,\Pi_j(k){(\bb{U}_B g_m)}_k\rangle_{L^2(\n{Y}_B)}\;
$$
with ${(\bb{U}_B g_m)}_k(y):=\frac{1}{2\sqrt{\pi}\ell_B}\expo{\ii\left(\frac{\sqrt{\pi}}{\ell_B}m-k\right)\cdot y}$, after a
 straightforward calculation. For every fixed $k\in \n{T}^2_B$ the vectors ${(\bb{U}_B g_m)}_k$ provides an  orthonormal basis of $L^2(\n{Y}_B)$. Since $\Pi_j(k)$ is finite rank, hence trace class, for all $k\in\n{T}^2_B$ (see Section \ref{subsect:VB-R-landau}) one gets the equality \begin{equation}\label{eq:smikz_2}
{\rm Tr}_{L^2(\R^2)}\big(\chi_{\n{Y}_B} \Pi_j \chi_{\n{Y}_B}\big)\;=\;\int_{\n{B}_B}\dd\mu(k)\; {\rm Tr}_{L^2(\n{Y}_B)}\big(\Pi_j(k)\big)\;.
\end{equation}
The claim follows by observing that
$$
\s{T}_B(\Pi_j)\;=\;\frac{1}{|\chi_{\n{Y}_B}|}{\rm Tr}_{L^2(\R^2)}\big(\chi_{\n{Y}_B} \Pi_j \chi_{\n{Y}_B}\big)
$$
in view of Lemma \ref{lemma:trace_unit_vol} and definition \ref{eq:T_P_proj}.
\qed

\medskip

\begin{corollary}\label{corol:rank}
The rank of the vector bundle $\bb{E}_{j}$ associated to the Landau projection $\Pi_j$ is given by de formula \eqref{eq:top_dix_intro_01}.
\end{corollary}
\proof
From Lemma \ref{lemma:trac1} it follows that  
${\rm rk}(\bb{E}_{j})=|\n{Y}_B|\s{T}_B(\Pi_j)$ and from Theorem \ref{teo:Dix_Tr_for} one gets ${\rm rk}(\bb{E}_{j})={\rm Tr}_{\rm Dix}(Q_{B,\xi}^{-1}\Pi_j)$ since $|\n{Y}_B|=2\Omega_B$. The last equality follows from Lemma \ref{lemma:meas_Q_2} (ii).
\qed

\subsection{The Chern numbers of the  Landau projections}
\label{subsect:VB-Chern-N}
In this section we want to compute the Chern class of the 
(complex) line bundle $\bb{E}_{j}$ associated to the Landau projection $\Pi_j$. 
More precisely,  we will  prove the second of formulas \eqref{eq:top_dix_intro}, \ie
\begin{equation}\label{eq:top_dix_intro_02}
c_1(\bb{E}_{j})\;=\; \frac{\ii}{\ell_B^2}{\rm Tr}_{\rm Dix}\big(Q_{B,\xi}^{-1}\Pi_j\big[\partial_1(\Pi_j),\partial_2(\Pi_j)\big]\big)\;=\;1
\end{equation}
completing, in this way, the proof of Theorem \ref{theo:main1}.

\medskip

The spectral line bundle  $\bb{E}_{j}$ is defined through the family of projections $\kappa\mapsto \Pi_j$ and in this case it is well known that the two-form which represents the Chern class of  $\bb{E}_{j}$ is given by
$$
\widetilde{c}_1(\bb{E}_{j})\;:=\;\frac{\ii}{2\pi}R_j(k) \dd k
$$ 
where $ \dd k:=\dd k_1\wedge \dd k_2$ is the (non normalized) volume-form and
\begin{equation}\label{eq:def_R_j}
R_j(k)\;:=\;{\rm Tr}_{L^2(\n{Y}_B)}\big(\Pi_j(k)\big[\partial_{k_1}\Pi_j(k),\partial_{k_2}\Pi_j(k)\big]\big)
\end{equation}
is the trace of the Grassmann-Berry curvature (see \cite[Section II.F]{denittis-gomi-15} and \cite[Section 12.5]{taubes-11}).
The Chern number of $\bb{E}_{j}$ is then given by
\begin{equation}\label{eq:def_R_j_2}
{c}_1(\bb{E}_{j})\;=\;\int_{\n{B}_B}\widetilde{c}_1(\bb{E}_{j})\;=\;\frac{\ii 2\pi}{|\n{Y}_B|}\int_{\n{B}_B}\dd\mu(k)\; R_j(k)
\end{equation}
where in the last equality the  normalized measure $\dd\mu(k)$ and the relation $|\n{B}_B||\n{Y}_B|=(2\pi)^2$ have been used.

\medskip

The next task is to rewrite the formula for ${c}_1(\bb{E}_{j})$ in the position space $L^2(\R^2)$.
Let $X_1$ and $X_2$ be the position operators on $L^2(\R^2)$ and $\Pi_j$ the $j$-th Lanadau projection. The two commutators $
\partial_i(\Pi_j):=-\ii[X_i,\Pi_j]$, $i=1,2$, define integral operators with kernels
\begin{equation}\label{eq:int_deriv_P}
\partial_i(\Pi_j)(x,y)\;=\;-\ii(x_i-y_i)\Pi_j(x,y)
\end{equation}
where $\Pi_j(x,y)$ is the kernel \eqref{L_proj_ker}. It turns out that $
\partial_i(\Pi_1)$ and $
\partial_i(\Pi_2)$ are
bounded operators that commute with the  magnetic translations. Therefore, they can be decomposed by means of transformation the {magnetic} Bloch-Floquet-Zak transform.
An explicit calculation provides 
\begin{equation}\label{eq:BFZ_deriv}
\partial_i(\Pi_j)\;\longmapsto\;\bb{U}_B\partial_i(\Pi_j)\bb{U}_B^{-1}:=\int^\oplus_{\n{B}_B} \dd\mu(k)\;  \partial_{k_i}\Pi_j(k)\;.
\end{equation}
\begin{lemma}\label{lemma:chern1}
Let $\Pi_j$ be the $j$-th Landau projection.
The bounded operator $\Pi_j[\partial_1(\Pi_j),\partial_2(\Pi_j)]$ admits the trace per unit volume and
\begin{equation}\label{eq:claim_chern}
{c}_1(\bb{E}_{j})\;=\;\ii 2\pi\;\s{T}_B\big(\Pi_j[\partial_1(\Pi_j),\partial_2(\Pi_j)]\big)\;.
\end{equation}
\end{lemma}
\proof
Let $\s{R}_j:=\Pi_j[\partial_1(\Pi_j),\partial_2(\Pi_j)]$. Since $\partial_1(\Pi_j)$ and $\partial_2(\Pi_j)$ are bounded and invariant under magnetic translations and the definition  it follows that also $\s{R}_j$ meets the same properties.
From Lemma \ref{lemma:trace_unit_vol} one deduces that $\chi_{\n{Y}_B}\Pi_j[\partial_1(\Pi_j),\partial_2(\Pi_j)]\chi_{\n{Y}_B}$ is trace class. The invariance under magnetic translations, along with the definition \eqref{eq:T_P_proj}, justifies the equality
$$
\s{T}_B\big(\s{R}_j\big)\;=\;\frac{1}{|\n{Y}_B|}
{\rm Tr}_{L^2(\R^2)}\big(\chi_{\n{Y}_B}\s{R}_j\chi_{\n{Y}_B}\big)\;.
$$
The same argument used to prove equation \eqref{eq:smikz_2} can be used to 
obtain
$$
{\rm Tr}_{L^2(\R^2)}\big(\chi_{\n{Y}_B}\s{R}_j\chi_{\n{Y}_B}\big)\;=\;
\int_{\n{B}_B}\dd\mu(k)\; {\rm Tr}_{L^2(\n{Y}_B)}\big(\s{R}_j(k)\big)
$$
where $\s{R}_j(k)$ is the Bloch-Floquet-Zak decomposition of the operator $\s{R}_j$. A comparison between the definition of $\s{R}_j$, the Bloch-Floquet-Zak decomposition of $\partial_i(\Pi_j)$ provided by \eqref{lemma:chern1}, and the definition of $R_j(k)$ in \eqref{eq:def_R_j}
gives that $R_j(k)= {\rm Tr}_{L^2(\n{Y}_B)}(\s{R}_j(k))$.
This implies that 
$$
\s{T}_B\big(\s{R}_j\big)\;=\;\frac{1}{|\n{Y}_B|}
\int_{\n{B}_B}\dd\mu(k)\;R_j(k)\;.
$$
By putting together the latter equation with \eqref{eq:def_R_j_2} one obtains equation \eqref{eq:claim_chern}.
\qed

\medskip

\begin{corollary}\label{corol:chern}
The Chern class of the vector bundle $\bb{E}_{j}$ associated to the Landau projection $\Pi_j$ is given by  formula \eqref{eq:top_dix_intro_02}.
\end{corollary}
\proof
Let us start simplifying the expression of the operator $\s{R}_j$ introduced in the proof of Lemma \ref{eq:claim_chern}.
The position operators can be expressed as
$X_1=\ell_B(K_2-G_1)$ and $X_2=\ell_B(G_2-K_1)$. Since the Landau projections commute with $G_1$ and $G_2$ one gets that $\partial_1(\Pi_j)=-\ii\ell_B[K_2,\Pi_j]$
and $\partial_2(\Pi_j)=\ii\ell_B[K_1,\Pi_j]$. Since $K_1=\frac{1}{\sqrt{2}}(\rr{a}^++\rr{a}^-)$
and $K_2=\frac{1}{\ii\sqrt{2}}(\rr{a}^+-\rr{a}^-)$ one obtains
\begin{equation}\label{eq:commut_proj_ident_I}
\begin{aligned}
\partial_1(\Pi_j)\;&=\;-\frac{\ell_B}{\sqrt{2}}\left([\rr{a}^+,\Pi_j]-[\rr{a}^-,\Pi_j]\right)\\
\partial_2(\Pi_j)\;&=\;\ii\frac{\ell_B}{\sqrt{2}}\left([\rr{a}^+,\Pi_j]+[\rr{a}^-,\Pi_j]\right)
\\
\end{aligned}
\end{equation}
and in turn
$$
[\partial_1(\Pi_j),\partial_2(\Pi_j)]\;=\;-\ii\ell_B^2\big[[\rr{a}^+,\Pi_j],[\rr{a}^-,\Pi_j]\big]\;.
$$
The relations \eqref{eq:shift_P}  along with the orthogonality of the Landau projections and the commutation relations $[\rr{a}^-,\rr{a}^+]={\bf 1}$ provide
$$
[\partial_1(\Pi_j),\partial_2(\Pi_j)]\;=\;-\ii\ell_B^2\big(\Pi_j+(j-1)\Pi_{j-1}-(j+1)\Pi_{j+1}\big)\;.
$$
Exploiting again  the orthogonality of the  projections one gets $
\s{R}_j=-\ii\ell_B^2\Pi_j$. It turns out that $\s{R}_j$ is in the ideal where Theorem \ref{teo:Dix_Tr_for} holds true. Therefore, from Lemma \ref{eq:claim_chern} and Lemma \ref{lemma:meas_Q_2} (ii) we deduce formula \eqref{eq:top_dix_intro_02}.
\qed

\begin{remark}[An integral identity]{\upshape
The operator $\s{R}_j$ defined in the  proof of Lemma \ref{lemma:chern1} has an integral kernel which can be  explicitly computed from 
\eqref{L_proj_ker} and \eqref{eq:int_deriv_P}. Along   the diagonal $x=y$ the kernel reads
$$
\varrho_j(x)\;=\;\int_{\R^2\times\R^2}\dd y\dd z\; \big[
x\wedge z +z\wedge y +y\wedge x
\big]\Pi_j(x,y)\Pi_j(y,z)\Pi_j(z,x)\;.
$$
On the other hand, in view of the equality $
\s{R}_j=-\ii\ell_B^2\Pi_j$, one gets $\varrho_j(x)=\frac{-\ii}{2\pi}$.
This leads to the following integral identity
$$
\int_{\R^2\times\R^2}\dd y\dd z\; f_x(y,z)\expo{\ii f_x(y,z)}\Psi_j(x-y)\Psi_j(y-z)\Psi_j(z-x)\;=\;\frac{\pi^2}{2\ii}
$$
where $ f_x(y,z):=x\wedge z +z\wedge y +y\wedge x$ and $\Psi_j(x):=\expo{-\frac{|x|^2}{2}}L^{(0)}_j(|x|)$.
}\hfill $\blacktriangleleft$
\end{remark}

\section{The geometry of the non-Abelian Landau levels}
\label{sec:mSo-no-ab}
The magnetic Hamiltonian with non-Abelian magnetic field acts on the Hilbert space $L^2(\R^2)\otimes\C^{2s+1}$ which describes particles of spin $s$ and is defined 
in a similar way  to  \eqref{eq:mSo1}
 by 
\begin{equation}\label{eq:mSo1-NA}
\s{H}_{\s{A}}\;:=\;\dfrac{1}{2m}\left(-\ii \hslash \nabla\otimes{\bf 1}_{2s+1}-\frac{q}{c}\s{A}\right)^2
\end{equation}
where  ${\bf 1}_{2s+1}$ is the identity matrix acting on  $\C^{2s+1}$  and $\s{A}$
is a
 \emph{non-abelian} vector potential. In this paper we will focus on  vector potential of the form
\begin{equation}\label{eq:vp-nonab}
\s{A}(x_1,x_2)\;:=\;A(x_1,x_2)\otimes{\bf 1}_{2s+1}\;+\;b\; {\bf 1}_{L^2}\otimes  (\gamma_1,\gamma_2)\;,
\end{equation}
where $A$ is a usual vector potential which generates an orthogonal magnetic field $B=\nabla\times A$, ${\bf 1}_{L^2}$ is the identity operator on the coordinate space $L^2(\R^2)$,
$b$ a coupling constant (with the dimension of a magnetic field times a  length),
 and $\gamma_1,\gamma_2\in{\rm Mat}_{2s+1}(\C)$ two hermitian matrices.
 We will use the short notation 
 ${\bf 1}:={\bf 1}_{L^2}\otimes{\bf 1}_{2s+1}$ for the identity operator on the full space.
 The non-abelian magnetic field (or curvature) associated with $\s{A}$ is given by the equation
 $$
 \s{B}\;:=\;\nabla\times \s{A}-\frac{\ii}{\hslash} \s{A}\times \s{A}\;.
 $$
We will refer to \cite{estienne-haaker-schoutens-11} (and references therein) for more details about non-Abelian magnetic fields. We just notice that with the choice \eqref{eq:vp-nonab} The non-abelian magnetic field turns out to be orthogonal, \ie 
$\s{B}=(0,0,\s{B}_\bot)$ with 
\begin{equation}\label{eq:orth_NA_MF}
\s{B}_\bot\;:=\;B\otimes{\bf 1}_{2}- \ii\frac{b^2}{\hslash}{\bf 1}_{L^2}\otimes [\gamma_1,\gamma_2]\;.
\end{equation}

\subsection{The non-Abelian Landau Hamiltonian}
\label{subsec:LH_NA}
Hereafter we will assume that the Abelian part of the vector potential  which enters in the definition of the non-Abelian magnetic Hamiltonian \eqref{eq:mSo1-NA} is given by the potential $A_L$ defined in \eqref{eq:vp}. We will denote with the symbol $H_{B,b}(\gamma_1,\gamma_2)$ the related 
non-Abelian Landau Hamiltonian.

\medskip

The introduction of the \emph{non-Abelian kinetic momenta}
\begin{equation}\label{eq:Kms-nonAb0}
\begin{aligned}
\s{K}_{1}\;&:=\;K_1\otimes {\bf 1}_{2s+1}\;-\;{c}_{B}\; {\bf 1}_{L^2}\otimes\gamma_1\\
\s{K}_{2}\;&:=\;K_2\otimes {\bf 1}_{2s+1}\;-\;{c}_{B}\; {\bf 1}_{L^2}\otimes\gamma_2\;,
\end{aligned}
\end{equation}
with $K_1$ and $K_2$ being defined by \eqref{eq:Kms} and 
$$
c_b\;:=\; \frac{b}{B\ell_B}\;
$$
allows to write
\begin{equation}\label{eq:LH-NA}
\s{H}_{B,b}(\gamma_1,\gamma_2)\;=\; \frac{\epsilon_B}{2}\left(\s{K}_{1}^2+\s{K}_{2}^2\right)\;.
\end{equation}
A simple computation shows that
$$
\s{H}_{B,b}(\gamma_1,\gamma_2)\;=\;H_B\otimes {\bf 1}_{2s+1}\;+\;c_b\s{W}_1\;+\;c_b^2\s{W}_2
$$
where $H_B$ is the Landau Hamiltonian described in Section \ref{subsec:LH}, and
$$
\s{W}_1\;:=\;-{\epsilon_B}\big(K_1\otimes \gamma_1+K_2\otimes \gamma_2\big)\;,\qquad \s{W}_2\;:=\;\frac{\epsilon_B}{2}{\bf 1}_{L^2}\otimes\big(\gamma_1^2+\gamma_2^2\big)
$$
are perturbations.

\medskip

The non-Abelian kinetic momenta obey the following commutation relation
$$
[\s{K}_{1},\s{K}_{2}]\;=\;-\ii\; {\bf 1}\;+\;c_b^2\;{\bf 1}_{L^2}\otimes[\gamma_1,\gamma_2]\;.
$$ 
They have a \virg{canonical} commutation relation when
$$
[\gamma_1,\gamma_2]\;=\;-\ii\delta {\bf 1}_{2s+1}\;,\qquad\quad \delta\in\R\;.
$$ 

\medskip

The dual momenta $G_1\otimes  {\bf 1}_{2s+1}$ and $G_2\otimes  {\bf 1}_{2s+1}$ commute with $\s{K}_{1}$, $\s{K}_{2}$ and consequently with $H_{B,b}(\gamma_1,\gamma_2)$. 
This implies that the magnetic translations $T_m\otimes {\bf 1}_{2s+1}$ are symmetries of the Hamiltonian $\s{H}_{B,b}(\gamma_1,\gamma_2)$. 
As a consequence, the magnetic Bloch-Floquet-Zak transform $\bb{U}_B \otimes {\bf 1}_{2s+1}$ decomposes $\s{H}_{B,b}(\gamma_1,\gamma_2)$ in a family of Hamiltonians acting on the fiber Hilbert space $ L^2(\n{Y}_B)\otimes\C^{2s+1}$
and parametrized by the points of the 
{Brillouin torus} $\n{B}_B$. The details of the construction can be recovered from Section \ref{subsect:magn_BFZ_tras}.

\medskip

Even the construction of Section \ref{subsect:VB-R-landau} can be extended to the non-Abelian case.
More precisely every isolated spectral region $\Sigma\subset \sigma(\s{H}_{B,b}(\gamma_1,\gamma_2))$ separated from the rest of the spectrum by gaps defines  a spectral projection $\s{P}_\Sigma$ and in turn a  projection-valued map $\n{B}_B\ni k\mapsto \s{P}_\Sigma(k)$. The latter provides a vector bundle $\bb{E}_{\Sigma}\to \n{B}_B$ according to the prescription \eqref{eq_E_j_bund}. In the next sections we will study the topology of the spectral bundles obtained from two distinct models of non-Abelian Landau Hamiltonians.

\medskip

In the following we will focus our attention on the special (but interesting) case of particles with spin
$s=\frac{1}{2}$. In this case a basis of the algebra  ${\rm Mat}_{2}(\C)$ is given by the identity ${\bf 1}_2$ and the three  Pauli matrices
$$
\sigma_1\;:=\;\left(\begin{array}{cc}
0 & 1 \\ 1 & 0\end{array}\right)\;,\quad
\sigma_2\;:=\;\left(\begin{array}{cc}
0 & -\ii \\ \ii & 0\end{array}\right)\;,\quad
\sigma_3\;:=\;\left(\begin{array}{cc}
1 & 0 \\ 0 & -1\end{array}\right)\;.
$$
%


\subsection{The Jaynes-Cummings model}
\label{subsect:Jaynes-Cummings} 
This model corresponds to the choice of
$$
\gamma_1\;=\;-\sigma_2\;,\qquad\quad \gamma_2\;=\;+\sigma_1
$$
in the non-abelian vector potential \eqref{eq:vp-nonab}. The associated orthogonal part of the non-Abelian magnetic field 
is given by
$$
\s{B}_\bot\;=\;B\otimes{\bf 1}_{2}- \ii\frac{b^2}{\hslash}{\bf 1}_{L^2}\otimes [\sigma_1,\sigma_2]\;=\;{\bf 1}_{L^2}\otimes\big(B{\bf 1}_{2}+2\frac{b^2}{\hslash} \sigma_z\big)\;.
$$
according to formula \eqref{eq:orth_NA_MF}.
This magnetic field is a $\n{U}(2)$ matrix, and is a superposition
of a $\n{U}(1)$  field $B {\bf 1}_{2}$ and a $\n{SU}(2)$ field  $2\frac{b^2}{\hslash} \sigma_z$ and for this reason one refers to it as a  $\n{U}(1)\times \n{SU}(2)$ gauge field (see \eg \cite{palmer-pachos-11}).

\medskip

The resulting non-Abelian Landau Hamiltonian 
is given by 
\begin{equation}\label{eq:Ham_JC}
\s{H}_{JC}\;=\;H_B\otimes{\bf 1}_2 \;+\;c_b\epsilon_B \s{W}_{JC}\;+\;c_b^2\epsilon_B\;{\bf 1}
\end{equation}
with the perturbation given by 
$$
\begin{aligned}
\s{W}_{JC}\;&=\; \big(K_1\otimes \sigma_2-K_2\otimes \sigma_1\big)\\
&=\;\ii\rr{a}^+\otimes\left(\frac{\sigma_1-\ii\sigma_2}{\sqrt{2}}\right)\;-\;\ii\rr{a}^-\otimes\left(\frac{\sigma_1+\ii\sigma_2}{\sqrt{2}}\right)\;.
\end{aligned}
$$
The first equality says that $\s{W}_{JC}$ is the \emph{Rashba spin-orbit coupling}  \cite{wang-vasilopoulos-05,zhang-06} while the second equality shows that $\s{W}_{JC}$ can be interpreted as the  celebrated \emph{Jaynes-Cummings potential} \cite{shore-93}. The latter observation justifies the use of the  expression Jaynes-Cummings model for the Hamiltonian $H_{JC}$.

\medskip

The Jaynes-Cummings model can be 
 solved exactly \cite{wang-vasilopoulos-05,zhang-06,estienne-haaker-schoutens-11,juarez-zuniga-moya-14}. In matricial form the Hamiltonian \eqref{eq:Ham_JC}  reads
$$
\s{H}_{JC}\;=\;
\epsilon_B\;\left(
\begin{array}{cc}
\rr{a}^{+}\rr{a}^{-}+\frac{1}{2}\left(1+2c_b^2\right){\bf 1} & -\ii\sqrt{2} c_b\;\rr{a}^- \\ 
\ii\sqrt{2} c_b\;\rr{a}^+ & \rr{a}^{+}\rr{a}^{-}+\frac{1}{2}\left(1+2c_b^2\right){\bf 1}\end{array}
\right)\;.
$$
Let us introduce the family of vectors
$$
\Phi_{(j,m)}^{\pm}\;:=\;
\left(\begin{array}{c}
\sin(\theta_j^\pm)\;\psi_{(j-1,m)} \\
\ii \cos(\theta_j^\pm)\; \psi_{(j,m)}
\end{array}\right)\;,\qquad\quad (j,m)\in\N_0^2
$$
where the $\psi_{(j,m)}$ are the elements of the Laguerre basis \eqref{eq:lag_pol} (with the convention $\psi_{(-1,m)}=0$) and  the angles $\theta_j^\pm$ are defined by the relation 
$$
\tan\big(\theta_j^\pm\big)\;:=\;\frac{\sqrt{8c_b^2j}}{1\pm\sqrt{1+8c_b^2j}}\;.
$$
A direct computation shows
that the family $\Phi_{(j,m)}^{\pm}$ defines a 
complete orthonormal system in $L^2(\R^2)\otimes\C^2$ which diagonalizes the 
Jaynes-Cummings model. Indeed, it holds that  $\s{H}_{JC}\Phi_{(j,m)}^{\pm}=E_{j}^{\pm}\Phi_{(j,m)}^{\pm}$ where the eigenvalues are given by the formula
\begin{equation}\label{eq:JC_spec}
\left\{
\begin{aligned}
&E_{0}\;:=\;\epsilon_B\left(\frac{1}{2}+c_b^2\right)&\quad&\text{if}\;\;j=0\\
&E_{j}^\pm\;:=\;\epsilon_B\left(j\pm\frac{1}{2}\sqrt{1+j8c_b^2}+c_b^2\right)&\quad&\text{if}\;\;j>0\;.
\end{aligned}
\right.
\end{equation}
Therefore, the spectrum of $\s{H}_{JC}$ is pure point and each eigenvalue is infinitely degenerate  as a consequence of the commutation relation with the dual momenta $G_1\otimes  {\bf 1}_{2}$ and $G_2\otimes  {\bf 1}_{2}$. Every eigenvalue $E_{j}^\pm$ defines a 
spectral projections $\s{P}_j^\pm$ given by the formula
\begin{equation}\label{eq_spec_proj_HCJ}
\s{P}_j^\pm\;=\;\left(\begin{array}{cc}\sin(\theta_j^\pm)^2\Pi_{j-1} & -\ii \frac{\sin(\theta_j^\pm)\cos(\theta_j^\pm)}{\sqrt{j}}\rr{a}^-\Pi_j\\
\ii \frac{\sin(\theta_j^\pm)\cos(\theta_j^\pm)}{\sqrt{j}}\Pi_j \rr{a}^+& \cos(\theta_j^\pm)^2\Pi_{j}\end{array}\right)\;
\end{equation}
where the $\Pi_j$ are the Landau projections \eqref{eq:Lan_proj}.

\medskip

The Hamiltonian $\s{H}_{JC}$ has a relevant discrete symmetry.
Let us define the \emph{twisted} flip operator 
$$
\s{F}\;:=\;F\;\otimes\;\vartheta\;=\;\left(\begin{array}{cc}
 F &0\\ 0& -\ii F
\end{array}\right)
$$
where  $F$ is the flip operator defined by \eqref{eqq:flip_ones} and 
and $\vartheta\in{\rm Mat}_2(\C)$
is (up to a phase) the unitary operator that meets the relations $\vartheta\sigma_1\vartheta^{-1}=-\sigma_2$ and $\vartheta\sigma_2\vartheta^{-1}=\sigma_1$. Starting from the matricial form of the  kinetic momenta
$$
\s{K}_{1}\;=\;
\left(\begin{array}{cc}
K_1 & -\ii\;{c}_{b} \\ 
+\ii\;{c}_{b} & K_1\end{array}\right)\;,\qquad
\s{K}_{2}\;=\;
\left(\begin{array}{cc}K_2 & -{c}_{b} \\ 
-{c}_{b} & K_2\end{array}\right)\;,
$$
and with the help of the relations \eqref{eq:disc_sym_1_abo2},
one easily verifies that
\begin{equation}\label{eq:spin_flip1}
\s{F}\s{K}_1\s{F}^{-1}\;=\;
\left(\begin{array}{cc}
G_1 &c_b\\ c_b & G_1
\end{array}\right)\;\qquad
\s{F}\s{K}_2\s{F}^{-1}
\;=\;\left(\begin{array}{cc}
G_2 &-\ii c_b\\ \ii c_b & G_2
\end{array}\right)\;.
\end{equation}
Let $\Xi:=\s{F}C$ be the  composition of the 
{twisted} flip operator $\s{F}$ and the complex conjugation $C$ on $L^2(\R^2)\otimes\C^2$. 
By combining the relations \eqref{eq:disc_sym_1_abo1}
and \eqref{eq:spin_flip1}, one gets
\begin{equation}
\Xi \s{K}_1\Xi^{-1}\;=\;-\s{K}_2\;,\qquad\quad \Xi \s{K}_2\Xi^{-1}\;=\;-\s{K}_1
\end{equation}
that in turn implies
\begin{equation}\label{eq:re_struc_HCJ}
\Xi \s{H}_{CJ}\Xi^{-1}\;=\;\s{H}_{CJ}
\end{equation}
 in view of the general structure \eqref{eq:LH-NA} for   non-Abelian Landau Hamiltonians.
Moreover, a direct computation shows that $\Xi^2=\s{F}\s{F}^{-1}={\bf 1}$, showing that  $\Xi$ is an anti-linear involution on $L^2(\R^2)\otimes\C^2$ or a generalized even TRS in the sense of Definition \ref{def:introG-TRS}.

\medskip

As a consequence of the invariance of $\s{H}_{JC}$ under the magnetic translations, and the consequent possibility of defining the  {magnetic} Bloch-Floquet-Zak transform, a vector bundle 
$\bb{E}_{j}^\pm\to \n{B}_B$ can be associated to  each spectral projector $\s{P}_j^\pm$ (see the construction of Section \ref{subsect:VB-R-landau}). Equation \eqref{eq:re_struc_HCJ} 
implies that the  anti-unitary operator $\Xi$ endows the vector bundle 
$\bb{E}_{j}^\pm$ with a \virg{Real} structure over the involutive torus $(\n{T}^2,\rr{f})$. 
 The resulting \virg{Real} vector bundle will be denoted with $(\bb{E}_{j}^\pm,\Xi)$. The rank $r$ of $\bb{E}_{j}^\pm$ can be deduced with the same argument used in the proof of Lemma \ref{lemma:deg-eigen_XX}. Let $k\mapsto \s{P}_j^\pm(k)$ be the fiber-decomposition of  $\s{P}_j^\pm$ obtained via the {magnetic} Bloch-Floquet-Zak transform. 
The rank $r_0$ of the vector bundle $\bb{E}_{j}^\pm$ must match
the dimension of the range  of $\s{P}_j^\pm(k)$ which is computed by the trace on the space $L^2(\n{Y}_B)\otimes\C^2$. The latter is  given by 
${\rm Tr}_{L^2(\n{Y}_B)\otimes\C^2}={\rm Tr}_{L^2(\n{Y}_B)}\otimes {\rm Tr}_{\C^2}$. A direct computation shows that
\begin{equation}\label{eq:rk_CJ_mod}
\begin{aligned}
r\;&=\;{\rm Tr}_{L^2(\n{Y}_B)\otimes\C^2}\big(\s{P}_j^\pm(k)\big)\\
&=\;
\sin(\theta_j^\pm)^2\;{\rm Tr}_{L^2(\n{Y}_B)}\big(\Pi_{j-1}(k)\big)\;+\; \cos(\theta_j^\pm)^2\; {\rm Tr}_{L^2(\n{Y}_B)}\big(\Pi_{j}(k) \big)
\end{aligned}
\end{equation}
where (the fibered version of) equation
\eqref{eq_spec_proj_HCJ} has been used for the computation of the trace ${\rm Tr}_{\C^2}$.
In view of Lemma \ref{lemma:deg-eigen_XX} one immediately gets $r=1$, namely each $(\bb{E}_{j}^\pm,\Xi)$ is a \virg{Real} line bundle over $(\n{T}^2,\rr{f})$. 

\medskip

The rank of $\bb{E}_{j}^\pm$ can be also computed  in terms of the Dixmier trace. Indeed,
 Lemma \eqref{lemma:tensor_dix} provides
$$
\begin{aligned}
{\rm Tr}_{\rm Dix}\left(\big(Q_{B,\xi}^{-1}\otimes{\bf 1}_2\big) \s{P}_j^\pm\right)\;=\;&
\sin(\theta_j^\pm)^2\;{\rm Tr}_{\rm Dix}\big(Q_{B,\xi}^{-1}\Pi_{j-1}\big)\\&+\; \cos(\theta_j^\pm)^2\; {\rm Tr}_{\rm Dix}\big(Q_{B,\xi}^{-1}\Pi_{j} \big)\;
\end{aligned}
$$
and comparing the latter equation with  \eqref{eq:rk_CJ_mod} and \eqref{eq:top_dix_intro_01}
 one gets
$$
{\rm rk}\big(\bb{E}_{j}^\pm\big)\;=\; {\rm Tr}_{\rm Dix}\left(\big(Q_{B,\xi}^{-1}\otimes{\bf 1}_2\big) \s{P}_j^\pm\right)\;=\;1\;.
$$

\medskip

To classify the topological phases of the Jaynes-Cummings model
we can invoke Proposition \ref{prop:cohom-R} which ensures that the topology of  
  $(\bb{E}_{j}^\pm,\Xi)$ is completely characterized by the Chern class of $\bb{E}_{j}^\pm$ as a complex vector bundle over $\n{T}^2$. 
Therefore, to complete the proof of Theorem   \ref{theo:main3} we need to compute $c_1(\bb{E}_{j}^\pm)$.  After repeating  step by step the arguments used in Section \ref{subsect:VB-Chern-N}, and in particular in Lemma \ref {lemma:chern1},  one can prove that  
\begin{equation}\label{eq:chr_dix_CL1}
c_1\big(\bb{E}_{j}^\pm\big)\;=\;\frac{\ii 2\pi}{|\n{Y}_B|}{\rm Tr}_{L^2(\R^2)\otimes\C^2}\big(\chi_{\n{Y}_B}\s{R}_j^\pm\chi_{\n{Y}_B}\big)\;=\;\ii 2\pi\s{T}_B\big({\rm Tr}_{\C^2}\big(\s{R}_j^\pm\big)\big)
\end{equation}
where $\s{R}_j^\pm:=\s{P}^\pm_j[\partial_1(\s{P}^\pm_j),\partial_2(\s{P}^\pm_j)]$
and ${\rm Tr}_{\C^2}$ is meant as a map from the bounded operators on $L^2(\R^2)\otimes\C^2$ into the bounded operators on $L^2(\R^2)$.
Let us assume for the moment that  ${\rm Tr}_{\C^2}(\s{R}_j^\pm)$ meets the condition of Theorem \ref{teo:Dix_Tr_for}.  By combining the latter with Lemma \ref{lemma:tensor_dix}
one can rewrite equation \eqref{eq:chr_dix_CL1} as follows
\begin{equation}\label{eq:chr_dix_CL2}
c_1\big(\bb{E}_{j}^\pm\big)
\;=\;\frac{\ii}{\ell^2_B}{\rm Tr}_{\rm Dix}\left(\big(Q_{B,\xi}^{-1}\otimes{\bf 1}_2\big) \s{R}_j^\pm\right)\;.
\end{equation}
Let us study the operator ${\rm Tr}_{\C^2}(\s{R}_j^\pm)$. By setting
$$
[\partial_1(\s{P}^\pm_j),\partial_2(\s{P}^\pm_j)]\;=\;
-\ii\left(\begin{array}{cc}D_1 & L \\L^* & D_2\end{array}\right)
$$
one gets
$$
\begin{aligned}
{\rm Tr}_{\C^2}(\s{R}_j^\pm)\;=\;&-\ii \sin(\theta_j^\pm)^2\Pi_{j-1}D_1-\ii \cos(\theta_j^\pm)^2\Pi_{j}D_2\\
&
+\frac{\sin(\theta_j^\pm)\cos(\theta_j^\pm)}{\sqrt{j}}\big(\Pi_j \rr{a}^+L- \rr{a}^-\Pi_j L^*\big)\;.
\end{aligned}
$$
A tedious calculation provides
$$
\begin{aligned}
D_1\;:=\;&\ii\sin(\theta_j^\pm)^4\big[\partial_1(\Pi_{j-1}),\partial_2(\Pi_{j-1})\big]\\
&+\;\ell_B^2\sin(\theta_j^\pm)^2\cos(\theta_j^\pm)^2\big((j-1)\Pi_{j-2}+\Pi_{j-1}-j\Pi_j\big)
\end{aligned}
$$
and
$$
\begin{aligned}
D_2\;:=\;&\ii\cos(\theta_j^\pm)^4\big[\partial_1(\Pi_{j}),\partial_2(\Pi_{j})\big]\\
&+\;\ell_B^2\sin(\theta_j^\pm)^2\cos(\theta_j^\pm)^2\big(j\Pi_{j-1}+\Pi_j-(j+1)\Pi_{j+1}\big)
\end{aligned}
$$
for the diagonal elements and

$$
L\;:=\;\frac{\sin(\theta_j^\pm)\cos(\theta_j^\pm)}{\sqrt{j}}\left(\rr{a}^-[\partial_1(\Pi_{j}),\partial_2(\Pi_{j})]-\ii\ell_B^2 \Pi_j\rr{a}^-\right)
$$
for the off-diagonal one. By putting all the pieces together, and after some more algebraic manipulations, one gets
$$
\begin{aligned}
{\rm Tr}_{\C^2}(\s{R}_j^\pm)\;=\;&\sin(\theta_j^\pm)^6(-\ii\ell_B^2\Pi_{j-1})+ \cos(\theta_j^\pm)^6(-\ii\ell_B^2\Pi_{j})\\
&-\ii\ell^2_B \sin(\theta_j^\pm)^2\cos(\theta_j^\pm)^2\left(\sin(\theta_j^\pm)^2\Pi_{j-1}+\cos(\theta_j^\pm)^2\Pi_{j}\right)\\
&
+\sin(\theta_j^\pm)^2\cos(\theta_j^\pm)^2
\left((-\ii\ell_B^2\Pi_{j})+(-\ii\ell_B^2\Pi_{j-1})\right)\\
\end{aligned}
$$
where the identity $\Pi_{j}[\partial_1(\Pi_{j}),\partial_2(\Pi_{j})]=-\ii\ell_B^2\Pi_{j}$ has been repeatedly used.
Finally, by using basic trigonometric identities, one obtains
$$
{\rm Tr}_{\C^2}(\s{R}_j^\pm)\;=\;-\ii\ell^2_B\left(\sin(\theta_j^\pm)^2\Pi_{j-1}+\cos(\theta_j^\pm)^2\Pi_j
\right)\;. 
$$
The last equation shows that ${\rm Tr}_{\C^2}(\s{R}_j^\pm)$ is in the right algebra for the application of Theorem \ref{teo:Dix_Tr_for}. Then, equation \eqref{eq:chr_dix_CL2} can be used and one immediately gets
\begin{equation}\label{eq:chr_dix_CL2_02}
c_1\big(\bb{E}_{j}^\pm\big)
\;=\;\frac{\ii}{\ell^2_B}{\rm Tr}_{\rm Dix}\left(Q_{B,\xi}^{-1} {\rm Tr}_{\C^2}(\s{R}_j^\pm)\right)\;=\;1\;.
\end{equation}
%

\subsection{The \virg{Quaternionic} model}
\label{eq:non-Ab-Q} 
 This model corresponds to the  elections of
$$
\gamma_1\;=\;-\alpha\;,\qquad\quad \gamma_2\;=\;\sigma_2\alpha\sigma_2\;,
$$
in the non-abelian vector potential \eqref{eq:vp-nonab},
with 
$$
 \alpha\;=\;\alpha^\ast\;=\;\overline{\alpha}
 $$
a real and hermitian element of ${\rm Mat}_{2}(\C)$. 
The matrices $\gamma_1$ and $\gamma_2$ can be parametrized by three real parameters $r_0,r_1,r_2\in\R$ as follows
$$
\begin{aligned}
\gamma_1\;&=\;
\left(\begin{array}{cc}
-r_0-r_2 & -r_1 \\
-r_1 & -r_0+r_2\end{array}\right)\;,\qquad
\gamma_2\;=\;
\left(\begin{array}{cc}
r_0-r_2 & -r_1 \\
-r_1 & r_0+r_2\end{array}\right)\;.
\end{aligned}
$$ 
Since $[\gamma_1,\gamma_2]=0$ it follows 
from equation \eqref{eq:orth_NA_MF} that
$$
\s{B}_\bot\;=\;B\otimes{\bf 1}_{2}\;.
$$
This magnetic field corresponds to a 
$\n{U}(1)$ gauge field. As a consequence the {non-Abelian kinetic momenta} meet the canonical commutation relation $[\s{K}_1,\s{K}_2]=-\ii{\bf 1}$.

\medskip

The resulting non-Abelian Landau Hamiltonian  $\s{H}_{Q}$
is given by \eqref{eq:LH-NA} or equivalently  by 
\begin{equation}\label{eq:Ham_Q}
\s{H}_{Q}\;:=\;\epsilon_B\left(\rr{A}^{+}\rr{A}^{-}+\frac{1}{2}{\bf 1}\right)
\end{equation}
where $\rr{A}^{\pm}:=\frac{1}{\sqrt{2}}(\s{K}_1\pm\ii\s{K}_2)$ are explicitly given by
$$
\begin{aligned}
\rr{A}^{+}\;&=\;
\left(\begin{array}{cc}\rr{a}^++c_b(\expo{-\ii\frac{\pi}{4}}r_0+\expo{\ii\frac{\pi}{4}}r_2) & c_b\expo{\ii\frac{\pi}{4}} r_1 \\ c_b\expo{\ii\frac{\pi}{4}} r_1 & \rr{a}^++c_b(\expo{-\ii\frac{\pi}{4}}r_0-\expo{\ii\frac{\pi}{4}}r_2) \end{array}\right)\\
\rr{A}^{-}\;&=\;\left(\begin{array}{cc}\rr{a}^-+c_b(\expo{\ii\frac{\pi}{4}}r_0+\expo{-\ii\frac{\pi}{4}}r_2) & c_b\expo{-\ii\frac{\pi}{4}} r_1 \\
 c_b\expo{-\ii\frac{\pi}{4}} r_1 & \rr{a}^-+c_b(\expo{\ii\frac{\pi}{4}}r_0-\expo{-\ii\frac{\pi}{4}}r_2) \end{array}\right)\;.
\end{aligned}
$$
It turns out that
\begin{equation}\label{eq:Ham_Q2}
\s{H}_{Q}\;=\;H_B\otimes{\bf 1}_2 \;+\;c_b\epsilon_B \s{W}_{Q}\;+\;c_b^2\epsilon_B|r|^2\;{\bf 1}
\end{equation}
with  $|r|^2:=r_0^2+r_1^2+r_2^2=1$. The perturbation $\s{W}_{Q}$ is given by 
\eqref{eq:pot_WQ} or can be equivalently expressed by 
$$
\begin{aligned}
\s{W}_{Q}\;=\;  &\expo{\ii\frac{\pi}{4}}\rr{a}^+\otimes\big(r_0{\bf 1}_2-\ii(r_1\sigma_1+r_2\sigma_3)\big)\\
\;&+\; \expo{-\ii\frac{\pi}{4}}\rr{a}^-\otimes\big(r_0{\bf 1}_2+\ii(r_1\sigma_1+r_2\sigma_3)\big)\;.
\end{aligned}
$$

\medskip

The equation \eqref{eq:Ham_Q}, along with the commutation relation $[\rr{A}^{-},\rr{A}^{+}]={\bf 1}$, might suggest at first sight to use the technique of ladder operators to compute the spectrum of $\s{H}_{Q}$. 
However, this simple approach does not work because the operator $\rr{A}^{+}$ has no ground state (\ie the kernel is empty). More specifically consider the orthonormal basis of $L^2(\R^2)\otimes\C^{2}$
given by
$$
\Phi_{(j,m)}^{\pm}\;:=\;\frac{1}{\sqrt{2(r_1^2+r_2^2)\pm2r_2\sqrt{r_1^2+r_2^2}}}
\left(\begin{array}{c}
r_1\;\psi_{(j,m)} \\
-\big(r_2\pm\sqrt{r_1^2+r_2^2}\big)\; \psi_{(j,m)}
\end{array}\right)\;.
$$
A direct computation shows that 
$$
\rr{A}^{-}\Phi_{(0,m)}^{\pm}\;=\;c_b\expo{\ii\frac{\pi}{4}}\left( r_0\pm\ii\sqrt{r_1^2+r_2^2} \right)\Phi_{(0,m)}^{\pm}\;.
$$
and  $\Phi_{(0,m)}^{\pm}$ are the only eigenvectors of $\rr{A}^{-}$. 

\medskip

The calculation of the  spectrum of $\s{H}_{Q}$ is beyond the scope of this work and will be left for future investigations. However, from the general structure \eqref{eq:LH-NA} we know that $\s{H}_{Q}$ has a positive spectrum and we  can denote with $\s{P}_E:=\chi_{(-\infty,E]}(\s{H}_{Q})$ the Fermi at energy $E>0$. If the energy $E>0$  lies in a spectral gap of $\s{H}_{Q}$ then $\s{P}_E$ define a vector bundle $\bb{E}_{E}\to \n{B}_B$ according to the prescription \eqref{eq_E_j_bund}.
In this case the rank and the Chern class of 
$\bb{E}_{E}$ can be computed  again by
formulas of the type \eqref{eq:top_dix_intro} for the projection $\s{P}_E$. 
Even though, we are not computing exactly these numbers, we can have access to some informations
 by examining the symmetries of $\s{H}_{Q}$.

\medskip

Let us introduce the twisted flip operator $\s{F}':=F\otimes\sigma_2$. A direct check shows  that
$$
\begin{aligned}
\s{F}\s{K}_1\s{F}\;&=\;FK_1F\otimes{\bf 1}_2-{c}_{b}\;{\bf 1}_{L^2}\otimes \sigma_2\gamma_1\sigma_2\;=\;G_1\otimes{\bf 1}_2+{c}_{b}\;{\bf 1}_{L^2}\otimes \gamma_2\;,\\
\s{F}\s{K}_2\s{F}\;&=\;FK_2F\otimes{\bf 1}_2-{c}_{b}\;{\bf 1}_{L^2}\otimes \sigma_2\gamma_2\sigma_2\;=\;G_2\otimes {\bf1}_2\;+\;{c}_{b}\;{\bf 1}_{L^2}\otimes \gamma_1\;.\\
\end{aligned}
$$ 
Let $\Xi':=\s{F}'C$ be the  composition of the 
{twisted} flip operator $\s{F}'$ with the complex conjugation $C$ on  $L^2(\R^2)\otimes\C^2$. 
By combining the last relations with  \eqref{eq:disc_sym_1_abo1}
 one gets
\begin{equation}
\Xi' \s{K}_1{\Xi'}^{-1}\;=\;-\s{K}_2\;,\qquad\quad \Xi' \s{K}_2{\Xi'}^{-1}\;=\;-\s{K}_1\;,
\end{equation}
and in turn
\begin{equation}\label{eq:re_struc_H_Q}
\Xi' \s{H}_{Q}{\Xi'}^{-1}\;=\;\s{H}_{Q}
\end{equation}
 in view of the general structure \eqref{eq:LH-NA} for   non-Abelian Landau Hamiltonians.
Moreover, a direct computation shows that ${\Xi'}^2=\s{F}'C\s{F}'C=-{\s{F}'^2}=-{\bf 1}$, namely $\Xi'$ is an anti-linear anti-involution on $L^2(\R^2)\otimes\C^2$. Therefore, ${\Xi'}$ provides $\s{H}_{Q}$ with a generalized TRS of \virg{Quaternionic} type according to Definition \ref{def:introG-TRS}. The latter fact justifies the name of \virg{Quaternionic} model for $\s{H}_{Q}$. 

\medskip

The generalized TRS $\Xi'$ endows the vector bundle $\bb{E}_E$ associated with $\s{P}_E$ with a  \virg{Quaternionic} structure over the involutive torus $(\n{T}^2,\rr{f})$. The resulting \virg{Quaternionic} vector bundle will be denoted with $(\bb{E}_{E},\Xi')$.
For the theory and the classification of \virg{Quaternionic} vector bundles we refer to 
\cite{denittis-gomi-14-gen,denittis-gomi-18,denittis-gomi-18-b}. Since the fixed point set of $(\n{T}^2,\rr{f})$ is not empty 
we can deduce from \cite[Proposition 2.1]{denittis-gomi-14-gen} that $\bb{E}_{E}$ has even rank. This translates into
$$
{\rm rk}\big(\bb{E}_{E}\big)\;=\; {\rm Tr}_{\rm Dix}\left(\big(Q_{B,\xi}^{-1}\otimes{\bf 1}_2\big) \s{P}_E\right)\;\in\;2\N\;
$$
by using the same arguments as that in Section \ref{subsect:Jaynes-Cummings}. The second topological information comes from the isomorphism ${\rm Vec}^{2m}_{\rr{Q}}(\n{T}^2,\rr{f})\simeq 2\Z$ (\cf Corollary \ref{corol:appen_fin}) 
which says that the topology of $(\bb{E}_{E},\Xi')$ is completely determined by the first Chern class and that
\begin{equation}
c_1\big(\bb{E}_{E}\big)
\;=\;\frac{\ii}{\ell^2_B}{\rm Tr}_{\rm Dix}\left(\big(Q_{B,\xi}^{-1}\otimes{\bf 1}_2\big) \s{P}_E[\partial_1(\s{P}_E),\partial_2(\s{P}_E)]\right)\;\in\;2\Z\;.
\end{equation}
%

\appendix

\section{Equivariant cohomology for the flip involution}
\label{sec:Eq-co-flip}

This section  provides the computation of the twisted equivariant cohomology of the two-dimensional $\n{T}^2:=\R^2/\Z^2$ endowed with the flip involution $\rr{f}:(k_1,k_2)\mapsto (-k_2,-k_1)$.
Sometimes, 
 it is also  useful to use complex coordinates for $\n{T}^{2}\simeq\n{S}^1\times\n{S}^1$ through the identification $\n{S}^1\simeq \{z\in\C\ |\ |z|=1\}$. With this parametrization $(z_1,z_2)\in\n{T}^2$ the flip involution reads $\rr{f}:(z_1,z_2)\mapsto (\overline{z_2},\overline{z_1})$.

\medskip

The next result is needed to complete the proof of Proposition \ref{prop:cohom-R}

\begin{lemma}\label{lemma:iso_forg}
The map $\imath$ which forgets the $\Z_2$ action induces the   isomorphism
$$
H^2_{\Z^2}\big(\n{T}^2,\Z(1)\big)\;\stackrel{\imath}{\simeq}\; H^2\big(\n{T}^2,\Z\big)\;.
$$
 \end{lemma}
\proof
Let $\mathfrak{f}'$ be the involution on $\n{T}^2$ given by $\mathfrak{f}':(k_1,k_2) \mapsto (k_2,k_1)$.
Then, there is a $\Z_2$-equivariant homeomorphism $\varphi : (\n{T}^2, \mathfrak{f}) \to (\n{T}^2, \mathfrak{f}')$ given by $\varphi:(k_1, k_2) \mapsto (k_1, -k_2)$. 
This means that we can compute the equivariant cohomology groups for the involutive space $(\n{T}^2, \mathfrak{f}')$ instead of the involutive space $(\n{T}^2, \mathfrak{f})$.
The involution $\mathfrak{f}'$ on $\n{T}^2$ agrees with that induced from a natural action of the \emph{wallpaper group} \texttt{\bf cm} on $\R^2$ (cf \cite[Section 2.4]{G2}).
The
 low degree equivariant cohomology groups $H^n_{\Z_2}(\n{T}^2, \Z(k))$ 
 for the involutive space $(\n{T}^2, \mathfrak{f}')$ have been 
  computed in \cite[Theorem 1.3 \& Theorem 1.6]{G2} (by using stable splittings)  or in \cite{G-T} 
 (by the Gysin exact sequences) and summarized in the following table:
$$
\begin{array}{|c|c|c|c|c|}
\hline
& n = 0 & n = 1 & n = 2 & n = 3 \\
\hline
H^n_{\Z_2}(\n{T}^2, \Z(0)) & \Z & \Z & \Z_2 & \Z_2 \\
\hline
H^n_{\Z_2}(\n{T}^2, \Z(1)) & 0 & \Z_2 \oplus \Z & \Z & \Z_2 \\
\hline
\end{array}
$$
The  equivariant cohomology   and the ordinary cohomology $H^n(\n{T}^2, \Z)$ fit into the long exact sequence \cite[Proposition 2.3]{gomi-13}:
$$
\begin{aligned}
\cdots\to
H^{n-1}_{\Z_2}\big(\n{T}^2, \Z(0)\big) \to 
H^n_{\Z_2}\big(\n{T}^2, \Z(1)\big) &\overset{\imath}{\to}
H^n\big(\n{T}^2, \Z\big)\\ &\to
H^n_{\Z_2}\big(\n{T}^2, \Z(0)\big) \to\cdots
\end{aligned}
$$
where $\imath$ is the homomorphism which forgets  the $\Z_2$-action. Knowing  $H^n(\T^2, \Z)$, one concludes that $\imath$ provides an isomorphism on the degree $n=2$.
\qed

\medskip

The next step is to compute the equivariant relative cohomology group $H^2_{\Z_2}(\T^2|(\T^2)^{\rr{f}},\Z(1))$ where $(\T^2)^{\rr{f}}\simeq\n{S}^1$ is the fixed point set of the torus $\T^2$ with respect to the flip involution.
\begin{lemma}\label{lemm_eq_cohom_Q}
There is an isomorphism of groups
$$
H^2_{\Z_2}\big(\T^2|(\T^2)^{\rr{f}},\Z(1)\big)\;\simeq\;  \Z\;.
$$
\end{lemma}
\proof
In general, if $X$ is a $\Z_2$-CW complex and $Z\subseteq Y \subseteq X$ are $\Z_2$-sub-complexes, then there is an exact sequence of groups
$$
\begin{aligned}
\cdots\to
H^{n-1}_{\Z_2}\big(Y|Z,\Z(1)\big) \to
H^n_{\Z_2}\big(X|Y,\Z(1)\big) &\to
H^n_{\Z_2}\big(X|Z,\Z(1)\big)\\ &\to
H^n_{\Z_2}\big(Y|Z,\Z(1)\big) \to
\cdots\;.
\end{aligned}
$$
 This is just a consequence of the excision axiom and the exactness axiom. We apply this exact sequence to $X := \T^2$, $Y := (\T^2)^{\rr{f}}$ and a fixed point $Z := \{\ast\}$:
$$
\begin{aligned}
\cdots \to
\tilde{H}^{n-1}_{\Z_2}\big((\T^2)^{\rr{f}},\Z(1)\big) &\to
H^n_{\Z_2}\big(\T^2| (\T^2)^{\rr{f}},\Z(1)\big)\\
& \to
\tilde{H}^n_{\Z_2}\big(\T^2,\Z(1)\big) \to
\tilde{H}^n_{\Z_2}\big((\T^2)^{\rr{f}},\Z(1)\big) \to
\cdots,
\end{aligned}
$$
where  $\tilde{H}^n_{\Z_2}(M, \Z(1)): = H^n_{\Z_2}(M| \{\ast\}, \Z(1))$ is the reduced cohomology of the $\Z_2$-space $M$ with a fixed point $\{\ast\}$. By using the  
 direct sum decomposition 
 $$
 \begin{aligned}
 H^n_{\Z_2}\big(M, \Z(1)\big)\; &\simeq\; \tilde{H}^n_{\Z_2}\big(M, \Z(1)\big)\; \oplus\; H^n_{\Z_2}\big(\{\ast\}, \Z(1)\big)\\
 &=\;
 \left\{
  \begin{aligned}
  &\tilde{H}^n_{\Z_2}\big(M, \Z(1)\big)&\quad& n\; \text{even} \\
  &\tilde{H}^n_{\Z_2}\big(M, \Z(1)\big)\oplus\Z_2&\quad& n\; \text{odd}
  \end{aligned}
 \right.
\end{aligned}
 $$
 (\cf \cite[Section 5]{denittis-gomi-14})
 and the 
computation of the 
 equivariant cohomology groups of $(\T^2,\rr{f})$ provided in the proof of  Proposition \ref{prop:cohom-R}, 
 one can compute $\tilde{H}^n_{\Z_2}(\T^2,\Z(1))$. The computation of $\tilde{H}^n_{\Z_2}((\T^2)^{\rr{f}},\Z(1))$ follows similarly by observing that
 $(\T^2)^{\rr{f}}$ coincides with $\n{S}^1$ with the trivial involution (the computation of the related equivariant cohomology groups is in  \cite{gomi-13}). The values of the various cohomology groups are displayed in the following table: 
$$
\begin{array}{|c|c|c|c|c|}
\hline
& n = 0 & n = 1 & n = 2 & n = 3 \\
\hline
\tilde{H}^n_{\Z_2}(\T^2, \Z(1))  & 0 & \Z & \Z & 0 \\
\hline
\tilde{H}^n_{\Z_2}((\T^2)^{\rr{f}},\Z(1)) & 0 & 0 & \Z_2 & 0 \\
\hline
\end{array}
$$
From the exact sequence above, one immediately gets the isomorphism $H^2_{\Z_2}(\T^2| (\T^2)^{\rr{f}}, \Z(1)) \simeq \Z$. 
 \qed

\medskip

Lemma \ref{lemm_eq_cohom_Q} provides the computation of the equivariant relative cohomology group 
$H^2_{\Z_2}(\T^2| (\T^2)^{\rr{f}}, \Z(1))$. However, in the  proof it is not specified whether the injective homomorphism 
\begin{equation}\label{eq:homo2maps}
\jmath\;:\;H^2_{\Z_2}\big(\T^2| (\T^2)^{\rr{f}}, \Z(1)\big) \;\to\; H^2_{\Z_2}\big(\T^2, \Z(1)\big)
\end{equation}
is the bijection $\jmath:n\mapsto n$  or the multiplication by two $\jmath:n\mapsto 2n$. 
\begin{lemma}\label{lemma:quat2Z}
The homomorphism \eqref{eq:homo2maps} coincides 
with the multiplication by two $\jmath:n\mapsto 2n$, $n\in\Z$. As a consequence
$$
H^2_{\Z_2}\big(\T^2| (\T^2)^{\rr{f}}, \Z(1)\big)\;\simeq\;2\Z\;. 
$$ 
\end{lemma}
\proof
By inspecting the
exact sequence used in the proof of Lemma
\ref{lemm_eq_cohom_Q}, it suffices to show that the restriction homomorphism
$$
\Z\;\simeq\; H^2_{\Z_2}\big(\T^2, \Z(1)\big)\; \to 
\;H^2_{\Z_2}\big((\T^2)^{\rr{f}}, \Z(1)\big)\;\simeq\;\Z_2
$$
is surjective. We will show this in a geometric manner. Since $H^2_{\Z_2}(X, \Z(1))$ classifies \virg{Real} line bundles over $X$, we shall construct a \virg{Real} line bundle over $\T^2$ whose restriction to $(\T^2)^{\rr{f}}$ is a non-trivial real line bundle. Let $\tilde{L}_{\T^2} := \R^2 \times \C$ be the product line bundle over $\R^2$. This is equipped with two actions
\begin{align*}
T_1 &: \tilde{L}_{\T^2} \to \tilde{L}_{\T^2}, & T_1(x_1, x_2, z) &:= (x_1 + 1, x_2 , \expo{\ii 2\pi x_1}z), \\
T_2 &: \tilde{L}_{\T^2} \to \tilde{L}_{\T^2}, & T_2(x_1, x_2, z) &:= (x_1, x_2+1, \expo{\ii 2\pi x_1}z).
\end{align*}
These two actions commute with each other, so that they make the line bundle $\tilde{L}_{\T^2} \to \R^2$ into a $\Z^2$-equivariant line bundle, where $\Z^2$ acts on $\R^2$ by translations. Furthermore, this $\Z^2$-action is free. Thus, taking the quotient, we get a complex line bundle $L_{\T^2} := \tilde{L}_{\T^2}/\Z^2$ over the torus $\T^2 = \R^2/\Z^2$. It can be proven that $L_{\T^2}$ is non-trivial and has Chern class $c_1(L_{\T^2})=1$. We can define a \virg{Real} structure on $\tilde{L}_{\T^2}$ by
\begin{align*}
\tilde{\Theta} &: \tilde{L}_{\T^2} \to \tilde{L}_{\T^2}, &
\tilde{\Theta}(x_1, x_2, z) &:= \left(x_2, x_1, \expo{\ii \pi  \big((x_1+x_2)^2 - (x_1+x_2)\big)} \overline{z}\right)\;.
\end{align*}
Because of the  relations
\begin{align*}
\tilde{\Theta}\circ T_1 &\;=\; T_2 \circ\tilde{\Theta}\;, &
\tilde{\Theta}^2\;=\;{\rm Id}_{\tilde{L}_{\T^2}}\;.
\end{align*}
$\tilde{\Theta}$ descends to  a \virg{Real} structure on $L_{\T^2}$. The restriction $L_{\T^2}|_{(\T^2)^{\rr{f}}}$ is isomorphic to the quotient $L_{\n{S}^1} = \tilde{L}_{\n{S}^1}/\Z$ over $\n{S}^1 = \R/\Z$ of the product bundle $\tilde{L}_{\n{S}^1} = \R \times \C$ over $\R$ under the free action of $\Z$ generated by 
\begin{align*}
T_0 &: \tilde{L}_{\n{S}^1} \to \tilde{L}_{\n{S}^1}\;, &
T(x, z) &\;=\; \big(x + 1, \expo{\ii 4\pi  x}z\big)\;.
\end{align*}
The \virg{Real} structure $\Theta_0$ on $L_{\n{S}^1} \simeq L_{\T^2}|_{(\T^2)^{\rr{f}}}$ is induced from the following \virg{Real} structure on $\tilde{L}_{\n{S}^1}$
\begin{align*}
\tilde{\Theta}_0 &: \tilde{L}_{\n{S}^1} \to \tilde{L}_{\n{S}^1}\;, &
\tilde{\Theta}_0(x, z) &\;:=\; \big(x, \expo{\ii 2\pi  (2x^2 - x)} \overline{z}\big)\;.
\end{align*}
As a complex line bundle, $L_{\n{S}^1} \to \n{S}^1$ admits a nowhere vanishing section $\sigma : \n{S}^1 \to L_{\n{S}^1}$ given by $\sigma([x]) = [x, \expo{\ii 2\pi  x^2}]$. Under the \virg{Real} action ${\Theta}_0$, this section behaves as ${\Theta}_0(\sigma([x])) = \expo{-\ii2\pi  x} \sigma([x])$. Then, by using this section, we can construct an isomorphism between $L_{\n{S}^1}$ and the product bundle $\n{S}^1 \times \C \to \C$ with the \virg{Real} structure $(x, z) \mapsto (x, \expo{-\ii2\pi  x} \overline{z})$. The (dual of the) latter \virg{Real} line bundle is shown to be non-trivial in \cite[Example 3.10]{denittis-gomi-15}.
\qed

\medskip

\begin{corollary}\label{corol:appen_fin}
The first Chern class provides the isomorphism
$$
{\rm Vec}^{2m}_{\rr{Q}}\big(\n{T}^2,\rr{f}\big)\;\stackrel{c_1}{ \simeq}\; 2\Z\;.
$$
\end{corollary}
\proof
The FKMM invariant $\kappa$ \cite{denittis-gomi-14-gen,denittis-gomi-18,denittis-gomi-18-b} and Lemma \ref{lemma:quat2Z} provide the isomorphisms
$$
{\rm Vec}^{2m}_{\rr{Q}}\big(\n{T}^2,\rr{f}\big)\;\stackrel{\kappa}{ \simeq}\; H^2_{\Z_2}\big(\T^2| (\T^2)^{\rr{f}}, \Z(1)\big)\;\simeq\;2\Z
$$
 The nature of the $2\Z$-valued invariant can be described by observing that 
$$
H^2_{\Z_2}\big(\T^2| (\T^2)^{\rr{f}}, \Z(1)\big) \;\stackrel{\jmath}{\to}\; H^2_{\Z_2}\big(\T^2, \Z(1)\big)\;\stackrel{\imath}{\simeq}\;H^2\big(\T^2, \Z\big)\;\simeq\;\Z
$$
where $\jmath$ is the injection described in 
Lemma \ref{lemma:quat2Z} and $\imath$ is the 
isomorphism which forgets  the $\Z_2$-action.
The resulting map ${\rm Vec}^{2m}_{\rr{Q}}(\n{T}^2,\rr{f}) \to H^2(\T^2, \Z)$
given by the composition of 
 these homomorphisms is nothing but the first Chern class of the complex vector bundle underlying the \virg{Quaternionic} vector bundle. 
\qed

\section{A primer on Dixmier trace}
\label{sec:Limit_formula}
This appendix is devoted to the construction of the \emph{Ma\u{c}aev ideals} and the \emph{Dixmier trace}. 
Useful references for these subjects  are \cite[Appendix A]{connes-moscovici-95}, \cite[Sect.~7.5 and App.~7.C]{gracia-varilly-figueroa-01}, \cite{alberti-matthes-02} and \cite{simon-05}.


\subsection{Trace, Schatten ideals and Ma\u{c}aev ideals}
\label{sect:schatten-macaev}
We will assume the familiarity of the 
reader with the theory of compact operators
 (see \cite[VI.5]{reed-simon1}).
Let $\s{H}$ be a separable Hilbert,  $\s{B}(\s{H})$  the $C^\ast$-algebra of the bounded  operators acting on $\s{H}$ and $\s{K}(\s{H})\subset\s{B}(\s{H})$
the two-sided ideal of  {compact operators}. We just  recall three important facts about compact operators: (i)  $T\in\s{K}(\s{H})$  if and only if $T$ is the norm limit of a sequence of finite rank operators; (ii)
the spectrum $\sigma(T)$ of a compact operator is a discrete set having no limit point, except perhaps the zero and every non-zero  eigenvalue has finite  multiplicity (\emph{Riesz-Schauder Theorem}); (iii) $\s{K}(\s{H})$ is the only non-trivial  norm-closed two-sided ideal in $\s{B}(\s{H})$. The latter property  implies that 
$T\in \s{K}(\s{H})$ if and only if $|T|:=\sqrt{T^* T}\in \s{K}(\s{H})$.

\medskip

The \emph{singular values} $\mu_n(T)$ of a compact operator $T$ are  the eigenvalues of  $|T|$. 
By convention  the singular values will be listed in decreasing order, repeated according to the multiplicity, \ie 
$$
\mu_0(T)\;\geqslant\;\mu_1(T)\;\geqslant\;\ldots\;\geqslant\;\mu_n(T)\;\geqslant\;\mu_{n+1}(T)\;\geqslant\;\ldots\;\geqslant\;0
$$
with  $\mu_0(T) =\||T|\|_{\s{B}(\s{H})}=\|T\|_{\s{B}(\s{H})}$.

\medskip

From an analytic point of view compact operators are in a sense \virg{small} or better \emph{infinitesimal}. Indeed, if $T$ is a compact operator then
 for all $\varepsilon>0$
there exists a  finite-dimensional subspace $V_\varepsilon\subset\s{H}$ such that
$\|T|_{V_\varepsilon^\bot}\|_{\s{B}(\s{H})} <\varepsilon$.
Moreover one has that
$$
\mu_n(T)\;=\;\inf\big\{
\|T|_{V^\bot}\|_{\s{B}(\s{H})}\ |\ \text{\upshape dim}\ V=n\ 
\big\}.
$$
Compact operators can be classified according to  their order of infinitesimal. One says  that 
$T\in\s{K}(\s{H})$ is of
\emph{order} $\alpha\in(0,+\infty)$   if 
 there exist a $C>0$ and a $N_0\in\N$ such that
$$
\text{Mult}\big[\mu_n(T)\big]\; \mu_n(T)\;  \leqslant\;  C n^{-\alpha} \qquad\quad \forall\ \ n\geqslant N_0
$$
where $\text{Mult}[\mu_n(T)]$ is the multiplicity of $\mu_n(T)$.
This definition is consistent: Indeed if $T_1$ is an infinitesimal of order $\alpha$  and $T_2$ an infinitesimal of order $\beta$ then
$T_1T_2$ is an infinitesimal of order at most $\alpha+\beta$ (as a consequence of the {submultiplicative property} for  singular values).
Therefore,  the set of the infinitesimals of order $\alpha$ is a (non-closed) two-sided ideal in $\s{B}(\s{H})$.

\medskip

Let us recall few standard facts about the notion of  \emph{trace} on $\s{H}$ (\cf \cite[Sect~VI.6]{reed-simon1}). 
Every  orthonormal basis $\{\phi_n\}_{n\in\N}$ of  $\s{H}$ defines 
 a linear functional on the cone of positive operators by
\begin{equation}\label{eq:def_trace1}
\text{\upshape Tr}_{\s{H}}(T)\;:=\;\sum_{n=0}^{+\infty}\langle\phi_n,T\phi_n\rangle_{\s{H}}\;.
\end{equation}
The linear functional  is {monotone} (with respect to the ordering of the positive operators) and its
range  is $[0,+\infty]$. Moreover, $\text{\upshape Tr}_{\s{H}}$ 
is independent of the particular choice of the orthonormal basis.
A bounded operator  $T$ is called \emph{trace class} if and only if $\text{\upshape Tr}_{\s{H}}(|T|)<+\infty$. The family of  trace class operators is denoted by $\s{L}^1(\s{H})$.
One has that $\s{L}^1(\s{H})\subset \s{K}(\s{H})$ and $T$ is trace class if and only if
$$
\text{\upshape Tr}_{\s{H}}(|T|)\;=\;\sum_{n=0}^{+\infty}\mu_n(T)\;<\;+\infty\;.
$$
The set $\s{L}^1(\s{H})$  is a  two-sided self-adjoint ideal of $\s{B}(\s{H})$ which is not closed with respect to the operator norm but which is closed with respect to  the \emph{trace-norm} 
$\|T\|_1:=\text{\upshape Tr}_{\s{H}}(|T|)$. The finite rank operators are $\|\cdot\|_1$-dense in $\s{L}^1(\s{H})$.
The ideal $\s{L}^1(\s{H})$ is the natural domain of definition for the trace functional. Indeed if $T\in \s{L}^1(\s{H})$ (not necessary positive) then the sum \eqref{eq:def_trace1}
converges absolutely and the limit is independent of the choice of a particular orthonormal basis. Therefore, 
$\text{\upshape Tr}_{\s{H}}:\s{L}^1(\s{H})\to\C$ defines a linear $\ast$-functional
bounded by the norm $\|\;\|_1$. Finally,  the \emph{tracial property}
$
\text{\upshape Tr}_{\s{H}}(TS)=\text{\upshape Tr}_{\s{H}}(ST)
$
holds  for all $T\in\s{L}^1(\s{H})$ and $S\in\s{B}(\s{H})$. 

\medskip

The definition of the trace class ideal $\s{L}^1(\s{H})$ can be generalized. For each $p\in[1,+\infty)$ one defines the \emph{$p$-th Schatten class} as the set 
$$
\s{L}^p(\s{H})\;:=\;\left\{T\in \s{K}(\s{H})\ \big|\ \text{\upshape Tr}_{\s{H}}\big(|T|^p\big)=\sum_{n=0}^{+\infty}\mu_n(T)^p<+\infty\right\}.
$$
Every $\s{L}^p(\s{H})$ is a two-sided self-adjoint ideal contained in $\s{K}(\s{H})$ which is not closed with respect to the operator norm but which is closed with respect the \emph{$p$-norm} 
$\|T\|_p:=\text{\upshape Tr}_{\s{H}}(|T|^p)^{\frac{1}{p}}$. Moreover,  $\s{L}^p(\s{H})\subset \s{L}^q(\s{H})$ for every $p\leqslant q$ and
 the finite rank operators are $\|\cdot\|_p$-dense in $\s{L}^p(\s{H})$. 
The $2$-nd Schatten class $\s{L}^2(\s{H})$ is usually called \emph{Hilbert-Schmidt ideal} and it can be endowed with the structure of a Hilbert space by means of  the inner product
$
\langle T_1,T_2\rangle_{\rm H.S.}:=\text{\upshape Tr}_{\s{H}}\big(T_1^* T_2\big).
$

\medskip

Operator in $\s{L}^p(\s{H})$, for $p\in[1,+\infty)$, are infinitesimal of order strictly greater than ${1}/{p}$. To see this, let us introduce the
partial sums 
\begin{equation}\label{eq:partial_sigma}
\sigma_N^p(T)\;:=\;\sum_{n=0}^{N-1}\mu_n(T)^p.
\end{equation}
Then, $T\in \s{L}^p(\s{H})$ if and only if $\|T\|_p^p=\lim_{N\to\infty}\ \sigma_N^p(T)<+\infty$ which is the same of
$\text{Mult}[\mu_n(T)] \mu_n(T)^p\leqslant C\ n^{-(1+\epsilon)}$ for some positive constants $C$ and $\epsilon$ and $n\geqslant N_0$. 
Then, the ideal of infinitesimal operator  of order $1/p$  is strictly larger than $\s{L}^p(\s{H})$. 
However, from $\sum_{n=1}^{N}{n^{-1}}\sim\log(N)$ one infers that infinitesimals of order $1$
have  partial sums $\sigma_N^1$ which are at most logarithmically divergent. This observation
suggests  to consider the \virg{regularized} partial sums
\begin{equation}\label{eq:partial_gamma}
\gamma^p_N(T)\;:=\;\frac{\sigma_N^p(T)}{\log(N)}\;=\;\frac{1}{\log(N)}\sum_{n=0}^{N-1}\mu_n(T)^p\;,\qquad N>1\;.
\end{equation}
 The positive sequence $\gamma_N^p(T)$, although bounded if $T$ is an infinitesimal of order ${1/p}$,  
does not converge in general when $N\to\infty$. This suggests to consider 
 the \emph{supremum limit}. 
The \emph{$(1^+)$-Ma\u{c}aev class} is the set
\begin{equation}\label{eq:norm_p+}
\s{L}^{1^+}(\s{H})\;:=\;\left\{T\in \s{K}(\s{H})\; \Big|\; \|T\|_{1^+}:=\sup_{N>1}\ \gamma^1_N(T)<+\infty\right\}.
\end{equation}
 In particular $T\in \s{L}^{1^+}(\s{H})$ when  \eqref{eq:partial_gamma} converges.
The above definitions generalizes for $p>1$ as follows 
\begin{equation}\label{eq:norm_p++}
\begin{aligned}
&\s{L}^{p^+}(\s{H})\;:=\;\left\{T\in \s{K}(\s{H})\; |\; \|T\|_{p^+}:=\sup_{N>1}\ N^{\frac{1-p}{p}}\sigma_N^1(T)<+\infty\right\}.
\end{aligned}
\end{equation}
These definitions encode the fact that  $\s{L}^{p^+}(\s{H})$ is the ideal of infinitesimal operator of order $1/p$. More precisely, one can prove that
$T\in\s{L}^{p^+}(\s{H})$ if and only if $\text{Mult}[\mu_n(T)] \mu_n(T)^p\sim O(n^{-1})$ if $p>1$. For $p=1$ this characterization fails as showed in \cite[Remark 1.1]{alberti-matthes-02}
 but the infinitesimals of order 1 are  anyway  in $\s{L}^{1^+}(\s{H})$ (\cf Lemma \ref{lemma:base_trace_mesurA} below).
For every $p\in[1,+\infty)$ the set
$\s{L}^{p^+}(\s{H})$ is a two-sided self-adjoint ideals which is not closed with respect to the operator norm but which is closed with respect to the \emph{(Calder\'on)} norm
$\|\cdot\|_{p^+}$. The $\|\cdot\|_{p^+}$-closure of the finite rank operators does not coincide with $\s{L}^{p^+}(\s{H})$ but defines  the \emph{$(p^+_0)$-Ma\u{c}aev (sub-)class} which is characterized by
\begin{equation}\label{eq:Lp+0}
\s{L}_0^{p^+}(\s{H})\;:=\;\left\{T\in \s{K}(\s{H})\; |\;  \text{lim}_{N\to\infty}\ \gamma^p_N(T)=0\right\}.
\end{equation}
As a consequence    the ideals $\s{L}^{p^+}(\s{H})$ are not separable. The spaces $\s{L}_0^{p^+}(\s{H})$ are two-sided self-adjoint ideals and the inclusions
$\s{L}^p(\s{H})\subset\s{L}^{p^+}_0(\s{H})\subset\s{L}^{p^+}(\s{H})\subset\s{L}^{p'}$ hold for every $p'>p\geqslant 1$.

\medskip

The ideal $\s{L}^{1^+}(\s{H})$, usually called the \emph{Dixmier ideal}, is of particular importance for our aims. Operators $T\in\s{L}^{1^+}(\s{H})$ with a convergent sequence $\gamma_N^1(T)$  
are called \emph{measurable}. Evidently, every operator in $\s{L}_0^{1^+}(\s{H})$ is measurable since in this case the sequence $\gamma_N^1(T)$ converges to zero. The set of measurable operators is
 a closed subspace of 
$\s{L}^{1^+}(\s{H})$ which is invariant under conjugation by bounded invertible operators \cite[Proposition 7.15]{gracia-varilly-figueroa-01}.  The following result 
 turns out to be a very useful criterion for the measurability of an operator:
\begin{lemma}[{\cite[Lemma 1.6]{alberti-matthes-02}}]\label{lemma:base_trace_mesurA}
Let $T\in\s{K}(\s{H})$ be a compact operator such that $\text{\upshape Mult}[\mu_n(T)] \mu_n(T)\sim C n^{-1}$. Then $T$ is a measurable element in $\s{L}^{1^+}(\s{H})$ and
$\lim_{N\to\infty}\gamma^1_N(T)= \alpha C$
where 
$$
\alpha\;:=\;\lim_{N\to\infty}\ \frac{\log(N)}{\log\left(\sum_{n=0}^{N-1}\text{\upshape Mult}[\mu_n(T)]\right)}.
$$
\end{lemma}
\medskip

\noindent
We point out that in the original statement of \cite[Lemma 1.6]{alberti-matthes-02} the terms $[\mu_n(T)]$ and $\alpha$ are both omitted. 
Of course this is possible when $\text{\upshape Mult}[\mu_n(T)]$ is a bounded sequence  with $\lim_{n\to\infty}\text{\upshape Mult}[\mu_n(T)]=1$.


\subsection{The Dixmier trace}
\label{sect:dixmier}
The aim of this section is to   define  an \virg{integral}  which neglects infinitesimal operators   of order greater than 1. More precisely, we want  a trace functional such that
 $\s{L}^{1^+}(\s{H})$ is in the domain of  such a trace and infinitesimal operator of order higher than 1 have vanishing trace. 
The usual trace is not appropriate since its domain $\s{L}^{1}(\s{H})$ is smaller than $\s{L}^{1^+}(\s{H})$. 
Dixmier \cite{dixmier-66} has shown that such a  trace exists and corresponds \virg{morally} to the operation of the extraction of the  limit 
$\lim_{N\to\infty}\gamma^1_N(T)$ defined by \eqref{eq:partial_gamma}.
However, this procedure does not define a trace since linearity and convergence
are not guaranteed. Then, one needs a more sophisticated object called
the \emph{Dixmier trace}. 

\medskip

Given a $T\in\s{K}(\s{H})$ let us define the following family of functions 
\begin{equation}\label{eq:dix02}
\sigma_\lambda(T)\;: =\; \inf\big\{\|R\|_1+\lambda\|S\|_{\s{B}(\s{H})}\; |\;  R,S\in\s{K}(\s{H}),\; \;  R+S=T\big\}
\end{equation}
indexed by the {scale parameter} $\lambda\in[1,\infty)$.
The functions $\sigma_\lambda$ are norms defined on $\s{K}(\s{H})$ and for integer values of the scale parameter $\sigma_{\lambda=N}(T)$ coincides with \eqref{eq:partial_sigma}. 
 The function $\lambda\mapsto \sigma_\lambda(T)$ is piecewise linear and concave and
the inequalities
\begin{equation}\label{eq:dix_ieq}
\sigma_\lambda(T_1+T_2)\;\leqslant\;\sigma_\lambda(T_1)+\sigma_\lambda(T_2)\;\leqslant\;\sigma_{2\lambda}(T_1+T_2)
\end{equation}
hold true for positive $T_1,T_2\in\s{K}(\s{H})$ and $\lambda\geqslant1$.

\medskip

The function
$\sigma_\lambda(T)$ can be interpreted as the trace of $T$ \virg{cutoff} at the inverse scale $\lambda$.
Equation \eqref{eq:dix_ieq} suggests that  for large $\lambda$, $\log(\lambda)^{-1}\sigma_\lambda$ is an \virg{almost additive} functional on the cone 
of the positive compact operators. If it were actually additive, it would be extended by linearity to a trace since the invariance under unitary operator due to $\sigma_\lambda(T)=\sigma_\lambda(UTU^{-1})$. 
However, a genuine trace can be obtained
 by suitably averaging the norms induced by $\log(\lambda)^{-1}\sigma_\lambda$. 
Observe that the norm $\|\cdot\|_{1^+}$ defined in \eqref{eq:norm_p+}  can be replaced by the equivalent norm (still denoted with the same symbol) 
$
\|T\|_{1^+}:=\sup_{\lambda\geqslant {\rm e}}\log(\lambda)^{-1}\ \sigma_\lambda(T)
$.
Consider the following \emph{Ces\`aro mean}
\begin{equation}\label{eq:dix03}
\tau_\lambda(T)\;: =\; \frac{1}{\log(\lambda)}\int_{\lambda_0}^\lambda\ \frac{\sigma_s(T)}{\log(s)}\ \frac{\dd s}{s}\;,\qquad\ \text{for}\ \lambda\geqslant \lambda_0>{\rm e}.
\end{equation}
This is still not an additive functional, but it has an \virg{asymptotic additivity} property as shown by the following inequality \cite[Lemma A.4]{connes-moscovici-95} 
\begin{equation}\label{eq:dix04}
\big|\tau_\lambda(T_1+T_2)-\tau_\lambda(T_1)-\tau_\lambda(T_2)\big|\;\leqslant\; C_\lambda \big(\|T_1\|_{1+}+\|T_2\|_{1+}\big)\;,
\end{equation}
where $C_\lambda=\log(2)\log(\lambda)^{-1}(2+\log(\log(\lambda)))$,
valid for any pair of positive operators $T_1,T_2\in\s{L}^{1^+}(\s{H})$. 
The function $\log(\lambda)^{-1}\log(\log(\lambda))$
is bounded on the interval $[\lambda_0,\infty)$ and falls to zero at infinity. 
Thus the function $\lambda\mapsto\tau_\lambda(T)$ lies in $\s{C}_\text{b}([\lambda_0,\infty))$ and the left hand side of \eqref{eq:dix04} lies in $\s{C}_0([\lambda_0,+\infty))$.
Let $\rr{B}_\infty:=\s{C}_\text{b}([\lambda_0,\infty))/\s{C}_0([\lambda_0,\infty))$ be the quotient $C^\ast$-algebra and denote by $
\tau:\s{L}^{1^+}(\s{H})\to \rr{B}_\infty$ the map which associates to any positive
 $T\in \s{L}^{1^+}(\s{H})$ the equivalence class $\tau(T):=[\lambda\mapsto\tau_\lambda(T)]\in\rr{B}_\infty$. One can prove that $\tau$ is additive and positive-homogeneous on the positive cone of  
$\s{L}^{1^+}(\s{H})$.  Moreover,
 $\tau$  extends by linearity to a linear map  $\tau:\s{L}^{1^+}(\s{H})\to \rr{B}_\infty$ defined on the full ideal $\s{L}^{1^+}(\s{H})$ 
and which verifies the trace property 
$\tau(ST)=\tau(TS)$ for all $T\in\s{L}^{1^+}(\s{H})$ and bounded $S$. Finally, $\tau:\s{L}^{1^+}_0(\s{H})\to \{0\}$.

\medskip

To define a trace functional with domain the Dixmier ideal $\s{L}^{1^+}(\s{H})$, all we have to do is to follow the map  
$\tau$ with a state $\omega:\rr{B}_\infty\to\C$. The latter is a positive linear form on
$\s{C}_\text{b}([\lambda_0,+\infty))$ which vanishes on $\s{C}_0([\lambda_0,+\infty))$, 
normalized by $\omega(1)=1$. Let $f\in \s{C}_\text{b}([\lambda_0,+\infty))$ such that $f$ has limit $L$ when $\lambda\to\infty$. Then $f-L\in\s{C}_0([\lambda_0,+\infty))$ and $\omega(f-L)=0$ which implies $\omega(f)=L$ independently of
the choice of the state $\omega$. On the other hand, if
$f$ has two distinct limit points, one gets two states $\omega_1$ and $\omega_2$ whose values on
$f$ are different. Then the states of  $\rr{B}_\infty$ correspond to \virg{generalized limits} as $\lambda\to\infty$, of bounded but not necessarily convergent functions.

\begin{definition}[Dixmier traces]
To each state $\omega$ on the commutative $C^\ast$-algebra $\rr{B}_\infty$ there corresponds a \emph{Dixmier trace} on $\s{L}^{1^+}(\s{H})$ define by $\text{\upshape Tr}_{\text{\upshape Dix},\omega}:=\omega\circ\tau$.
\end{definition}

\medskip

The Dixmier trace has many of the properties of a usual trace. Indeed, each $\text{\upshape Tr}_{\text{\upshape Dix},\omega}$ is
   a positive linear functional on $\s{L}^{1^+}(\s{H})$ such that 
$\text{\upshape Tr}_{\text{\upshape Dix},\omega}(ST)=\text{\upshape Tr}_{\text{\upshape Dix},\omega}(TS)$ for all $T\in \s{L}^{1^+}(\s{H})$ and $S\in\s{B}(\s{H})$.
Moreover,
$$
|\text{\upshape Tr}_{\text{\upshape Dix},\omega}(TS)|\leqslant\text{\upshape Tr}_{\text{\upshape Dix},\omega}(|TS|)\leqslant
\|S\|_{\s{B}(\s{H})}\ \text{\upshape Tr}_{\text{\upshape Dix},\omega}(|T|)\leqslant\|S\|_{\s{B}(\s{H})}\ \|T\|_{1^+}
$$  
and the kernel  of $\text{\upshape Tr}_{\text{\upshape Dix},\omega}$ coincide with    
$\s{L}^{1^+}_0(\s{H})$. The \emph{(abstract) H\"{o}lder inequality} holds true, namely
if $T_1,T_2\in \s{K}(\s{H})$ such that $T_1^p,T_2^q\in\s{L}^{1^+}(\s{H})$ with $p,q\in[0,\infty]$ and $1={1}/{p}+{1}/{q}$, then
$$
\text{\upshape Tr}_{\text{\upshape Dix},\omega}(|T_1T_2|)\;\leqslant\;\big(\text{\upshape Tr}_{\text{\upshape Dix},\omega}(|T_1|^p)\big)^{\frac{1}{p}}\ \big(\text{\upshape Tr}_{\text{\upshape Dix},\omega}
(|T_2|^q)\big)^{\frac{1}{q}}\;.
$$ 
As a consequence, one has that if $T\in \s{L}^{1^+}(\s{H})$ is positive then also $T^{\frac{1}{2}}S T^{\frac{1}{2}}\in \s{L}^{1^+}(\s{H})$ for every $S\in\s{B}(\s{H})$ and 
\begin{equation}\label{eq:dix_trac_rad}
\text{\upshape Tr}_{\text{\upshape Dix},\omega}(TS)\;=\;\text{\upshape Tr}_{\text{\upshape Dix},\omega}\left(T^{\frac{1}{2}}\ S\ T^{\frac{1}{2}}\right)\;.
\end{equation}

\medskip

In general no explicit general formula for $\text{\upshape Tr}_{\text{\upshape Dix},\omega}(T)$ can be given without specifying the state $\omega$. However, we can at least rewrite 
it as a generalized  limit of a sequence. If $\{a_N\}\in\ell^\infty(\N)$ is a bounded sequence
we can extend it  piecewise-linearly to a function in $\s{C}_\text{b}([\lambda_0,\infty))$. Let $a_\infty$ be the image of this function in 
$\rr{B}_\infty$. We write $\lim_{\omega}(a_N):=\omega(a_\infty)$. Clearly $\lim_{\omega}$ defines a positive linear functional on the space 
$\ell^\infty(\N)$, coinciding with the ordinary limit  on the subspace of convergent sequences. 
Moreover, $\lim_{\omega}$ has the \emph{scale invariance property}, \ie $\lim_{\omega}(\{a_1,a_2,a_3,a_4\ldots\})=\lim_{\omega}(\{a_1,a_1,a_2,a_2\ldots\})$. 
With this identification in mind one has that
$$
\text{\upshape Tr}_{\text{\upshape Dix},\omega}(T)\;=\;\lim_{\omega}\left(\gamma_N^1(T)\right),\qquad\qquad T\in \s{L}^{1^+}(\s{H})\;,\quad T\geqslant0\;.
$$
Then,
the value of $\text{\upshape Tr}_{\text{\upshape Dix},\omega}(T)$ is independent of the choice of the particular state $\omega$ if and only if the operator $T$ is  {measurable}. Indeed, in this case
the function $\lambda\mapsto\tau_\lambda(T)$ converges when $\lambda\to\infty$   and the common value of all the Dixmier traces is given by
$$
\text{\upshape Tr}_{\text{\upshape Dix}}(T)=\lim_{\lambda\to\infty}\tau_\lambda(T)\;=\;\lim_{\lambda\to \infty}\frac{\sigma_\lambda(T)}{\log(\lambda)}\;=\; \lim_{N\to\infty}\gamma_N^1(T)
$$
if $T$ is positive, or by a linear combination of the above formula in the generic case.   
When  $T\in\s{L}^{1^+}(\s{H})$ is a measurable operator we  will use the short notation $\text{\upshape Tr}_{\text{\upshape Dix}}(T)$ instead of $\text{\upshape Tr}_{\text{\upshape Dix},\omega}(T)$
in order to emphasize the independence of $\omega$.

\medskip

There are no general criteria for the calculation of the Dixmier trace. Nevertheless, the following result 
 turns out to be very useful:
\begin{lemma}[{\cite[Lemma 7.17]{gracia-varilly-figueroa-01}}]\label{lemma:base_trace}
Let $T\in\s{L}^{1^+}(\s{H})$ be a positive operator and let $S\in\s{B}(\s{H})$ be some bounded operator. Let $\{\psi_n\ |\ n\in\N\}$  be an orthonormal basis of eigenvectors of $T$ (ordered according to 
the decreasing sequence of eigenvalues). Then
$$
\text{\upshape Tr}_{\text{\upshape Dix},\omega}(TS)\;=\;\lim_{\omega}\left( \frac{1}{\log(N)}\sum_{n=0}^{N-1} \langle\psi_n;TS\psi_n\rangle_{\s{H}}\right)\;.
$$
\end{lemma}

\medskip

The following result provides a useful criterion to determine whether
 $\text{\upshape Tr}_{\text{\upshape Dix}}(T)$ is independent of $\omega$ and to compute its value.

\begin{theorem}[{\cite[Appendix~A]{connes-moscovici-95}}]\label{theo:zeta_connes}
Let $T\in \s{L}^{1^+}(\s{H})$ be a positive operator
and define the  \emph{zeta function}
$\zeta_{T}(s):=\text{\upshape Tr}_{\s{H}}(T^s)$. Then the following convergence conditions are equivalent
$$
\lim_{s\to1^+}\ (s-1)\zeta_{T}(s)\;=\;L\ \ \ \ \Leftrightarrow\ \ \ \ \lim_{N\to\infty}\frac{1}{\log(N)}\sum_{n=0}^{N-1}\mu_n(T)\;=\;L
$$
and $\text{\upshape Tr}_{\text{\upshape Dix}}(T)=L$ independently of the choice of the state $\omega$.
\end{theorem}

\noindent
For a detailed demonstration the reader can refer to \cite[Lemmas 7.19, 7.20]{gracia-varilly-figueroa-01} and references therein.

\medskip

As a corollary of the Lemma \ref {lemma:base_trace} we  prove a technical result which  will be usefull in the following.
\begin{lemma}\label{lemma:tensor_dix}
Let $T\in\s{L}^{p+}(\s{H})$ and ${M}\in\text{\upshape Mat}_\ell(\C)$. Then $T\otimes M\in \s{L}^{p+}(\s{H}\otimes\C^\ell)$. Moreover, if $T\in\s{L}^{1+}(\s{H})$
then
$$
\text{\upshape Tr}_{\text{\upshape Dix},\omega}(T\otimes M)=\text{\upshape Tr}_{\text{\upshape Dix},\omega}(T)\ \text{\upshape Tr}_{\C^\ell}(M)\;.
$$
\end{lemma}
\proof
 Since both $\s{L}^{p+}(\s{H})$ and $\text{\upshape Mat}_\ell(\C)$
are generated by its positive elements, there is no loss of generality in assuming that $T\geqslant0$ and $M\geqslant0$. In this case we can use the characterization 
$$
\sigma_N^1(T\otimes M)=\text{sup}\big\{\text{Tr}_{\s{H}}\left(P_VTP_V\right)\ \text{\upshape Tr}_{\C^\ell}(P_UMP_U)\ |\ \text{dim}V+\text{dim}U=N\big\}
$$
where $V\subset\s{H}$ and $U\subset\C^\ell$ are finite dimensional subspaces and $P_V$ and $P_U$ are the related orthogonal projections \cite[Lemma 7.32]{gracia-varilly-figueroa-01}. This leads immediately to
the inequality $\sigma_N^1(T\otimes M)\leqslant \sigma_N^1(T)\ \text{\upshape Tr}_{\C^\ell}(M)$ and consequently $\|T\otimes M\|_{p+}\leqslant \|T\|_{p+}\ \text{\upshape Tr}_{\C^\ell}(M)$.
Now, let us assume that $T\in\s{L}^{1+}(\s{H})$ and let $\{\varphi_n\}_{n\in\N}\subset \s{H}$ be the orthonormal basis of eigenvectors of $T$ (ordered according to 
the decreasing series of eigenvalues) and $\{e_j\}_{j=1,\ldots,\ell}$ the canonical basis of $\C^\ell$. Since $T\otimes {\bf 1}_\ell\in \s{L}^{1+}(\s{H}\otimes\C^\ell)$
for the above argument we can apply Lemma \ref{lemma:base_trace}  in order to prove that
$$
\text{\upshape Tr}_{\text{\upshape Dix},\omega}(T\otimes M)=\lim_{\omega}\left(\frac{\log(N)}{\log(\ell N)} \frac{1}{\log(N)}\sum_{n=0}^{N-1} 
\sum_{j=1}^\ell\langle\psi_n,T\psi_n\rangle_{\s{H}}\ \langle e_j,M e_j\rangle_{\C^\ell}\right).
$$
This equality concludes the proof.
\qed

\subsection{Dixmier trace and trace per unit volume}
\label{subsect:ttrace_unit_N_dix}
Let  $\s{M}_B$ be the von Neumann algebra generated by the (spectral) Landau projections $\Pi_j$ of the Landau Hamiltonian $H_B$.
The algebra $\s{M}_B$ is abelian and its Gelfand spectrum is given by  the pure states $\{\delta_k\}_{k\in\N_0}$ defined by $\delta_k(\Pi_j)=\delta_{k,j}$. As a consequence one has the Gelfand isomorphism $\s{M}_B\simeq\ell^{\infty}(\N_0)$. Inside $\s{M}_B$ there is the ideal $\s{M}^1_B\subset \s{M}_B$ defined by sequence in $\ell^{1}(\N_0)$, namely $T\in \s{M}^1_B$ if and only if $\{t_k\}_{k\in \N_0}\in \ell^{1}(\N_0)$ where $t_j:=\delta_k(T)$. 
It turns out that $\s{M}^1_B$ has the \emph{Schur's property} \cite[Example 2.5.24]{megginson-83}, namely weakly convergent sequences  automatically converge  in norm.
The map
\begin{equation}\label{eq:int_l1}
\int(T)\;:=\;\sum_{k=0}^\infty\delta_k(T)\;=\;\sum_{k=1}^\infty t_k\;,\qquad T\in \s{M}^1_B
\end{equation}
defines a faithful, semi-finite normal (FSN) trace on $\s{M}_B$ with domain $\s{M}^1_B$ which coincides with the usual (discrete) integral on $\ell^{1}(\N_0)$. 
In this section we will provide two different formulas to compute this trace.

\medskip

To construct the first formula we need some preliminary results.
Let $Q_B$ be the unbounded operator defined by \eqref{eq:Q_op1} and $Q_{B,\xi}^{-s}$ be  the compact operator defined by  \eqref{eq:Q_op2} for all $s>0$ and $\xi\geqslant 0$. The next result concerns with the measurability properties of $Q_{B,\xi}^{-s}$.
\begin{lemma}\label{lemma:meas_Q}
Let $Q_{B,\xi}^{-s}$ be  the compact operator defined by \eqref{eq:Q_op1} and \eqref{eq:Q_op2}. Then:
\begin{itemize}\label{eq:comp_Trac_Q}
\item[(i)] $Q_{B,\xi}^{-s}$ is trace class for every $s>2$,  $\xi\geqslant 0$ and
\begin{equation}\label{eq:traXX}
{\rm Tr}_{L^2(\R^2)}\big(Q_{B,\xi}^{-s}\big)\;=\;\rr{Z}(s-1,1+2\xi)-(1+2\xi)\;\rr{Z}(s,1+2\xi)
\end{equation}
where $\rr{Z}$ is the \emph{Hurwitz zeta function}\footnote{The Hurwitz zeta function is defined by the absolutely convergent series $\rr{Z}(s,\xi):=\sum_{j=0}^\infty(j+\xi)^{-s}$ for every $s>0$ and $\xi>0$. The  Riemann zeta function is $\rr{Z}_0(s):= \rr{Z}(s,1)$.};
\vspace{1mm}
\item[(ii)] $Q_{B,\xi}^{-s}$ is a measurable element of the Dixmier ideal and
\begin{equation}\label{eq:traXXX}
{\rm Tr}_{\rm Dix}\big(Q_{B,\xi}^{-2}\big)\;=\;\frac{1}{2}
\end{equation}
independently of $\xi\geqslant 0$.
\vspace{1mm}
\end{itemize}
\end{lemma}
\proof
(i) The trace of $Q_{B,\xi}^{-s}$, when it exists, is given by the limit of the increasing sequence $\sigma_N^{1}(Q_{B,\xi}^{-s})$ of the first $N$ eigenvalues of $Q_{B,\xi}^{-s}$  counted with their multiplicity. This sequence can only converge (whenever it is bounded) or diverge. Both situations can be controlled by the subsequence $\sigma_{\jmath_N}^{1}(Q_{B,\xi}^{-s})$ of the first $\jmath_N:=\frac{1}{2}N(N+1)$ eigenvalues of $Q_{B,\xi}^{-s}$. An explicit computation provides
$$
\begin{aligned}
\sigma_{\jmath_N}^{1}\big(Q_{B,\xi}^{-s}\big)\;&=\;\sum_{j=0}^{N-1}\frac{\text{Mult}[\lambda_j]}{(\lambda_j+2\xi)^s}\;=\;\sum_{j=1}^{N}\frac{j}{(j+1+2\xi)^s}\\
&=\;\sum_{j=1}^{N}\frac{1}{(j+1+2\xi)^{s-1}}-(1+2\xi)\sum_{j=1}^{N}\frac{1}{(j+1+2\xi)^{s}}\;.
\end{aligned}
$$
This series is absolutely convergent whenever $s>2$ and $\xi\geqslant 0$ and in this case 
${\rm Tr}_{L^2(\R^2)}(Q_{B,\xi}^{-s})=\lim_{N\to\infty}\sigma_{\jmath_N}^{1}(Q_B^{-s})$ is given by \eqref{eq:traXX}.\\
(ii) To prove that $Q_{B,\xi}^{-2}$ is in the Dixmier ideal we need to study the sequence $\gamma_{N'}^1(Q_{B,\xi}^{-2}):=\log(N')^{-1}\sigma_{N'}^{1}(Q_{B,\xi}^{-2})$ with $N'\in\N$. By observing that for every $N'\in\N$ there is a $N:= N(N')$ such that $\jmath_N\leqslant N'\leqslant \jmath_{N+1}$ one has that
$$
\frac{\log(\jmath_N)}{\log(\jmath_{N+1})}\gamma_{\jmath_N}^1\big(Q_{B,\xi}^{-2}\big)\;\leqslant\;\gamma_{N'}^1\big(Q_{B,\xi}^{-2}\big)\;\leqslant\;\frac{\log(\jmath_{N+1})}{\log(\jmath_N)}\gamma_{\jmath_{N+1}}^1\big(Q_{B,\xi}^{-2}\big)\;.
$$
 Since $\lim_{N\to \infty}\frac{\log(\jmath_N)}{\log(\jmath_{N+1})}=1$ one infers that the convergence of the sequence $\gamma_{N'}^1(Q_{B,\xi}^{-2})$ is equivalent to the convergence of the subsequence 
 $$
 \gamma_{\jmath_N}^1\big(Q_{B,\xi}^{-2}\big)\;=\;\frac{1}{\log(\jmath_N)}\left(\sum_{j=1}^{N}\frac{1}{j+1+2\xi}-(1+2\xi)\sum_{j=1}^{N}\frac{1}{(j+1+2\xi)^{2}}\right)\;.
 $$
From the absolute convergence of the second series inside the brackets and by observing that $\lim_{N\to \infty}\frac{\log(N)}{\log(\jmath_{N})}=\frac{1}{2}$ one gets that 
$$
\lim_{N\to \infty} \gamma_{\jmath_N}^1\big(Q_{B,\xi}^{-2}\big)\;=\;\frac{1}{2}\lim_{N\to \infty}\frac{1}{\log(N)}\sum_{j=1}^{N}\frac{1}{j+1+2\xi}\;=\;\frac{1}{2}\;.
$$
The equality $
{\rm Tr}_{\rm Dix} (Q_{B,\xi}^{-2})=\lim_{N\to \infty} \gamma_{N}^1 (Q_{B,\xi}^{-2} )=\lim_{N\to \infty} \gamma_{\jmath_N}^1 (Q_{B,\xi}^{-2})
$
concludes the proof.
\qed

\medskip

\noindent
Item (i) of Lemma \ref{lemma:meas_Q} can be equivalently stated by saying that  $Q_{B,\xi}^{-1}$ is an element of the  Schatten ideal $\s{L}^s(L^2(\R^2))$ for all $s>2$. The measurability of $Q_{B,\xi}^{-2}$ implies that the value of the Dixmier trace in Lemma \ref{lemma:meas_Q} (ii) is defined unambiguously. Moreover, form \eqref{eq:traXX} one gets 
$$
\lim_{s\to2^+}(s-2) {\rm Tr}_{L^2(\R^2)}\big(Q_{B,\xi}^{-s}\big)\;=\;\lim_{s\to1^+}(s-1)\rr{Z}(s,1+2\xi)\;=\;1
$$
which, along with  \eqref{eq:traXXX}, implies 
$$
{\rm Tr}_{\rm Dix}\big(Q_{B,\xi}^{-2}\big)\;=\;\frac{1}{2}\lim_{s\to2^+}(s-2) {\rm Tr}_{L^2(\R^2)}\big(Q_{B,\xi}^{-s}\big)
$$
in accordance with the Connes-Moscovici residue formula described in Theorem \ref{theo:zeta_connes}.

\medskip

The measurability properties of $Q_{B,\xi}^{-s}$ change when $Q_{B,\xi}^{-s}$ is multiplied by a Landau projection $\Pi_j$.
\begin{lemma}\label{lemma:meas_Q_2}
Let $Q_{B,\xi}^{-s}$ be  the compact operator defined by \eqref{eq:Q_op1} and \eqref{eq:Q_op2} and $\Pi_j$ the $j$-th Landau projection. Then:
\begin{itemize}
\item[(i)] $Q_{B,\xi}^{-s}\Pi_j$ is trace class for every $s>1$,  $\xi\geqslant 0$ and
\begin{equation}\label{eq:traXX_II}
{\rm Tr}_{L^2(\R^2)}\big(Q_{B,\xi}^{-s}\Pi_j\big)\;=\;{\rr{Z}\big(s,j+2(1+\xi)\big)}
\end{equation}
where $\rr{Z}$ is the {Hurwitz zeta function};
\vspace{1mm}
\item[(ii)] $Q_{B,\xi}^{-1}\Pi_j$ is a measurable element of the Dixmier ideal and
\begin{equation}\label{eq:traXXX_II}
{\rm Tr}_{\rm Dix}\big(Q_{B,\xi}^{-1}\Pi_j\big)\;=\;1
\end{equation}
independently of $\xi\geqslant 0$.
\vspace{1mm}
\end{itemize}
\end{lemma}
\proof
(i) The spectrum of $Q_{B,\xi}^{-s}\Pi_j$ is given by 
$$
\sigma\big(Q_{B,\xi}^{-s}\Pi_j\big)\;=\;\left\{(k+j+2(1+\xi))^{-s}\ |\ k\in\N_0\right\}
$$
and all the eigenvalues are simple. As a consequence one has that 
$$
\sigma_N^1\big(Q_{B,\xi}^{-s}\Pi_j\big)\;=\;\sum_{k=0}^{N-1}\frac{1}{[k+j+2(1+\xi)]^s}\;.
$$
This series is absolutely convergent whenever $s>1$  and in this case 
${\rm Tr}_{L^2(\R^2)}(Q_{B,\xi}^{-s}\Pi_j)=\lim_{N\to\infty}\sigma_N^1(Q_{B,\xi}^{-s}\Pi_j)$ is given by \eqref{eq:traXX_II}.\\
(ii) To compute the Dixmier trace we need to analyze the sequence
 \begin{equation}\label{eq_ax_dix_001}
 \gamma_{N}^1\big(Q_{B,\xi}^{-1}\Pi_j\big)\;=\;\frac{1}{\log(N)}\sum_{k=1}^{N}\frac{1}{k+(j+1+2\xi)}\;.
\end{equation}
This series converges to $1$ proving the formula \eqref{eq:traXXX_II}.
\qed

\medskip

We are now in position to provide the first formula to compute the integral \eqref{eq:int_l1}.
This  involves the Dixmier trace and the operator $Q_{B,\xi}^{-1}$.
\begin{lemma}\label{lemm:tra_aux1}
	The equality
	$$
	\int(T)\;=\;{\rm Tr}_{\rm Dix}\big(Q_{B,\xi}^{-1} T\big)
	$$
	holds true for all $T\in\s{M}^1_B$, independently of $\xi\geqslant0$.
\end{lemma}
\proof
In the proof we will use that  the 
Dixmier ideal $\s{L}^{1^+}(\s{H})$ endowed with the Calder\'on norm $\| \; \|_{1^+}$ is a Banach (hence closed) space. Moreover, from  definition \eqref{eq:norm_p+} it follows that
\begin{equation}
\label{normRelation}
\| A \|_{\s{B}(\s{H})}\;=\;\mu_0(A)\; \leqslant\; \log(2)\frac{\mu_0(A)+\mu_1(A)}{\log(2)}\;\leqslant\; \|A\|_{1^+} 
\end{equation}
 for every $A \in \s{L}^{1^+}(\s{H})$.
From Lemma \ref{lemma:meas_Q_2} (ii) one gets that $Q_{B,\xi}^{-1}\Pi_j$ is a measurable element of $\s{L}^{1^+}(L^2(\R^2))$.
Moreover, from \eqref{eq_ax_dix_001} it follows that 
$$
\begin{aligned}
 \gamma_{N}^1\big(Q_{B,\xi}^{-1}\Pi_j\big)\;&\leqslant\;
 \frac{1}{\log(N)}\sum_{k=1}^{N}\frac{1}{k+(1+
2 \xi)}\;=\;\gamma_{N}^1\big(Q_{B,\xi}^{-1}\Pi_0\big)
 \end{aligned}
 $$ 
and in turn
\begin{equation}\label{eq:diseg_norm}
\|Q_{B,\xi}^{-1}\Pi_j\|_{1^+}\;\leqslant\;
\|Q_{B,\xi}^{-1}\Pi_0\|_{1^+}\;,\qquad\quad\forall\; j\in\N_0\;.
\end{equation}
Now, let $T\in\s{M}^1_B$. Then, there is a  $\{t_k\}_{k\in\N_0}\in\ell^1(\N_0)$ such that
$T=\sum_{k=0}^\infty t_k\Pi_k$. The Schur's property assures that
the operator $Q_{B,\xi}^{-1}T$ is the uniform norm limit of the partial sums $Q_{B,\xi}^{-1}T_N=\sum_{k=0}^{N} t_jQ_{B,\xi}^{-1}\Pi_k$.
From \eqref{eq:diseg_norm} one can prove that the sequence $\{ Q_{B,\xi}^{-1}T_N\}_{N \in \N_0}$ is indeed  Cauchy  in $\s{L}^{1^+}(L^2(\R^2))$, hence it converges to an element in the Dixmier ideal. From \eqref{normRelation} and the uniqueness of the limit, one gets 
that this limit coincide with $Q_{B,\xi}^{-1}T$ proving that  $Q_{B,\xi}^{-1}T\in \s{L}^{1^+}(L^2(\R^2))$ for all  $T\in\s{M}^1_B$.
Finally, from the linearity of the Dixmier trace and the continuity of the Dixmier trace with respect to the $\|\;\|_{1^+}$-norm, one easily gets that 
$$
{\rm Tr}_{\rm Dix}\big(Q_{B,\xi}^{-1} T\big)\; =\; \lim_{N \to \infty}  \sum_{k=0}^{N} {\rm Tr}_{\rm Dix}\left(t_kQ_{B,\xi}^{-1}\Pi_k\right)\; =\;  \sum_{k=0}^\infty t_k\;
$$
and this concludes the proof.
\qed

\medskip

The second formula for \eqref{eq:int_l1}  needs  the concept of  \emph{trace per unit volume} $\s{T}_B$. Let us start with a  preliminary result.
Let $\Lambda\subset\R^2$ be any compact subset and $\chi_{\Lambda}$ the characteristic function of the set $\Lambda$. The  function $\chi_{\Lambda}$ acts  on $L^2(\R^2)$ as  a (multiplication) self-adjoint projection.
\begin{lemma}\label{lemma:trace_unit_vol}
Let $\Lambda\subset\R^2$ be a compact subset and $\Pi_j$ the $j$-th Landau projection. Then $\Pi_j \chi_{\Lambda}$ and $\chi_{\Lambda}\Pi_j$ are trace class and
$$
{\rm Tr}_{L^2(\R^2)}\big(\Pi_j \chi_{\Lambda}\big)\;=\;{\rm Tr}_{L^2(\R^2)}\big(\chi_{\Lambda}\Pi_j \big)\;=\;{\rm Tr}_{L^2(\R^2)}\big(\chi_{\Lambda}\Pi_j \chi_{\Lambda}\big)\;=\;\frac{|\Lambda|}{2\pi\ell_B^2}
$$
with $|\Lambda|$ the volume of the set $\Lambda$.
\end{lemma}
\proof
From \eqref{L_proj_ker} one gets
$$
|\Pi_{j}(x,y)|\;=\;\frac{1}{2\pi\ell_B^2}\expo{-\frac{|x-y|^2}{4\ell_B^2}} \left| L_{j}^{(0)}\left(\frac{|x-y|^2}{2\ell_B^2}\right)\right|\;\leqslant\;C_j\; \expo{-\frac{|x-y|^2}{8\ell_B^2}}
$$
where $C_j:=\frac{\alpha_j}{2\pi\ell_B^2}$ and $\alpha_j:=\max_{\zeta\geqslant0}\left|\expo{-\frac{\zeta}{4}}L^{(0)}_j(\zeta)\right|$. Let $M_g$ be the operator of multiplication by the function $g(x):=\expo{-\frac{|x|^2}{16\ell_B^2}}$. $M_{g}^{-1}$ is the multiplication operator  by the function $1/g(x)$. One  has the identity
$$
\chi_{\Lambda}\Pi_j\;=\;(\chi_{\Lambda} \Pi_j M_{g}^{-1})(M_{g}\Pi_j)\;.
$$
A direct inspection shows that both the operators $\chi_{\Lambda} \Pi_j M_{g}^{-1}$ and $M_{g}\Pi_j$ have an integral kernel which is in $L^2(\R^2\times\R^2)$. Therefore  $\chi_{\Lambda} \Pi_j M_{g}^{-1}$ and $M_{g}\Pi_j$
are Hilbert-Schmidt and   $\chi_{\Lambda}\Pi_j$ is trace class.
The trace of $\chi_{\Lambda}\Pi_j$ can be computed by integrating along the diagonal the integral kernel \eqref{L_proj_ker}. 
The result is exactly $\frac{|\Lambda|}{2\pi\ell_B^2}$.
Since trace class operators form an ideal it follows that $\chi_{\Lambda}\Pi_j \chi_{\Lambda}$ is also trace class. The equality ${\rm Tr}_{L^2(\R^2)}(\chi_{\Lambda}\Pi_j)={\rm Tr}_{L^2(\R^2)}(\chi_{\Lambda}\Pi_j \chi_{\Lambda})$  follows from the cyclicity of the trace.
The claim for $\Pi_j\chi_{\Lambda}$ can be proved with a similar argument.
\qed

\medskip

Let $\Lambda_n\subseteq\R^2$ be any increasing sequence of compact subsets such that $\Lambda_n\nearrow\R^2$ which meet the \emph{F{\o}lner} condition (see \eg \cite{greenleaf-69} for more details). Typical examples for the $\Lambda_n$ are  an increasing sequence of concentric cubes or disks. From Lemma \ref{lemma:trace_unit_vol} it follows that
\begin{equation}\label{eq:T_P_proj}
\s{T}_B( \Pi_j )\;:=\;\lim_{n\to\infty}\frac{1}{|\Lambda_n|}{\rm Tr}_{L^2(\R^2)}\big(\chi_{\Lambda_n} \Pi_j  \chi_{\Lambda_n}\big)\;=\;\frac{1}{2\pi\ell_B^2}\;.
\end{equation}
The quantity $\s{T}_B$ is by definition the trace   {trace per unit volume}.
\begin{remark}[IDOS of $H_B$]{\upshape
The  trace per unit volume is usually used to define the 
\emph{integrate density of states} (IDOS) \cite{veselic-08}.
In the case of the   Hamiltonian $H_B$ the IDOS  is given by \cite[Appendix B]{nakamura-01}
$$
N_B(E)\;:=\;\s{T}_B\big(\chi_E(H_B)\big)\;=\;\frac{1}{2\pi\ell_B^2}\sum_{j=0}^{+\infty}\Theta(E-E_j)
$$
where $\chi_E$ is the characteristic function of the interval $[0,E]$,  $\Theta$ is the Heaviside step function
and the $E_j$ are the energy levels \eqref{eq:spec_H_B}. This result follows immediately from \eqref{eq:T_P_proj}.
}\hfill $\blacktriangleleft$
\end{remark}

\medskip

All the elements of $\s{M}^1_B$ admit a {trace per unit volume}. This is shown in the following result which provides the second formula for the computation fo the integral \ref{eq:int_l1}.
\begin{lemma}\label{lemm:tra_aux2}
The equality
$$
\int(T)\;=\;2\pi\ell_B^2\; \s{T}_B\big( T\big)
$$
holds true for all $T\in\s{M}^1_B$.
\end{lemma}
\proof
The strategy of the proof is the same as in Lemma \ref{lemm:tra_aux1}. 
Let $T\in\s{M}^1_B$. 
There exists a $\{t_k\}_{k\in\N_0}\in\ell^1(\N_0)$ such that
 $T=\sum_{k=0}^\infty t_k\Pi_k$. Moreover, the Schur's property assures that
  $T$ can be obtained as the uniform norm limit of the partial sums $T_N:=\sum_{k=0}^N t_k \Pi_k$.  Let $\Lambda\subset\R^2$ be any compact subsetset.
  From Lemma \ref{lemma:trace_unit_vol} we know that 
  $$
 \|\chi_{\Lambda} \Pi_k \chi_{\Lambda}\|_1\; =\; \frac{|\Lambda|}{2\pi\ell^2_B}
 $$ 
  independently of $k$. This fact can be used to conclude that $\chi_{\Lambda} T \chi_{\Lambda}$ is  a trace class operator. Indeed by linearity $\chi_{\Lambda} T_N \chi_{\Lambda}$ is trace class for all $N\in\N_0$ and  for every $M>N$
 $$
 \left\| \chi_{\Lambda} T_M \chi_{\Lambda} - \chi_{\Lambda} T_N \chi_{\Lambda} \right\|_1 \;\leqslant\; \frac{|\Lambda|}{2\pi\ell^2_B}\sum_{N+1}^M |t_k| 
 $$
  showing that $\{\chi_{\Lambda} T_N \chi_{\Lambda}\}_{N\in\N_0}$ is a Cauchy sequence  with respect to the trace-norm $\|\;\|_1$. Since the space of trace class operators is a Banach (hence closed) space with respect to the trace-norm  it follows that the 
  $\|\;\|_1$-limit of $\chi_{\Lambda} T_N \chi_{\Lambda}$ defines a trace class operator.
  From the uniqueness of the limit and the fact that trace-norm dominates the operator norm
   one gets that $\chi_{\Lambda}T\chi_{\Lambda}$ is indeed a trace class operator. Moreover 
 since the trace is $\|\;\|_1$-continuous one obtains
 $$
 \begin{aligned}
 {\rm Tr}_{L^2(\R^2)}\left(\chi_{\Lambda}T\chi_{\Lambda}\right) \;&=\; \lim_{N \to \infty}{\rm Tr}_{L^2(\R^2)}\left(\sum_{k=0}^N t_k \chi_{\Lambda}\Pi_k \chi_{\Lambda}\right) \\
 &=\; \lim_{N \to \infty} \frac{|\Lambda|}{2\pi\ell_B^2}\sum_{k=0}^N t_k \;=\; \frac{|\Lambda|}{2\pi\ell_B^2}\int(T)\; .
 \end{aligned}
 $$
The last equality, along with the arbitrariness of $\Lambda$, implies the claim.

\medskip
Let $\Omega_B:=\pi\ell_B^2$ be the  area of the magnetic disk of radius $\ell_B$. As a consequence of Lemma \ref{lemm:tra_aux1} and Lemma \ref{lemm:tra_aux2} one obtains the following result:
\begin{theorem}\label{teo:Dix_Tr_for}
The equality
$$
\s{T}_B(T)\;=\;\frac{1}{2\Omega_B} {\rm Tr}_{\rm Dix}\big(Q_{B,\xi}^{-1} T\big)
$$
holds true for all $T\in\s{M}^1_B$, independently of $\xi\geqslant0$.
\end{theorem}


\medskip
\medskip


\end{document}